\documentclass[11pt]{article}
\usepackage[left=1in,top=1in,right=1in,bottom=1in,head=.1in,nofoot]{geometry}

\usepackage{setspace,url,bm,amsmath} 

\usepackage{titlesec} 
\titlelabel{\thetitle.\quad} 

\usepackage[margin=20pt]{subcaption}

\usepackage{amsmath,amsfonts,amsthm,amssymb}
\usepackage{graphicx}
\usepackage{xcolor}
\usepackage[colorlinks=true,linkcolor=black,citecolor=blue,urlcolor=blue]{hyperref}
\usepackage{booktabs}
\usepackage{comment}
\usepackage{multirow}
\usepackage{threeparttable}
\usepackage{arydshln}

\theoremstyle{definition}
\newtheorem{assumption}{Assumption}
\newtheorem*{theorem*}{Theorem}
\newtheorem{theorem}{Theorem}
\newtheorem{proposition}{Proposition}
\newtheorem{lemma}{Lemma}
\newtheorem{remark}{Remark}

\newtheorem{corollary}{Corollary}
\newtheorem*{corollary*}{Corollary}

\newcommand\numberthis{\addtocounter{equation}{1}\tag{\theequation}}

\DeclareFontFamily{U}{mathx}{\hyphenchar\font45}
\DeclareFontShape{U}{mathx}{m}{n}{
      <5> <6> <7> <8> <9> <10>
      <10.95> <12> <14.4> <17.28> <20.74> <24.88>
      mathx10
      }{}
\DeclareSymbolFont{mathx}{U}{mathx}{m}{n}
\DeclareFontSubstitution{U}{mathx}{m}{n}
\DeclareMathAccent{\widecheck}{0}{mathx}{"71}
\DeclareMathAccent{\wideparen}{0}{mathx}{"75}

\usepackage{float}
\usepackage{natbib}
\usepackage{indentfirst}

\def\ncl{n_{\textnormal{cl}}}
\def\st{\mathcal{S}_1}
\def\sc{\mathcal{S}_0}
\def\sth{\mathcal{S}_{h1}}
\def\sch{\mathcal{S}_{h0}}
\def\sumicl{\sum_{i=1}^m}
\def\sumjcl{\sum_{j=1}^{n_i}}
\def\sumicrt{\sum_{i=1}^{n}}
\def\sumicrs{\sum_{i=1}^N}
\def\sumis{\sum_{i\in\mathcal{S}}}
\def\sumist{\sum_{i\in\mathcal{S}_1}}
\def\sumlsc{\sum_{i\in\mathcal{S}_0}}
\def\sumisth{\sum_{i\in\mathcal{S}_{h1}}}
\def\sumlsch{\sum_{i\in\mathcal{S}_{h0}}}
\def\sumistr{\sum_{i=1}^{n_h}}
\def\sumh{\sum_{h=1}^{H}}

\def\nht{n_{h1}}
\def\nhc{n_{h0}}

\def\pih{\pi_{h}}

\def\pht{p_{h1}}
\def\phc{p_{h0}}
\def\phz{p_{hz}}
\def\zhi{Z_{hi}}
\def\whi{w_{hi}}
\def\wht{w_{h}(1)}
\def\whc{w_{h}(0)}
\def\whione{w_{hi}^{*}}
\def\whitwo{w_{hi}^{(2)}}

\def\bhv{\hat{\bar{v}}}

\def\bhyt{\hat{\bar{Y}}(1)}
\def\bhyc{\hat{\bar{Y}}(0)}

\def\bx{\bar{x}}

\def\bhxt{\hat{\bar{x}}(1)}
\def\bhxc{\hat{\bar{x}}(0)}
\def\bhut{\hat{\bar{u}}(1)}
\def\bhuc{\hat{\bar{u}}(0)}
\def\bhrt{\hat{\bar{r}}(1)}
\def\bhrc{\hat{\bar{r}}(0)}

\def\bhxht{\hat{\bar{x}}_{h}(1)}
\def\bhxhc{\hat{\bar{x}}_{h}(0)}
\def\bhyht{\hat{\bar{Y}}_{h}(1)}
\def\bhyhc{\hat{\bar{Y}}_{h}(0)}

\def\bhuht{\hat{\bar{u}}_{h}(1)}
\def\bhuhc{\hat{\bar{u}}_{h}(0)}

\def\bxh{\bar{x}_{h}}
\def\bhvt{\hat{\bar{v}}(1)}
\def\bhvc{\hat{\bar{v}}(0)}
\def\bv{\bar{v}}

\def\byz{\bar{Y}(z)}

\def\byhz{\bar{Y}_{h}(z)}

\def\tXstr{\breve{X}_{\textnormal{str}}}
\def\Xstr{X_{\textnormal{str}}}
\def\tXcrs{\breve{X}_{\textnormal{crs}}}
\def\Xcrs{X_{\textnormal{crs}}}
\def\txi{\breve{x}_i}

\def\txl{\breve{x}_i}

\def\tvi{\breve{v}_i}
\def\tyi{\breve{Y}_i}
\def\tyidz{\check{Y}_{i\cdot}(z)}
\def\tyidt{\check{Y}_{i\cdot}(1)}
\def\tyidc{\check{Y}_{i\cdot}(0)}
\def\txid{\check{x}_{i\cdot}}
\def\txhi{\breve{x}_{hi}}

\def\tehi{\tilde{e}_{hi}}
\def\tel{\tilde{e}_{i}}
\def\tei{\tilde{e}_{i}}
\def\tehl{\tilde{e}_{hi}}
\def\txhl{\breve{x}_{hi}}
\def\tyhi{\breve{Y}_{hi}}

\def\Shx{S^2_{hx}}
\def\taux{\hat{\tau}_x}
\def\htau{\hat{\tau}}
\def\htaustr{\hat{\tau}_{\textnormal{str}}}
\def\taustr{\tau_{\textnormal{str}}}
\def\taucl{\tau_{\textnormal{cl}}}
\def\tauxstr{\hat{\tau}_{\textnormal{str},x}}
\def\taucrs{\tau_{\textnormal{crs}}}
\def\htaucrs{\hat{\tau}_{\textnormal{crs}}}
\def\tauxcrs{\hat{\tau}_{\textnormal{crs},x}}
\def\deltav{\hat{\delta}_{v}}
\def\betacr{\beta^{\textnormal{opt}}_{\textnormal{cr}}}
\def\betastr{\beta^{\textnormal{opt}}_{\textnormal{str}}}
\def\betacrs{\beta^{\textnormal{opt}}_{\textnormal{crs}}}
\def\gammacrs{\gamma^{\textnormal{opt}}_{\textnormal{crs}}}
\def\hbetacr{\hat{\beta}_{\textnormal{cr}}}
\def\hbetat{\hat{\beta}_1}
\def\hbetac{\hat{\beta}_0}
\def\hbetacr{\hat{\beta}_{\textnormal{cr}}}

\def\hbetastrone{\hat{\beta}^{(1)}_{\textnormal{str}}}

\def\hbetacrs{\hat{\beta}_{\textnormal{crs}}}
\def\hgammacrs{\hat{\gamma}_{\textnormal{crs}}}

\def\taulin{\hat{\tau}^{\textnormal{lin}}}

\DeclareMathOperator*{\argmin}{arg\,min}

\def\tautyr{\hat\tau^{\textnormal{tom}}}
\def\taustrtyr{\hat{\tau}^{\textnormal{tom}}_{\textnormal{str}}}
\def\taucrstyr{\hat{\tau}^{\textnormal{tom}}_{\textnormal{crs}}}
\def\taucltyr{\hat{\tau}^{\textnormal{tom}}_{\textnormal{cl}}}

\def\var{\operatorname{var}}

\def\syt{s^2_{1}}
\def\syc{s^2_{0}}
\def\syz{s^2_{z}}
\def\sxt{s^2_{x(1)}}
\def\sxc{s^2_{x(0)}}
\def\sxz{s^2_{x(z)}}
\def\svt{s^2_{v(1)}}
\def\svc{s^2_{v(0)}}
\def\svz{s^2_{v(z)}}
\def\sxvt{s_{xv(1)}}
\def\sxvc{s_{xv(0)}}

\def\svxt{s_{vx(1)}}
\def\svxc{s_{vx(0)}}
\def\svxz{s_{vx(z)}}

\def\sxyc{s_{x0}}
\def\sxyt{s_{x1}}
\def\sxyz{s_{xz}}
\def\svyc{s_{v0}}
\def\svyt{s_{v1}}
\def\svyz{s_{vz}}

\def\shyt{s^2_{h1}}
\def\shyc{s^2_{h0}}
\def\shyz{s^2_{hz}}
\def\shxt{s^2_{hx(1)}}
\def\shxc{s^2_{hx(0)}}
\def\shet{s^2_{he(1)}}
\def\shec{s^2_{he(0)}}
\def\set{s^2_{e(1)}}
\def\sec{s^2_{e(0)}}
\def\shxz{s^2_{hx(z)}}

\def\shxyc{s_{hx0}}
\def\shxyt{s_{hx1}}
\def\shxyz{s_{hxz}}

\def\Syt{S^2_{1}}
\def\Syc{S^2_{0}}
\def\Syz{S^2_{z}}
\def\Sx{S^2_{x}}
\def\Sv{S^2_{v}}
\def\Svx{S_{vx}}
\def\Svyt{S_{v1}}
\def\Svyc{S_{v0}}
\def\Svyz{S_{vz}}
\def\Syxc{S_{0x}}
\def\Syxt{S_{1x}}

\def\Sxyc{S_{x0}}
\def\Sxyt{S_{x1}}
\def\Sxyz{S_{xz}}
\def\Shyt{S^2_{h1}}
\def\Shyc{S^2_{h0}}
\def\Shyz{S^2_{hz}}
\def\Shx{S^2_{hx}}

\def\Shxyc{S_{hx0}}
\def\Shxyt{S_{hx1}}
\def\Shxyz{S_{hxz}}

\def\op{o_{\mathbb{P}}}
\def\Op{O_{\mathbb{P}}}

\def\hbetacr{\hat{\beta}_{\textnormal{cr}}}
\def\hbetastr{\hat{\beta}_{\textnormal{str}}}
\def\hbetacrs{\hat{\beta}_{\textnormal{crs}}}

\def\nstr{n_{\textnormal{str}}}
\def\vttstr{V_{\textnormal{str},\tau\tau}}
\def\vstrxt{V_{\textnormal{str},x\tau}}
\def\vstrtx{V_{\textnormal{str},\tau x}}
\def\vstrxx{V_{\textnormal{str},xx}}
\def\vtxstr{V_{\textnormal{str},\tau x}}

\def\vtx{V_{\tau x}}
\def\vxt{V_{x\tau}}
\def\vxx{V_{xx}}

\def\vtt{V_{\tau\tau}}
\def\vtxstr{V_{\textnormal{str},\tau x}}
\def\vxtstr{V_{\textnormal{str},x\tau}}
\def\vxxstr{V_{\textnormal{str},xx}}
\def\hvxxstr{\hat{V}_{\textnormal{str},xx}}
\def\hvstr{\hat{V}_{\textnormal{str}}}
\def\hvcrs{\hat{V}_{\textnormal{crs}}}
\def\vttstr{V_{\textnormal{str},\tau\tau}}
\def\vcrstt{V_{\textnormal{crs},\tau \tau}}
\def\vcrsxx{V_{\textnormal{crs},xx}}
\def\vcrstx{V_{\textnormal{crs},\tau x}}
\def\vcrsxt{V_{\textnormal{crs},x \tau}}
\def\vcrsvx{V_{\textnormal{crs},v x}}
\def\hvxvxv{\hat{V}_{\textnormal{crs},(x,v)}}
\def\vcrsvt{V_{\textnormal{crs},v \tau}}
\def\vcrsvv{V_{\textnormal{crs},vv}}
\def\vcrsxv{V_{\textnormal{crs},xv}}
\def\vcrstv{V_{\textnormal{crs},\tau v}}

\def\vhwtyrj{\hat{V}^{\textnormal{tom}}_{\textnormal{HC}j}}
\def\vhwlinj{\hat{V}^{\textnormal{lin}}_{\textnormal{HC}j}}
\def\vhwstrj{\hat{V}_{\textnormal{HC}j,\textnormal{str}}}
\def\vhwcrsj{\hat{V}_{\textnormal{HC}j,\textnormal{crs}}}
\def\vhwclj{\hat{V}_{\textnormal{HC}j,\textnormal{cl}}}
\def\hei{\hat{e}_{i}}
\def\hel{\hat{e}_{i}}
\def\hehi{\hat{e}_{hi}}
\def\hehl{\hat{e}_{hl}}

\newcommand\liu[1]{\textcolor{black}{#1}}

\begin{document}
\title{\bf
Tyranny-of-the-minority regression adjustment in randomized experiments}
\author{Xin Lu and Hanzhong Liu\thanks{Corresponding author: lhz2016@tsinghua.edu.cn}\\
{\small Center for Statistical Science, Department of Industrial Engineering, Tsinghua University, Beijing, China}\\
}
\date{}
\maketitle


\begin{abstract}
Regression adjustment is widely used for the analysis of randomized experiments to improve the estimation efficiency of the treatment effect. This paper reexamines a weighted regression adjustment method termed as \textbf{t}yranny-\textbf{o}f-the-\textbf{m}inority (ToM), wherein units in the minority group are given greater weights. We demonstrate that the ToM regression adjustment is more robust than \cite{Lin2013Agnostic}'s regression adjustment with treatment-covariate interactions, even though these two regression adjustment methods are asymptotically equivalent in completely randomized experiments. Moreover, we extend ToM regression adjustment to stratified randomized experiments, completely randomized survey experiments, and cluster randomized experiments. We obtain design-based properties of the ToM regression-adjusted average treatment effect estimator under such designs. In particular, we show that  ToM regression-adjusted estimator improves the asymptotic estimation efficiency compared to the unadjusted estimator even when the regression model is misspecified, and is optimal in the class of linearly adjusted estimators. We also study the asymptotic properties of various heteroscedasticity-robust standard error estimators  and provide recommendations for practitioners. Simulation studies and real data analysis demonstrate ToM regression adjustment’s superiority over existing methods.

\end{abstract}

\section{Introduction}

Since the seminal work of \cite{fisher1935design}, randomized experiments have been the gold standard for drawing causal inference. Complete randomization balances confounding factors on average such that the treatment effects can be identified without untestable assumptions as in observational studies. Different experimental designs have been proposed to improve the efficiency or address practical concerns regarding completely randomized experiments. For example, stratified randomized experiments further balance important discrete covariates and improve the efficiency of treatment effect estimation \citep{Fisher1926,imai2008misunderstandings,imbens2015causal}. Cluster randomized experiments are conducted when the individual-level treatment assignment is logistically unrealistic or when there are concerns regarding interference within clusters \citep{hayes2017cluster}. Completely randomized survey experiments address the lack of generalizability of completely randomized experiments \citep{imai2008misunderstandings,yang2021rejective}.


Regression adjustment is widely used during the analysis stage to utilize covariate information to improve  efficiency. \cite{fisher1935design} used covariates by adding them directly in the linear regression of outcome on treatment indicator and estimated the average treatment effect using the ordinary least squares (OLS). However, \cite{freedman2008regression} criticized this practice by demonstrating that this may degrade efficiency compared to the simple difference-in-means estimator under an unbalanced design or in the presence of heterogeneity between treatment and control groups. Echoing the critique of \cite{freedman2008regression} and to fix the efficiency loss issue, \cite{Lin2013Agnostic} recommended the addition of both covariates and treatment-covariate interactions in the regression adjustment. Since then, Lin's with-interaction regression adjustment has witnessed significant advances in the field of causal inference \citep{bloniarz2016lasso,liu2020regression,2020Rerandomization,ma2020regression,zhao2021covariate,su2021modelassisted,liu-yang-ren-factorial,lei2021regression,zdrep,liu2022lasso,Lu2022,zhao2021reconciling,zdfact}.

However, practitioners may be wary of using the with-interaction regression adjustment because it doubles the degrees of freedom used for the coefficients of covariates \citep{schochet2021design,negi2021revisiting}.
Although this regression adjustment method can be extended to other experimental designs, it may degrade the efficiency compared to the unadjusted estimator \citep{ma2020regression,liu2020regression,liu-yang-ren-factorial,liu2022lasso}. One strategy to remedy this issue is to approach covariate adjustment from the perspective of projection or conditional inference and plug-in unknown projection coefficients using several regressions \citep{yang2021rejective,liu-yang-ren-factorial,wang2021rerandomization}; however, this is more complicated and less robust than the weighted regression adjustment introduced later (see our simulation results). Additionally, heteroskedasticity-robust variance estimators from Lin's with-interaction regression adjustment can be anti-conservative, under the superpopulation framework \citep{negi2021revisiting, zhao2021covariate}, completely randomized survey experiments \citep{yang2021rejective}, or when the dimension of covariates is relatively large compared to the sample size \citep{lei2021regression}.

\cite{Lin2013Agnostic} discussed a weighted regression adjustment method named \textbf{t}yranny-\textbf{o}f-the-\textbf{m}inority (ToM), which embodies the principle of giving more weights to the units in the minority group. This method saves half of the degrees of freedom and is asymptotically equivalent to the with-interaction regression adjustment in completely randomized experiments \citep{Lin2013Agnostic}. 
However, \cite{Lin2013Agnostic} and other follow-up research have not assessed the robustness of the method in completely randomized experiments and potential application in other experimental designs.

To address the gap and drawbacks of the with-interaction regression adjustment, we re-examine the ToM regression adjustment method in completely randomized experiments. We demonstrate the robustness of the ToM regression-adjusted average treatment effect estimator using theoretical justifications and simulation studies. Simulation results reveal that ToM regression adjustment dramatically enhances the estimation efficiency and inference reliability when the design is away from balance or the number of covariates is relatively large compared to the sample size.

ToM regression adjustment can be applied under other experimental designs to enhance the efficiency. We illustrate its use and design-based properties in stratified randomized experiments, completely randomized survey experiments, and cluster randomized experiments. Under mild moment conditions, we show that the ToM regression-adjusted average treatment effect estimator is asymptotically normal and optimal in the class of linearly adjusted estimators for each of the aforementioned experimental designs. Moreover, we study the asymptotic properties of various heteroscedasticity-robust standard error estimators. 
Our analysis is design-based, that is, the analysis is conducted by conditioning on the potential outcomes and covariates, along with treatment assignment as the only source of randomness. Our theoretical results allow the linear regression model to be arbitrarily misspecified. 
Finally, we conduct simulation to evaluate the finite-sample performance of the ToM regression-adjusted estimator. Simulation results demonstrate the superiority of the ToM regression-adjusted estimator compared to existing estimators. Based on the theoretical and finite-sample results, we provide practical suggestions for choosing point and variance estimators to analyze the experimental results. These suggested estimators can be conveniently obtained using off-the-shelf statistical software packages.

The remaining paper is structured as follows. In Section~2, we introduce ToM regression adjustment in the context of completely randomized experiments and compare it with Lin's with-interaction regression adjustment to assess its robustness.
In Section~3, we extend the application of ToM regression adjustment under stratified randomized experiments, demonstrating its optimality for this design. 
In Section~4, we extend ToM regression adjustment and demonstrate its optimality for completely randomized survey experiments. 
In Section~5, we conduct simulation to compare the finite-sample performance of ToM regression adjustment with that of the existing methods. In Section~6, we use ToM regression adjustment to analyze two real datasets. We discuss the combination of ToM regression adjustment and rerandomization in Section~7 and conclude the paper in Section~8. The application of ToM regression adjustment under cluster randomized experiments and proofs are provided in the Supplementary Material.

\section{ToM regression adjustment in completely randomized experiments}
\label{sec:crt}

\subsection{Notation and framework}

Consider a completely randomized experiment with $n$ units. We randomly assign $n_1$ units to the treatment group and $n_0$ to the control group, with $n_0+n_1=n$. Let $Z_i$ be the treatment indicator of the $i$th unit with $Z_i=0$ when it is assigned to the control group and $Z_i=1$ when it is assigned to the treatment group. By design, $\sum_{i=1}^n Z_i=n_1$. Let $\st$ and $\sc$ be the set of units in the treatment and control groups, respectively. We use $Y_i(z)$ to denote the potential outcome of unit $i$
under treatment $z$, for $z=0,1$, with $Y_i=Z_iY_i(1)+(1-Z_i)Y_i(0)$ as the 
observed outcome. Let $x_i = (x_{i1},\ldots,x_{ik})^\top$ be the  
covariates of unit $i$ of length $k$. In a realized experiment, we observe $\{(Y_i,x_i,Z_i)\}_{i=1}^n$. 
We consider a design-based or randomization-based inference framework, under which $\{(Y_i(1),Y_i(0),x_i)\}_{i=1}^n$ are 
all fixed finite-population quantities and treatment assignment, $Z=(Z_1,\ldots,Z_n)$, is the only source of randomness. Throughout the study, we assume the validity of the stable unit treatment value assumption (SUTVA) \citep{Rubin1980}.

Let $\tau_i=Y_i(1)-Y_i(0)$ be the  unit-level treatment
effect. We are interested in the population average treatment effect $\tau=\sum_{i=1}^n \tau_i/n$. An unbiased estimator of $\tau$ is 
the difference in the observed means of the potential outcomes in the treatment and control groups \citep{imbens2015causal}, which is referred
to as the ``difference-in-means" estimator:
\begin{align*}
    \hat\tau = \sum_{i=1}^n Z_i Y_i/n_1-\sum_{i=1}^n (1-Z_i)Y_i/n_0.
\end{align*}

We use the following notation. Let $\byz=n^{-1}\sumicrt Y_i(z)$ $(z=0,1)$ and $\bx=n^{-1}\sumicrt x_i$ be the population means of potential outcomes and covariates, respectively. The population variances and covariances are defined as $$S_{x}^2 = (n-1)^{-1} \sum_{i=1}^n (x_i-\bx)(x_i-\bx)^\top,\quad
S_{xz}=S_{zx}^\top = (n-1)^{-1} \sum_{i=1}^n (x_i-\bx)\{Y_i(z)-\byz\} ,$$
$$S_{z}^2 = (n-1)^{-1}\sum_{i=1}^n \{Y_i(z)-\byz\}^2,\quad S_{\tau}^2 = (n-1)^{-1} \sum_{i=1}^n (\tau_i-\tau)^2,  \quad z=0,1.$$  
Let $\|\cdot\|_\infty$ be the infinity norm of a vector.
Let $Y_i \sim 1 + x_i$ denote the ordinary least squares (OLS) regression of $Y_i$ on $x_i$ with an intercept. Let $ Y_i\stackrel{w_i}{\sim} 1 + x_i $ denote the weighted least squares (WLS) regression of $Y_i$ on $x_i$ with an intercept and weight $w_i$.



\subsection{Regression without and with treatment-covariate interactions}

The covariates may be predictive of the potential outcomes. The difference-in-means estimator does not use the covariate information, which negatively affects the efficiency. Regression adjustment is widely used at the analysis stage to improve the efficiency by adjusting for the covariate imbalance between the treatment and control groups.

The difference-in-means estimator can be derived as the OLS estimator of the coefficient 
of $Z_i$ in the regression $Y_i \sim 1 + Z_i$.  Thus,
the easiest way of 
using covariates, which dates back to \cite{fisher1935design}, is to directly add $x_i$ in the 
regression formula,
$
Y_i\sim 1+Z_i+x_i.
$
The resulting regression-adjusted average treatment effect estimator is the OLS estimator of the coefficient of $Z_i$. We refer to this
regression method as ``Fisher's regression." Fisher's regression has been constantly used in observational studies
 \citep{sloczynski2018general}, completely randomized experiments \citep{negi2021revisiting}, 
 cluster randomized experiments \citep{schochet2021design}, and so on.

\cite{freedman2008regression} criticized Fisher's regression for its lack of guarantee regarding the improvement in efficiency compared to the difference-in-means estimator under unbalanced
design or in the presence of heterogeneity between treatment and control groups. Echoing the critique of \cite{freedman2008regression}, \cite{Lin2013Agnostic}
discussed the possibility of remedying this problem by adding the
treatment-covariate interactions in the regression,
\begin{align}
  \label{formula:cr-lin}
  Y_i\sim 1+Z_i +  ( x_i - \bx )  +Z_i ( x_i - \bx ).
\end{align}
The OLS estimator of the coefficient of $Z_i$, denoted by $\taulin$, is used as the average treatment effect estimator. Note that the covariates must be centered in the interaction term.

\cite{schochet2021design} pointed out that, to include the interaction term, 
 we risk the loss of the degrees of freedom that could seriously reduce power. Researchers may feel uncomfortable in the absence of sufficient degrees of 
 freedom in a with-interaction model that analyzes experiments with 20--100 units, such as clinics and schools, which is very common in development
 economics \citep{negi2021revisiting}. In the same paper,
 \cite{Lin2013Agnostic} commented on  the ToM
regression and demonstrated its asymptotic equivalence to the with-interaction regression for point estimation. However, most of the follow-up work focused on Fisher's regression and Lin's with-interaction regression. ToM 
regression was barely studied. Consequently, it is essential to re-examine 
 ToM regression because it saves half of the degrees of freedom with respective to covariates.

\subsection{ToM regression}

ToM regression accounts for the drawback in Fisher's regression by giving larger weights to the units in the minority group. This regression-adjusted estimator $\tautyr$ is derived as the WLS estimator of the coefficient of $Z_i$ in the regression 
of $Y_i$ on $(1,Z_i,x_i)$ with weights $w_i = Z_i/p_1^2+(1-Z_i)/p_0^2$, 
 where $p_1 = n_1/n$ and $p_0=n_0/n$ are the proportions of units assigned to the
 treatment and control groups, respectively.  
 We denote the WLS regression as 
 \begin{align}
  \label{formula:cr-tyr}
   Y_i\stackrel{w_i}{\sim} 1+Z_i+x_i. 
 \end{align}

\begin{remark}
  \cite{Lin2013Agnostic} used the following weights:
 $Z_i p_0/p_1+(1-Z_i)p_1/p_0$. These are equivalent to the weights $w_i$'s.
 We use $w_i$'s because they can be conveniently extended to other experimental designs.
\end{remark}

\cite{Lin2013Agnostic} observed that $\taulin$ and $\tautyr$ have the same asymptotic distribution. In the remaining of this section, we demonstrate the optimality of $\tautyr$, derive the asymptotic property of its heteroskedasticity-robust standard error, and show that $\tautyr$ is more robust than $\taulin$ through the perspectives of calibrated estimator and leverage score.

In fact, all regression-adjusted estimators are linearly adjusted
estimators, with the following form in completely randomized experiments:
$
\hat{\tau}(\beta) = \hat{\tau}-\beta^\top\taux,$
where $\hat{\tau}_x = \sum_{i=1}^n Z_i x_i/n_1-\sum_{i=1}^n (1-Z_i)x_i/n_0
$ and $\beta$ is some adjusted vector. Let
$\betacr$ correspond to linearly adjusted estimator with minimum sampling
variance, that is, $\var \{\hat{\tau}(\betacr) \} = \min_{\beta \in \mathbb{R}^k} \var\{ \hat{\tau}(\beta) \}$. As shown by \cite{li2017general}, the covariance of  $\sqrt{n}(\hat{\tau}-\tau,\hat{\tau}_x^\top)^\top$ is   
$$
\begin{pmatrix}
  V_{\tau\tau}&V_{\tau x}\\
  V_{x\tau} & V_{xx}
  \end{pmatrix} = 
\begin{pmatrix}
  p_1^{-1}S_{1}^2+p_0^{-1}S_0^2-S_{\tau}^2 & p_1^{-1}S_{1x}+p_0^{-1}S_{0x} \\
  p_1^{-1}S_{x1}+p_0^{-1}S_{x0} & (p_0p_1)^{-1}S_x^2
  \end{pmatrix}.
$$
Simple calculation gives $\betacr = \vxx^{-1}\vxt$ and $\var \{\hat{\tau}(\betacr)\}=\vtt-\vtx\vxx^{-1}\vxt$.
It has been shown that $\taulin$ has the same asymptotic distribution as the optimal 
linearly adjusted estimator $\hat{\tau}(\betacr)$ 
\citep{Lin2013Agnostic,Li9157,2020Rerandomization}. Under mild  conditions, \cite{Lin2013Agnostic} showed that $\taulin$ and $\tautyr$ have the same asymptotic
distribution, and therefore are both optimal.  Proposition~\ref{prop:CRT-consistency} presented below indicates this property.

\begin{assumption}
  \label{a:crt}
  As $n \rightarrow \infty$, for $z=0,1$, (i) $p_z$ has a positive limit; (ii)  $S_{z}^2$,  $S_{x}^2$, $S_{xz}$, $S^2_{\tau}$ have finite limits, the limit of $\var \{\hat{\tau}(\betacr)\}$ is positive and the limit of $S_{x}^2$ is nonsingular; and (iii) $\max_{1 \leq i \leq n} |Y_i(z)-\byz|^2 = o(n)$, $\max_{1 \leq i \leq n} \|x_i-\bx\|_{\infty}^2 = o(n)$. 
\end{assumption}

\begin{proposition}
  \label{prop:CRT-consistency}
  Under Assumption~\ref{a:crt}, both $n^{1/2}(\tautyr-\tau)$ and $n^{1/2}(\taulin-\tau)$ are asymptotically normal with zero mean and variance $\vtt-\vtx\vxx^{-1}\vxt$.
\end{proposition}

The heteroskedasticity-robust standard errors \citep{Huber1967,White1980} are frequently used to approximate the true asymptotic standard errors and can be conveniently obtained by standard statistical software packages. The classical linear regression literature suggests different ways of correcting the degrees of freedom loss, which leads to $\textnormal{HC}_j$ $(j=0,1,2,3)$. $\textnormal{HC}_0$ corresponds to the heteroskedasticity-robust standard error without correction. We have included the explicit formulas of $\textnormal{HC}_j$ $(j=0,1,2,3)$ in the Supplementary Material.

Let $(\vhwtyrj)^{1/2}$ $(j=0,1,2,3)$ be the heteroscedasticity-robust standard error of $\tautyr$ of regression~\eqref{formula:cr-tyr} corresponding to $\textnormal{HC}_j$. Theorem~\ref{prop:valid-CRT-confidence-interval} below depicts the conservativeness of the heteroscedasticity-robust standard error.

\begin{theorem}
  \label{prop:valid-CRT-confidence-interval}
  Under Assumption~\ref{a:crt}, for $j=0,1,2,3$, 
  \begin{align*}
    \vhwtyrj = n^{-1}\min_{\beta}\left[p_1^{-1}S^2_1(\beta)+
    p_0^{-1}S^2_0(\beta)\right] + o_{\mathbb{P}}(n^{-1}),
  \end{align*}
  where $ S^2_z(\beta) = (n-1)^{-1}\sum_{i=1}^n \{Y_i(z)-\byz-(x_i-\bx)^\top \beta\}^2,\ z=0,1.$
\end{theorem}

Let $q_\varsigma $ be the $\varsigma $th quantile of a standard normal distribution. We can construct Wald-type $1 - \alpha$ ($0 < \alpha < 1$) confidence intervals of $\tau$:
$$
\left[\tautyr+(\vhwtyrj)^{1/2} q_{\alpha/2}, \tautyr+(\vhwtyrj)^{1/2}q_{1-\alpha/2}\right], \quad j = 0,1,2,3,
$$
whose asymptotic coverage rates are greater than or equal to $1 - \alpha$.


\begin{remark}
Let  $(\vhwlinj)^{1/2}$ be the heteroscedasticity-robust standard error of $\taulin$ in the with-interaction regression $Y_i\sim 1+Z_i+ ( x_i - \bx ) +Z_i ( x_i - \bx )$. \cite{2020Rerandomization} and \cite{lei2021regression} showed that,
under Assumption~\ref{a:crt},
  \begin{align*}
    \vhwlinj = n^{-1} 
    \left\{p_1^{-1}\min_{\beta_1}S^2_1(\beta_1)+p_0^{-1}\min_{\beta_0}S^2_0(\beta_0)\right\} + o_{\mathbb{P}}(n^{-1}).
  \end{align*}
Since
  $$
  p_1^{-1} 
  \min_{\beta_1}S^2_1(\beta_1)+p_0^{-1}\min_{\beta_0}S^2_0(\beta_0) \leq
  \min_{\beta}\left[p_1^{-1}S^2_1(\beta)+
    p_0^{-1}S^2_0(\beta)\right],
  $$
$\vhwtyrj$ produces a more conservative inference than $(\vhwlinj)^{1/2}$. However, $(\vhwlinj)^{1/2}$ may produce a finite-sample  confidence interval with coverage probability lower than the nominal level when the design is not balanced or the number of covariates is relatively large compared to the sample size; see \cite{lei2021regression} and Section~5. Meanwhile, the classic Neyman-type variance estimator for the difference-in-means estimator is asymptotically equal to $p_1^{-1}S^2_1+
p_0^{-1}S^2_0$. Since
$$
p_1^{-1}S^2_1+
    p_0^{-1}S^2_0 \geq \min_{\beta}\left[p_1^{-1}S^2_1(\beta)+
    p_0^{-1}S^2_0(\beta)\right],
$$
$\vhwtyrj$ still improves the inference efficiency compared to the classic Neyman-type variance estimator. 
\end{remark}

ToM regression is more robust than the with-interaction regression because of the following two reasons. First, both $\tautyr$ and $\taulin$ are special cases of calibrated estimators of the form $\hat{\tau}^{\textnormal{cal}} = \sumist c_i Y_i-\sumlsc c_i Y_i$, where $c_i$'s are the calibrated weights \citep{deville1992calibration,deville1993generalized}.
Let $\hat{\bar{x}}(z)$ be the sample mean of $x_i$ under treatment $z$. As presented in the proof of Theorem~\ref{thm:total-distance-tyr-smaller-than-lin}, the calibrated weights for $\tautyr$ are
  \begin{align*}
    c^{\textnormal{tom}}_i = \begin{cases}
      n_1^{-1} -\taux^\top\left\{p_0^{-2}(n_0-1)\sxc+p_1^{-2}(n_1-1)\sxt\right\}^{-1}\{ x_i-\bhxt \} p_1^{-2},\quad i\in \mathcal{S}_1,\\
      n_0^{-1} + \taux^\top\left\{p_0^{-2}(n_0-1)\sxc+p_1^{-2}(n_1-1)\sxt\right\}^{-1} \{ x_i-\bhxc \} p_0^{-2}, \quad i \in \mathcal{S}_0.
    \end{cases}
  \end{align*}
In contrast, the calibrated weights for $\taulin$ are
  \begin{align*}
    c^{\textnormal{lin}}_i = \begin{cases}
      n_1^{-1} -p_0\taux^\top\left\{(n_1-1)\sxt\right\}^{-1}\{x_i-\bhxt \},\quad i\in \mathcal{S}_1,\\
      n_0^{-1} +p_1\taux^\top\left\{(n_0-1)\sxc\right\}^{-1}\{ x_i-\bhxc \},\quad i\in \mathcal{S}_0.
    \end{cases}
  \end{align*}
We use $c^{\textnormal{tom}}$ and $c^{\textnormal{lin}}$ to denote the vector of $c_i^{\textnormal{tom}}$'s and $c_i^{\textnormal{lin}}$'s, respectively.

The non-calibrated weights used for the difference-in-means estimator are $n_z^{-1}$ $(z=0,1)$ for units in the treatment arm $z$. Although both $\tautyr$ and $\taulin$ are asymptotically optimal, the calibrated weights of $\taulin$ are not satisfactory. For example, negative or large weights may occur, which affect the robustness of the regression-adjusted treatment effect estimator. \cite{deville1992calibration} proposed a distance between the calibrated and non-calibrated weights to measure the calibrated weights' robustness. For complete randomization, the distance measure is derived as 
  \begin{align*}
   F(c) =\sumist G(c_in_1) + \sumlsc G(c_in_0), \quad \text{where}\quad G(x) = (x-1)^2/2.
  \end{align*}
Here $G(c_in_z)$ is the distance between the ratio of the calibrated and non-calibrated weights and $1$. Large value of $F(c)$ suggests the existence of extreme calibrated weights. Theorem~\ref{thm:total-distance-tyr-smaller-than-lin} below indicates that $\tautyr$ is better than $\taulin$ in the sense of embodying non-extreme calibrated weights. In other words, $\tautyr$ makes fewer changes to the calibrated weights than $\taulin$ to achieve the same level of efficiency improvement.

  \begin{theorem}
    \label{thm:total-distance-tyr-smaller-than-lin}
   $
    F(c^{\textnormal{tom}}) \leq F(c^{\textnormal{lin}}).
   $ 
  \end{theorem} 
  
Second, for model-based inference, \cite{huber2004robust} observed that the inverse of leverage score measures the number of units required to determine the fitted value of $Y_i$. The gross error 
is not reflected in the residuals of high leverage score points. Leverage score also plays an important role for design-based inference. High leverage score negatively affects the asymptotic theory and corresponding inferences \citep{dorfman1991sound,lei2021regression}.
Theorem~\ref{thm:leverage-score-tyr-smaller-than-lin} below indicates that $\tautyr$ is  better than $\taulin$ in terms of having smaller leverage score.

  \begin{theorem}
    \label{thm:leverage-score-tyr-smaller-than-lin} Leverage scores in the with-interaction regression are
  \begin{align*}
   h_{i}^{\textnormal{lin}} = \begin{cases}
     n_1^{-1} + \{x_i-\bhxt\}\left\{(n_1-1)\sxt\right\}^{-1}\{x_i-\bhxt\}, \quad  i\in\mathcal{S}_1,\\
     n_0^{-1} + \{x_i-\bhxc\}^\top\left\{(n_0-1)\sxc\right\}^{-1}\{x_i-\bhxc\}, \quad  i\in\mathcal{S}_0.
   \end{cases}
  \end{align*}
In contrast, leverage scores in ToM regression are
\begin{align*}
 h_{i}^{\textnormal{tom}} = \begin{cases}
   n_1^{-1} + \{x_i-\bhxt\}\left\{(n_1-1)\sxt+(p_1/p_0)^2(n_0-1)\sxc\right\}^{-1}\{x_i-\bhxt\}, \quad  i\in\mathcal{S}_1,\\
   n_0^{-1} + \{x_i-\bhxc\}^\top\left\{(n_0-1)\sxc+(p_0/p_1)^2(n_1-1)\sxt\right\}^{-1}\{x_i-\bhxc\}, \quad  i\in\mathcal{S}_0.
 \end{cases}
\end{align*}
Moreover, for $i=1,\ldots,n$,
 $
 h_{i}^{\textnormal{tom}} \leq h_{i}^{\textnormal{lin}}.
 $
  \end{theorem}

Because cluster randomized experiments can be viewed as complete randomized experiments at the cluster level,
we obtained results that correspond to cluster randomized experiments; see the Supplementary Material for more details. In the following two sections, we extend ToM regression to stratified randomized experiments and completely randomized survey experiments, respectively.

\section{ToM regression adjustment in stratified randomized experiments}

Stratified randomized experiments are a combination of 
several completely randomized experiments conducted independently in each stratum. It is natural to extend
ToM regression adjustment to this experimental design. For simplicity, we use the same $k$ to denote covariate dimension. Consider a stratified randomized experiment with $H$ strata.
We use index $h$ to denote quantities with respect to population in stratum $h$ $(h=1,\ldots,H)$,
which leads to the stratum-specific analogs of
$n$, $n_{z}$, $p_z$, $\byz$, $\bx$, $S_{x}^2$, $S_{xz}$, $S_{z}^2$, $z=0,1$, denoted by
$n_h$, $n_{hz}$, $p_{hz}$, $\byhz$, $\bxh$, $S_{hx}^2$, $S_{hxz}$, $S_{hz}^2$, $z=0,1$. Throughout this section, we assume that $2 \leq n_{hz} \leq n_{h} - 2$ for all $h=1,\dots,H$.
We use double index $hi$ $(h=1,\ldots,H,\ i=1,\ldots,n_h)$ to denote unit $i$ in stratum $h$. 
Let $Y_{hi}(z)$ $(z=0,1)$, $Y_{hi}$, $\tau_{hi}$, $x_{hi}$ and $Z_{hi}$ be the potential outcomes, observed outcome, unit-level treatment effect, covariates, and treatment indicator of unit $hi$, respectively. Denote the total population size by $n_{\textnormal{str}}= \sum_{h=1}^H n_{h}$ and
 the proportion of population size of stratum $h$ by $\pi_{h} = n_h/n_{\textnormal{str}}$.
The average treatment effect is
\begin{equation}
  \label{equ:strata-tau}
  \taustr = \nstr^{-1} \sum_{h=1}^H\sum_{i=1}^{n_h} \{Y_{hi}(1)-Y_{hi}(0)\}= \nstr^{-1} \sumh\sumistr \tau_{hi}=\sum_{h=1}^H \pi_h \tau_h,
\end{equation}
where $\tau_h = \sum_{i=1}^{n_h} \tau_{hi}/n_h$ is the average treatment
effect in stratum $h$.

Replacing $\tau_h$ in equation~\eqref{equ:strata-tau} by its unbiased estimator
$$
\hat{\tau}_h = \sum_{i=1}^{n_{h}} Z_{hi}Y_{hi}/n_{h1} - \sum_{i=1}^{n_h} (1-Z_{hi})Y_{hi}/n_{h0},
$$
we obtain an unbiased estimator of $\taustr$, $\htaustr = \sum_{=1}^H \pi_h \hat{\tau}_h$.
As demonstrated by \cite{liu2020regression}, $\htaustr$ is the OLS estimator of the coefficient of $Z_{hi}$ in the following regression:
$$
Y_{hi} \sim 1 + Z_{hi}+\sum_{q=2}^{H} (\delta_{hq}-\pi_q)
+Z_{hi} \sum_{q=2}^{H} (\delta_{hq}-\pi_q),
$$
where $\delta_{hq}$ is the stratum indicator, $\delta_{hq}=1$ if $q=h$ and $\delta_{hq} = 0$ otherwise.

The straightforward extension of Lin's with-interaction regression to stratified randomized experiments is as follows:
$$
Y_{hi} \sim 1 + Z_{hi}+\sum_{q=2}^{H} (\delta_{hq}-\pi_q)
+Z_{hi} \sum_{q=2}^{H} (\delta_{hq}-\pi_q) + (x_{hi}-\bxh)+Z_{hi}(x_{hi}-\bxh).
$$
However, it can be showed that this regression-adjusted estimator, that is, the OLS estimator of the coefficient of $Z_{hi}$, guarantees the improvement of efficiency if the following Assumption~\ref{assum:equal-lin} is true. Otherwise, it may degrade the efficiency.

\begin{assumption}
\label{assum:equal-lin}
(i) Propensity scores are the same across strata, that is, $p_{hz}=p_{1z}$ for all $h=1,\dots,H$, (ii) $n_{h} = n_{1}$ or $n_{h} \rightarrow \infty$ for all $h=1,\dots,H$.
\end{assumption}

Equal propensity scores can be ensured across strata through the design; however, Assumption~\ref{assum:equal-lin}(ii) may be unrealistic in many stratified randomized experiments. To remedy this condition, \cite{liu2020regression} proposed the following weighted regression:
$$
Y_{hi} \stackrel{w^{\textnormal{liu}}_{hi}}{\sim} 1 + Z_{hi}+\sum_{q=2}^{H} (\delta_{hq}-\pi_q)
+Z_{hi} \sum_{q=2}^{H} (\delta_{hq}-\pi_q) + (x_{hi}-\bxh)+Z_{hi}(x_{hi}-\bxh),
$$
where $w^{\textnormal{liu}}_{hi} = Z_{hi}n_h/(n_{h1}-1)+(1-Z_{hi})n_h/(n_{h0}-1)$. 
They demonstrated that the resulting regression-adjusted estimator can guarantee the improvement of efficiency without Assumption~\ref{assum:equal-lin}(ii); however, Assumption~\ref{assum:equal-lin}(i) must still hold true. In this section, we apply ToM regression adjustment to stratified randomized experiments and demonstrate that this regression-adjusted estimator, denoted by $\hat{\tau}_{\textnormal{str}}^{\textnormal{tom}}$, improves the efficiency without Assumption~\ref{assum:equal-lin}.

We define $\hat{\tau}_{\textnormal{str}}^{\textnormal{tom}}$ as the WLS estimator of the coefficient of $Z_{hi}$ in the following weighted regression:
\begin{align}
  \label{formula:stratified}
  Y_{hi} \stackrel{\whi}{\sim} 1 + Z_{hi}+\sum_{q=2}^{H} (\delta_{hq}-\pi_q)
+Z_{hi} \sum_{q=2}^{H} (\delta_{hq}-\pi_q)+
x_{hi},
\end{align}
with weights
$$\whi=\zhi\pht^{-2}\frac{\nht}{\nht-1}+(1-\zhi)\phc^{-2}\frac{\nhc}{\nhc-1}.$$

\begin{remark}
Although the weights $\whione=Z_{hi}/p_{h1}^2 + (1-Z_{hi})/p_{h0}^2$ seem like a straightforward
extension of $w_i$ to stratified randomized experiments, only $\whi$ can guarantee the improvement of  efficiency and optimality of $\hat{\tau}_{\textnormal{str}}^{\textnormal{tom}}$. Moreover, when $\min\{ \nht, \nhc \} \rightarrow \infty$ for $h=1,\dots,H$, $\whi$'s are asymptotically equivalent to $\whione$'s.
\end{remark}

Let $\tauxstr = \sumh \pi_h\hat{\tau}_{hx}$ with $\hat{\tau}_{hx}=\sum_{i=1}^{n_{h}} 
Z_{hi}x_{hi}/n_{h1} - \sum_{i=1}^{n_h} (1-Z_{hi})x_{hi}/n_{h0}$. We define linearly adjusted estimator as $\htaustr(\beta) = \htaustr-\beta^\top\tauxstr$ for some adjusted vector $\beta$. 
By \citet[Proposition 2]{wang2021rerandomization}, $\nstr^{1/2}(\htaustr-\taustr,\tauxstr^\top)^\top$ has mean zero and covariance matrix
\begin{align*}
\begin{pmatrix}
  V_{\textnormal{str},\tau\tau} & V_{\textnormal{str},\tau x}\\
  V_{\textnormal{str},x\tau} & V_{\textnormal{str},xx}
\end{pmatrix} = 
\begin{pmatrix}
  \sumh \pi_h p_{h1}^{-1}S_{h1}^2+\pi_h p_{h0}^{-1}S_{h0}^2-\pi_h S_{h\tau}^2&\sumh \pi_h p_{h1}^{-1}S_{h1x}+\pi_h p_{h0}^{-1}S_{h0x} \\
  \sumh \pi_h p_{h1}^{-1}S_{hx1}+\pi_h p_{h0}^{-1}S_{hx0} & \sumh\pi_h (p_{h0}p_{h1})^{-1}S_{hx}^2
\end{pmatrix}.
\end{align*}
Let $\betastr$ be the optimal linear projection coefficient defined as $\betastr = \argmin_{\beta}\var\{\htaustr(\beta)\}$. Through simple calculation, we obtain $\betastr = \vstrxx^{-1}\vstrxt$, with $\var\{\htaustr(\betastr)\}= V_{\textnormal{str},\tau\tau}-\vstrtx\vstrxx^{-1}\vstrxt$.

To investigate the asymptotic normality and optimality of $\taustrtyr$, we require Assumption~\ref{a:strata} below.
\begin{assumption}
  \label{a:strata}
  As $\nstr \rightarrow \infty$, for $z=0,1$,
  \begin{itemize}
    \item[(i)]  $c\leq  \min_{1 \leq h \leq H} p_{h1}\leq \max_{1 \leq h \leq H} p_{h1}\leq 1-c$ for some constant $c \in (0,0.5]$ independent of $\nstr$;
    \item[(ii)]  $\sum_{h=1}^H \pi_{h}p_{hz}^{-1}S^2_{hz}$, $\sum_{h=1}^H \pi_{h}(p_{h1}p_{h0})^{-1}S^2_{hx}$, 
    $\sum_{h=1}^H \pi_{h}p_{hz}^{-1} S_{hxz}$, $\sum_{h=1}^H\pi_{h}S^2_{h\tau}$ have limiting values, the limit of $\var\{\htaustr(\betastr)\}$ is positive and 
    the limit of $\sum_{h=1}^H \pi_{h}(p_{h1}p_{h0})^{-1}S^2_{hx}$ is nonsingular;
    \item[(iii)] $\max_{1 \leq h \leq H} \max_{1 \leq i \leq n_{h}} |Y_{hi}(z)-\byhz|^2=o(\nstr)$, $\max_{1 \leq h \leq H} \max_{1 \leq i \leq n_{h}} \|x_{hi}-\bxh\|^2_{\infty}=o(\nstr)$.
  \end{itemize}
\end{assumption}




Assumption~\ref{a:strata} is quite general, with few requirements related to the number of strata, stratum sizes, and propensity scores across strata.

\begin{theorem}
  \label{thm:tyr-strata-consistensy}
  Under Assumption~\ref{a:strata}, $\nstr^{1/2}(\taustrtyr-\taustr)$ is asymptotically normal 
  with zero mean and variance $\vttstr-\vstrtx\vstrxx^{-1}\vstrxt$. Moreover, $\taustrtyr$ is optimal with minimum asymptotic variance in the class of linearly adjusted estimators $\{\htaustr(\beta):\beta \in \mathbb{R}^{k}\}$.
\end{theorem}

Let $\vhwstrj$ $(j=0,1,2,3)$ denote the variance estimator of $\taustrtyr$ from the regression formula~\eqref{formula:stratified}  corresponding to $\textnormal{HC}_j$. Theorem~\ref{thm:HC-str-limit} below presents the asymptotic property of $\hat{V}_{\textnormal{HC}2,\textnormal{str}}$. 
\begin{theorem}
  \label{thm:HC-str-limit}
  Under Assumption~\ref{a:strata}, 
  \begin{align}
    \label{eq:vhwstr-limit}
    \hat{V}_{\textnormal{HC}2,\textnormal{str}}  =\min_{\beta}\nstr^{-1}\sum_{h=1}^H\left\{ 
\pi_h p_{h1}^{-1}S_{h1}^2(\beta)+\pih p^{-1}_{h0}S_{h0}^2(\beta)\right\} +o_{\mathbb{P}}(\nstr^{-1}),
  \end{align}
  where
$$
S^2_{hz}(\beta) = (n_h-1)^{-1}\sumistr \{Y_{hi}(z)-\byhz-(x_{hi}-\bxh)^\top \beta\}^2.
$$
\end{theorem}


The variance of $\htaustr(\betastr)$ can be derived by replacing $Y_{hi}(z)$ by $Y_{hi}(z)-x_i^\top\betastr$ 
  in the formula of $\var(\htaustr)$.
   The optimality of $\betastr$ implies that
   \begin{equation}
     \label{eq:optimal-CRT-variance}
     \var\{\htaustr(\betastr)\}=\min_{\beta}\nstr^{-1}\sum_{h=1}^H\left\{\pih p_{h1}^{-1}S_{h1}^2(\beta)+\pih p^{-1}_{h0}S_{h0}^2(\beta)-S_{h\tau}^2\right\}.
   \end{equation} 
  Equations~\eqref{eq:vhwstr-limit} and \eqref{eq:optimal-CRT-variance} indicate that $\hat{V}_{\textnormal{HC}2,\textnormal{str}}$ is an asymptotic conservative 
  estimator of $\var\{\htaustr(\betastr)\}$.
  Since $\hat{V}_{\textnormal{HC}2,\textnormal{str}}\leq\hat{V}_{\textnormal{HC}3,\textnormal{str}}$, $\hat{V}_{\textnormal{HC}3,\textnormal{str}}$ is also a conservative estimator. 
  Therefore, the Wald-type confidence intervals
  $$
  \left[\tautyr+(\vhwstrj)^{1/2} q_{\alpha/2}, 
  \tautyr+(\vhwstrj)^{1/2}q_{1-\alpha/2}\right], \quad j=2,3,
  $$
  have asymptotic coverage rates greater than or equal to $ 1-\alpha$.

\begin{remark}
  \label{remark:anti-conservative-stratified}
    $\hat{V}_{\textnormal{HC}j,\textnormal{str}}$ $(j=0,1)$ can be anti-conservative and produce invalid confidence intervals. See the Supplementary Material for more details.
\end{remark}


\section{ToM regression adjustment in completely randomized survey experiments}

Survey experiments usually comprise two stages: random sampling of units from a target population and random assignment of treatments to the sampled units. These experiments are widely used for estimating treatment effect of a target population \citep{imai2008misunderstandings}. 
The standard survey experiments, completely randomized survey experiments, conduct simple random sampling without replacement to obtain a subset of units before assignment of sampled units through complete randomization
into different treatment arms; see, for example \citet[chap. 6]{imbens2015causal} and \citet{yang2021rejective}.

In a completely randomized survey experiment, suppose
 $n$ units in the experiment are a simple random sample without replacement from a target population of size $N$, with sampling fraction $f=n/N$. When $f=1$, it reduces to the completely randomized experiment.
   Let $R_i$ and $Z_i$ be the sampling indicator and treatment assignment
  indicator with $R_i=1$ if unit $i$ is sampled, and $0$ otherwise, and $Z_i=1$ if unit $i$ is assigned to 
  the treatment group, and $0$ otherwise. Denote the set of the sampled units by $\mathcal{S}=\{i \in\{1,\ldots,N \}:R_i=1\}$.
  By design, $Z_i$ is not defined if $i\not\in\mathcal{S}$. Let $Y_i=Z_iY_i(1)+(1-Z_i)Y_i(0)$ be the 
  observed potential outcome for the sampled unit $i$.
  The average treatment effect of interest in completely randomized survey experiments is $\taucrs = \sum_{i=1}^N \{Y_i(1)-Y_i(0)\}/N$. The difference-in-means estimator
  $\htaucrs=\sum_{i\in\mathcal{S}}Z_iY_i/n_1- \sum_{i\in\mathcal{S}} (1-Z_i)Y_i/n_0$ is an unbiased estimator of $\tau$ \citep{imbens2015causal,yang2021rejective}. Here $n_1 = \sum_{i \in \mathcal{S}}Z_i$ and $n_0 = \sum_{i \in \mathcal{S}}(1-Z_i)$ are the (fixed) numbers of treated and control units, respectively. Let $p_z = n_z / n$ ($z=0,1$).

We can observe two kinds of covariates: $v_i \in \mathbb{R}^{k_1}$ ($1\leq i\leq N$) which is available at the sampling stage and usually collected from baseline survey conducted by some investigators or previous studies on the same target population, and $x_i \in \mathbb{R}^{k_2}$ $(i\in \mathcal{S})$ which is available at the treatment assignment stage and usually collected after the experiment units are sampled. Here, $v_i$ can be a subset of $x_i$.


By a slight abuse of the notation, we define the following finite-population quantities of the $N$ units. 
  We use $S^2_{z}$, $S^2_x$, $\Sv$, $S^2_{\tau}$ to denote
  corresponding finite-population variances and $S_{xz}=S_{zx}^\top$, $\Svyz = S_{zv}^\top$, $\Svx$, $S_{x \tau} = S_{\tau x}^\top$, $S_{v \tau} = S_{ \tau v}^\top$
  to denote the corresponding finite-population covariances. Let $\byz$, $\bx$, $\bv$ 
  be the finite-population means. 

  To motivate the form of weighted regression adjustment, we consider a general form of linearly adjusted 
  estimator proposed by \cite{yang2021rejective}:
  $$
  \htaucrs(\beta,\gamma) = \htaucrs-\beta^\top\tauxcrs-\gamma^\top\deltav,
  $$
  where
  $$
  \tauxcrs = \sum_{i\in\mathcal{S}}Z_ix_i/n_1- \sum_{i\in\mathcal{S}}(1-Z_i)x_i/n_0,\quad 
  \deltav = \bhv-\bv,\quad \bhv = \sum_{i\in\mathcal{S}}v_i/n.
  $$
  The linearly adjusted estimator adjusts two kinds of covariate imbalances: 
  the difference between the sample mean and population mean of the covariates measured by $\deltav$, and the
  difference between the covariate means in the treatment and control groups measured by $\tauxcrs$. Note that 
  $\htaucrs(\beta,\gamma)$ is equal to the difference-in-means estimator applied to the observed
  adjusted potential outcomes,
  \begin{equation}
    \label{eq:adjusted-outcome-2stage}
    Y_i(z;\beta,\gamma) = Y_i(z)-(z-p_0)(v_i-\bv)^\top\gamma-x_i^\top\beta.
  \end{equation}
  
  Equation~\eqref{eq:adjusted-outcome-2stage} catalyzes the use of covariates $(z-p_0)(v_i-\bv)$ and $x_i$ in the regression adjustment. Therefore, 
  we propose a WLS regression adjustment of the following form
\begin{align}
  \label{formula:crs}
  Y_i \stackrel{w_i}{\sim} 1+ Z_i + x_i + (Z_i-p_0)(v_i-\bv)
\end{align}
  with weights $w_i = Z_i/p_1^2+(1-Z_i)/p_0^2$. Define
  $\taucrstyr$ as the estimated coefficient of $Z_i$ through the WLS.

 \begin{remark}
    Note that the regression formula only needs to center $v_i$ at its finite-population mean $\bv$. In practice, it is very difficult to collect $v_i$ and $x_i$ for the units that are not in the sample, that is, $i \notin \mathcal{S}$. Fortunately, $\bv$ is still available from some baseline surveys. Thus, ToM regression adjustment is still applicable.
  \end{remark}

  By \citet[][Lemma B1]{yang2021rejective}, $n^{1/2}(\htaucrs-\tau_{\textnormal{crs}}, \tauxcrs^\top, \deltav^\top )^{\top}$ has mean zero and covariance
  $$
  \left(\begin{array}{ccc}
  \vcrstt & \vcrstx & \vcrstv \\
  \vcrsxt & \vcrsxx  & \vcrsxv\\
  \vcrsvt & \vcrsvx & \vcrsvv
  \end{array}\right)=\left(\begin{array}{ccc}
  p_{1}^{-1} \Syt+p_{0}^{-1} \Syc-f S_{\tau}^{2} & p_{1}^{-1} \Syxt+p_{0}^{-1} \Syxc & (1-f) S_{\tau v} \\
  p_{1}^{-1} \Sxyt+p_{0}^{-1} \Sxyc & \left(p_{1} p_{0}\right)^{-1} \Sx & 0 \\
  (1-f) S_{v\tau} & 0& (1-f) \Sv
  \end{array}\right).
  $$
  The optimal projection coefficients $\betacrs = \vcrsxx^{-1}\vcrsxt$
  and $\gammacrs = \vcrsvv^{-1}\vcrsvt$ produce the minimum variance,
  $$\var\{\htaucrs(\betacrs,\gammacrs)\}
  =\vcrstt-\vcrstx\vcrsxx^{-1}\vcrsxt-\vcrstv\vcrsvv^{-1}\vcrsvt.$$
  Under Assumption~\ref{a:crs} below, we demonstrate the asymptotic normality and optimality of $\taucrstyr$
  in Theorem~\ref{thm:tyr-crs-consistensy}. 

  \begin{assumption}
    \label{a:crs}
    As $n \rightarrow \infty$, for $z=0,1$,
    \begin{itemize}
      \item[(i)] $f$ has a limit in $[0,1)$ and $p_1$ has a limit in $(0,1)$;
      \item[(ii)] $S^2_{z}$, $\Svyz$, $S_{xz}$,  $S^2_\tau$, $\Sv$, $S^2_{x}$ have finite limits, and the limit of $\var\{\htaucrs(\betacrs,\gammacrs)\}$ is positive 
      while the limits of $\Sv$ and $S^2_{x}$ are nonsingular;
      \item[(iii)] $\max_{i=1}^N |Y_i(z)-\byz|^2 = o(n)$, 
      $\max_{i=1}^N \|x_i-\bx\|_{\infty}^2 = o(n)$, $\max_{i=1}^N \|v_i-\bv\|_{\infty}^2 = o(n)$.
    \end{itemize}
  \end{assumption}

  \begin{theorem}
    \label{thm:tyr-crs-consistensy}
    Under Assumption~\ref{a:crs}, $n^{1/2}(\taucrstyr-\taucrs)$ is asymptotic normal with zero mean and variance $\vcrstt-\vcrstx\vcrsxx^{-1}\vcrsxt-\vcrstv\vcrsvv^{-1}\vcrsvt$. Moreover, $\taucrstyr$ is optimal with minimum asymptotic variance in the class of linearly adjusted estimators $\{\htaucrs(\beta,\gamma): \beta \in \mathbb{R}^{k_2}, \gamma \in \mathbb{R}^{k_1}\}$.
  \end{theorem}

 We can estimate the variance of $\taucrstyr$ by the heteroscedasticity-robust standard error.
  Let  $\vhwcrsj$ $(j=0,1,2,3)$ denote the variance estimator of $\taucrstyr$ from the regression formula~\eqref{formula:crs}  corresponding to $\textnormal{HC}_j$.

\begin{theorem}
  \label{prop:vhwcrs-consistent}
Under Assumption~\ref{a:crs}, for $j=0,1,2,3$, 
\begin{equation}
  \label{eq:vcrstyr-limit}
  \vhwcrsj =  n^{-1} \min_{\beta,\gamma} \left\{ p_{1}^{-1}S_{1}^2(\beta,\gamma)+p^{-1}_{0}S_{0}^2(\beta,\gamma)\right\} +o_{\mathbb{P}}(n^{-1}),
\end{equation}
where
$$
S^2_{z}(\beta,\gamma) = (N-1)^{-1}\sumicrs \{Y_{i}(z)-\byz-(x_{i}-\bx)^\top \beta-(z-p_0) (v_i-\bv)^\top \gamma\}^2.
$$
\end{theorem}

It is easy to show that  $\var\{\htaucrs(\betacrs,\gammacrs)\}$ can be derived by replacing $Y_{i}(z)$ by the adjusted potential outcome 
  $Y_i(z;\betacrs,\gammacrs)$
  in the formula of $\vttstr$.
   The optimality of $(\betacrs,\gammacrs)$ implies that
   \begin{equation}
     \label{eq:optimal-crs-variance}
     \var\{\htaucrs(\betacrs,\gammacrs)\}=n^{-1} \min_{\beta,\gamma} \left\{ p_{1}^{-1}S_{1}^2(\beta,\gamma)+p^{-1}_{0}S_{0}^2(\beta,\gamma)-fS^2_{\tau}(\gamma)\right\},
   \end{equation} 
   where 
   $$
   S^2_{\tau}(\gamma) = (N-1)^{-1}\sumicrs \{\tau_i-\tau-(v_i-\bv)^\top \gamma \}^2.
   $$
  Equations~\eqref{eq:vcrstyr-limit} and~\eqref{eq:optimal-crs-variance}  indicate that  
  $\vhwcrsj$ is an asymptotic conservative 
  estimator of $\var\{\htaucrs(\betacrs,\gammacrs)\}$, and thus an asymptotic conservative 
  estimator of $\var(\taucrstyr)$. Therefore, the Wald-type confidence intervals
  $$
  \left[\taucrstyr+(\vhwcrsj)^{1/2} q_{\alpha/2}, 
  \taucrstyr+(\vhwcrsj)^{1/2}q_{1-\alpha/2}\right], \quad j=0,1,2,3,
  $$
  have asymptotic coverage rates greater than or equal to $ 1-\alpha$.

With the assumption that the units are a random sample from an infinite superpopulation, \cite{negi2021revisiting} demonstrated that the variance estimator constructed by the with-interaction regression is anti-conservative if the covariates $x_i$ are not centered at their finite-population mean but at their sample mean which introduces an extra variability. This conclusion holds for completely randomized survey experiments with $ 0 < f<1$. In practice, $\bx$ is often not available; consequently, the with-interaction regression adjustment is not applicable for $ 0 < f<1$.  In contrast, ToM regression adjustment does not require the centering of covariates $x_i$ at $\bx$. The resulting point estimator
  is consistent and asymptotically normal and the variance estimator is asymptotically conservative 
  regardless of $f$.

\section{Numerical studies}
In this section, we compare the finite-sample performance of the point estimator and variance estimator derived by 
ToM regression adjustment with existing competitors in the literature in completely randomized experiments, stratified randomized experiments, and completely randomized survey experiments.

\subsection{Complete randomized experiments}

In his seminal paper, \cite{Lin2013Agnostic} demonstrated the equivalence of with-interaction regression adjustment and ToM
 regression adjustment in a low-dimensional and large-sample setting that the asymptotic theory works perfectly.
  In this section, we consider a relatively large dimension of covariates compared to the sample size. We further investigate how ``imbalance in information" between treatment and control groups can influence the performance of the estimators, which is reflected by $p_z$ and the signal-to-noise ratio defined later.

We set $n=100$ and generate data using the following model:
\begin{equation}
  \label{model:crt}
  Y_i(z) = f_z(x_i)+e_{i}(z),~ \text{with}~ f_z(x_i)=\alpha_z + x_{i}^
\top\beta_z,\quad z=0,1,\quad i=1,\ldots,n,
\end{equation}
where $(\alpha_z,\beta_z)$ has independent and identically distributed (i.i.d.) entries generated from $t_3$, $t$-distribution with three degrees of freedom, for $z=0,1$.  Thus, there is heterogeneity between 
treatment and control groups. The covariates $x_i$'s
are realizations of independent random vectors drawn from $\mathcal{N}(0,\Sigma)$ with 
$\Sigma_{ij} = 0.6\delta_{ij}+0.4$ ($1\leq i,j\leq k$), where $\delta_{ij}=1$ if $i=j$, and $0$ otherwise. The errors $e_{i}(z)$'s are realizations of i.i.d. normal random variables with zero mean and variance fulfilling a given 
signal-to-noise ratio $\textnormal{SNR}z$, that is, ratio of the finite-population variance of 
$f_z(x_i)$ to that of $e_{i}(z)$. 
After generation, $\{(Y_i(1), Y_i(0), x_{i})\}_{i=1}^N$ are fixed. 
The treatment assignment stage assigns $n_1=p_1n$ units to the
treatment group and a completely randomized experiment is simulated $1000$ times.

We focus on the root mean squared errors (RMSE) of point estimators and empirical coverage probabilities of $95\%$ confidence intervals.
We vary the $\textnormal{SNR}z$ $(z=0,1)$, $k$, and $p_1$ in each scenario. Table~\ref{tab:params} presents the values of the factors 
considered in the simulation. Each scenario
is repeated under $100$ different random seeds.

\begin{table}
  \caption{\label{tab:params}Parameters in simulation}
  \centering
  \begin{tabular}{ccc}
    \toprule
    \multirow{5}{*}{completely  randomized  experiments} & $p_1$ & $0.3,0.4,0.5$ \\
    &random seed      &  $1:100$ \\
    &$k$        & $\left\{1, 5, 9 ,13, 17, 21, 25, 29\right\}$ \\
    &$\textnormal{SNR}0$        &       $\left\{0.25,0.5,1,2\right\}$    \\
    &$\textnormal{SNR}1$          &     $\left\{0.25,0.5,1,2\right\}$    \\
    \midrule
   \multirow{5}{*}{stratified randomized experiments} &strata & $\left\{\textnormal{MS},\textnormal{FL}, \textnormal{MS+FL}\right\}$\\
    &random seed      &  $1:100$ \\
    &$k$        & $\left\{1, 5, 9 ,13, 17, 21, 25, 29\right\}$ \\
    &$\textnormal{SNR}0$        &       $\left\{0.25,0.5,1,2\right\}$    \\
    &$\textnormal{SNR}1$          &     $\left\{0.25,0.5,1,2\right\}$    \\
    \midrule
    \multirow{4}{*}{completely randomized survey experiments}
    &random seed      &  $1:100$ \\
    &$k$        & $\left\{2,5,8,11,14,17\right\}$ \\
    &$\textnormal{SNR}0$        &       $\left\{0.25,0.5,1,2\right\}$    \\
    &$\textnormal{SNR}1$          &     $\left\{0.25,0.5,1,2\right\}$    \\
    \bottomrule
  \end{tabular}
\end{table}

\begin{figure}[ht]
  \includegraphics[width = \textwidth]{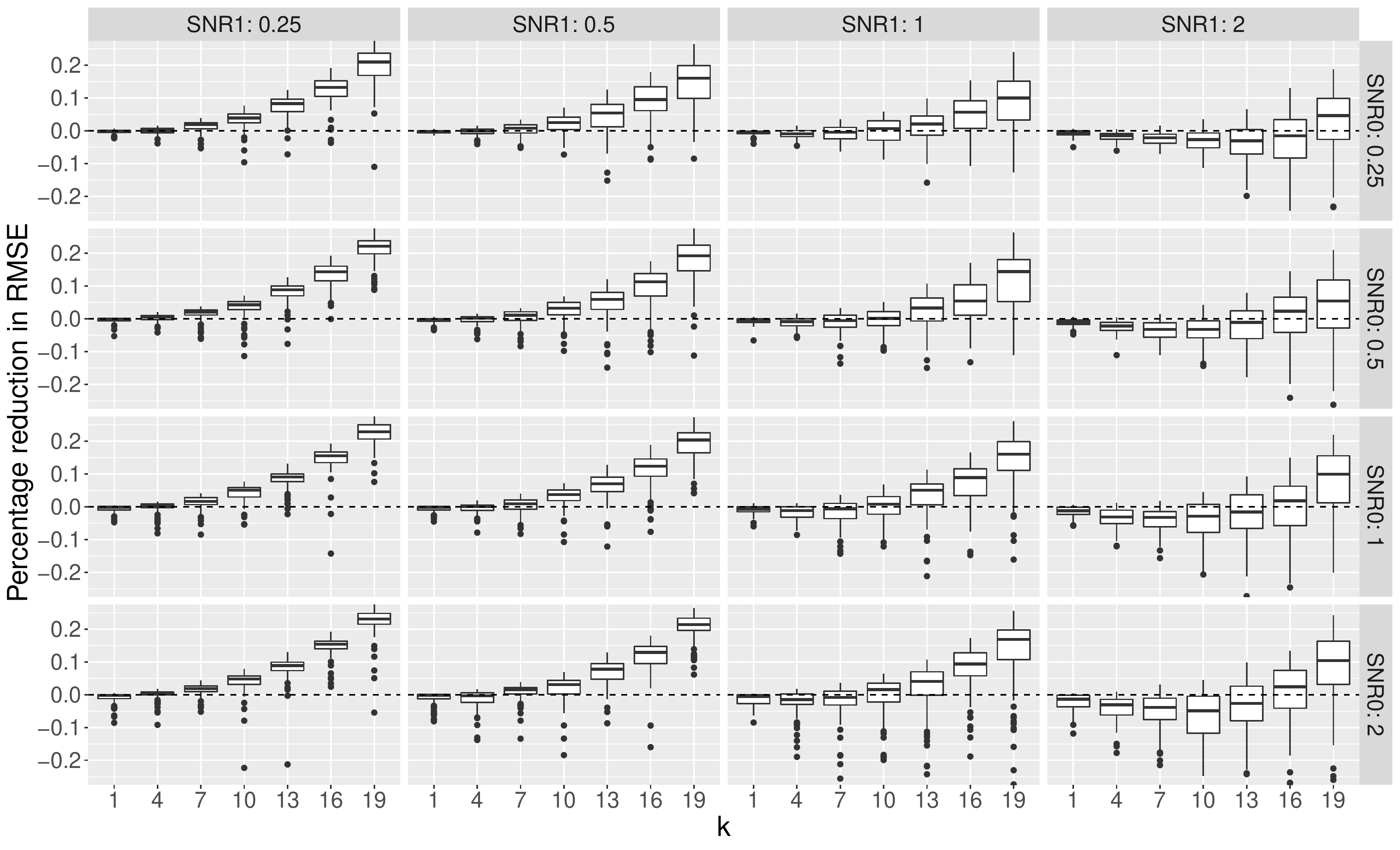}
  \caption{Percentage reduction in RMSE of $\tautyr$ versus $\taulin$ when $p_1=0.3$.}
  \label{figure:prmse0.3}
\end{figure}

Figure~\ref{figure:prmse0.3} depicts the percentage
reduction in RMSE of $\tautyr$ versus $\taulin$, that is, $\operatorname{RMSE}(\taulin)/\operatorname{RMSE}(\tautyr)-1$ when $p_1=0.3$. The results for $p_1=0.4,0.5$ are provided in the Supplementary Material. It can be observed that the RMSE of $\tautyr$ is overall smaller than that of $\taulin$ and the percentage
reduction in RMSE increases as $k$ becomes larger. ToM regression-adjusted estimator $\tautyr$ is clearly advantageous when the majority group (control group) has a larger SNR and the minority group (treatment group) has a smaller SNR. 
This is because $\tautyr$ uses the data from both groups
in a pooled fashion, with larger weights bestowed to the minority group and $\taulin$ in a separate fashion with equal weights. Therefore, the performance of $\taulin$ 
heavily depends on how well it estimates the adjusted coefficient in the minority group. 
When the minority group has a small SNR, the adjusted coefficient may be poorly estimated by $\taulin$.

\begin{figure}[ht]
  \includegraphics[width = \textwidth]{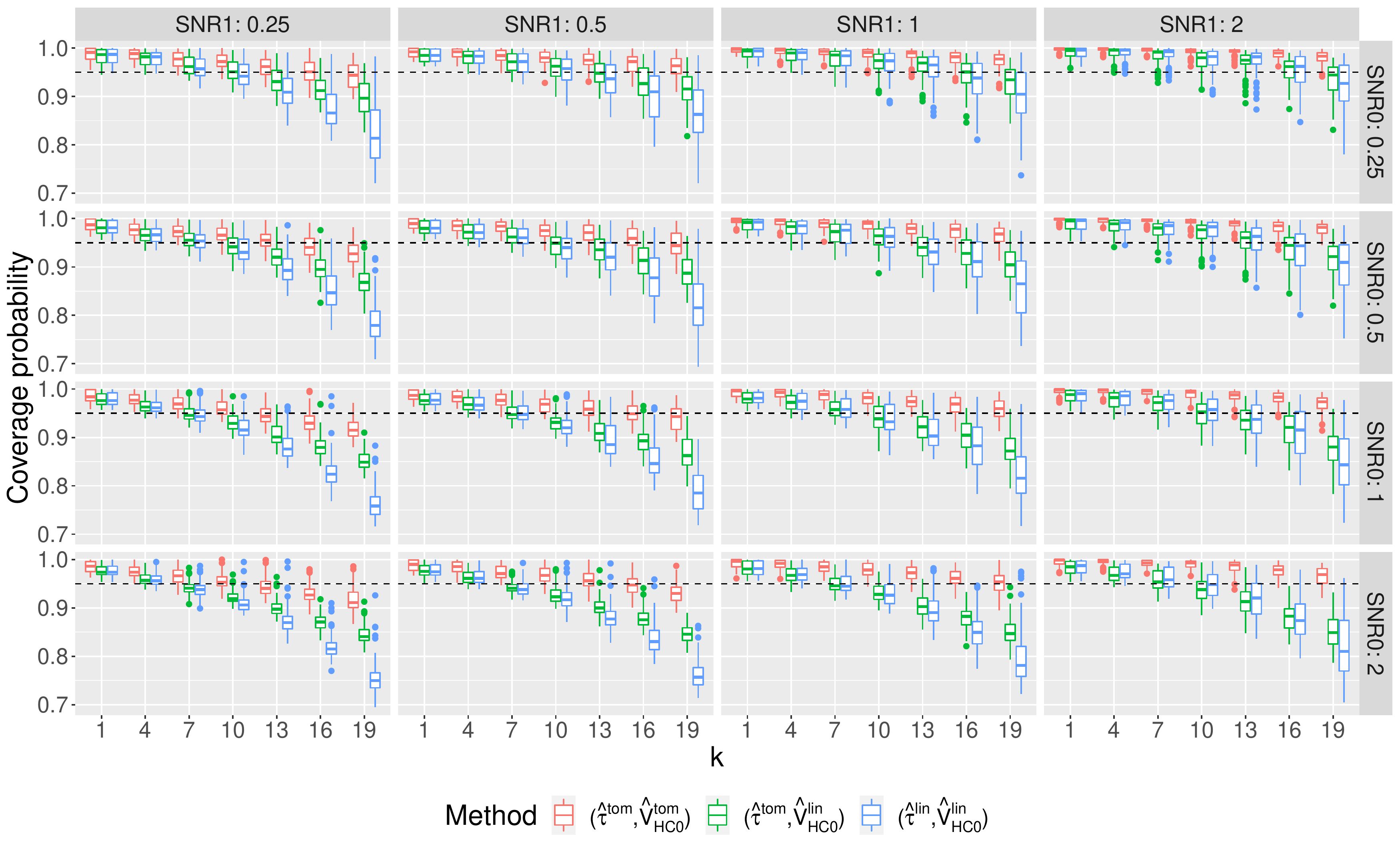}
  \caption{Coverage probabilities for $p_1=0.3$ in completely randomized experiments.}
  \label{figure:coverage0.3}
\end{figure}

Figure~\ref{figure:coverage0.3} depicts the coverage probabilities of $95\%$ confidence intervals constructed by 
$(\hat{\tau}^{\textnormal{tom}},\hat{V}_{\textnormal{HC}0}^{\textnormal{tom}})$, 
$(\hat{\tau}^{\textnormal{tom}},\hat{V}_{\textnormal{HC}0}^{\textnormal{lin}})$,
and
$(\hat{\tau}^{\textnormal{lin}},\hat{V}_{\textnormal{HC}0}^{\textnormal{lin}})$
when $p=0.3$. 
It can be observed that these three methods tend to have worse coverage probabilities when $k$ becomes larger. 
Combination of $(\tautyr,\hat{V}_{\textnormal{HC}0}^{\textnormal{tom}})$ is the most robust under all scenarios. Similar results were observed by \cite{lei2021regression}: $\hat{V}_{\textnormal{HC}0}^{\textnormal{lin}}$ tends to underestimate the
variance for large $k$. In contrast, $\hat{V}_{\textnormal{HC}0}^{\textnormal{tom}}$ provides a better variance estimation for large $k$. Combination of 
$(\hat{\tau}^{\textnormal{tom}},\hat{V}_{\textnormal{HC}0}^{\textnormal{tom}})$ has larger coverage probabilities on average than
$(\hat{\tau}^{\textnormal{lin}},\hat{V}_{\textnormal{HC}0}^{\textnormal{lin}})$ when $k$ is large. Therefore, its use is recommended.
If one prefers a less conservative inference when $k$ is small, combination of 
$(\hat{\tau}^{\textnormal{tom}},\hat{V}_{\textnormal{HC}0}^{\textnormal{lin}})$ is recommended. 
Moreover, all combinations have better coverage probabilities
if the minority group has a larger SNR and majority group has a smaller SNR. In contrast,
all combinations have worse coverage probabilities
if the minority group has a smaller SNR and majority group has a larger SNR. 


\subsection{Stratified randomized experiments}
We consider three kinds of strata 
size distributions: (1) many small strata (MS); (2) a few large strata (FL); and (3) many small
strata compounded with a few large strata (MS+FL). For each scenario, strata sizes $\left\{n_{h}\right\}_{h=1}^H$ are generated 
as independent samples with
(1) $H=20$ from uniform distribution on $\{10,11,\dots,20\}$; (2)
$H=2$ from uniform distribution on $\{140,141,\dots,160\}$; and (3) $H=12$ with $10$ strata sizes from uniform distribution on $\{10,11,\dots,20\}$
and $2$ strata sizes from uniform distribution on $ \{140,141,\dots,160\}$.


The potential outcomes are generated from the following random effect model:
$$
Y_{hi}(z) = f_{hz}(x_{hi})+e_{hi}(z),~ \text{with}~ f_{hz}(x_{hi})=\alpha_{hz} + x_{hi}^
\top\beta_{hz},
$$
$$ z=0,1,\quad h=1,\ldots,H, \quad i=1,\ldots,n_h, 
$$
where the intercepts and slopes are generated by
$\beta_{hz}=\beta_{z}+\zeta_{hz}$ and $\alpha_{hz} = \alpha_{z}+\eta_{hz}$ with $(\alpha_{z},\beta_{z})$ and $(\eta_{hz},\zeta_{hz})$ $(z=0,1)$ embodying i.i.d. entries generated from $t_3$ and standard normal distribution, respectively. The covariates $x_i$'s
are realizations of independent random vectors of length $k$ from ${N}(0,\Sigma)$ with 
$\Sigma_{ij} = 0.6\delta_{ij}+0.4$. $e_{hi}(z)$'s are realizations of i.i.d. normal random variables with zero mean and variance fulfilling a given signal-to-noise ratio $\textnormal{SNR}z$, that is, the ratio of the finite-population variance of 
$f_{hz}(x_{hi})$ to that of $e_{hi}(z)$. 

We ensure at least two units in each treatment arm for each stratum. 
The number of units assigned to treatment $n_{h1}$'s are generated by
$
n_{h1} = \lfloor c_{h} n_{h}\rfloor, 
$
truncated at $2$ and $n_h-2$,
where $c_{h}$'s are i.i.d. samples from Beta distribution $\textnormal{Beta}(4,5)$. We vary the strata size distribution, $\textnormal{SNR}z$, and $k$ in each 
scenario. Values of these factors are presented in Table~\ref{tab:params}. Each scenario is repeated under $100$ random seeds. For each seed and each scenario,
we simulate the stratified randomized experiments $1000$ times and compute the empirical RMSE of point estimators and empirical 
coverage probabilities of $95\%$ confidence intervals.

So far, Lin's with-interaction regression adjustment has not been extended to stratified randomized experiments. Therefore, we consider constructing point and variance estimators from the conditional inference or projection perspective and using a plug-in principle \citep{yang2021rejective,wang2021rerandomization,liu-yang-ren-factorial}. Recall that the optimal linearly adjusted coefficient is $\betastr = \vstrxx^{-1}\vstrxt$. Let $\shxyt$, $\shxyc$, $\shyt$, and $\shyc$ be the sample analogs
of $\Shxyt$, $\Shxyc$, $\Shyt$, and $\Shyc$. 
We estimate $\betastr$ by $\hat{\beta}^{\textnormal{plg}}_{\textnormal{str}} = \vstrxx^{-1}\hat{V}_{\textnormal{str},x\tau}$ with
$$
\hat{V}_{\textnormal{str},x\tau}=\sumh \pi_h p_{h1}^{-1}s_{h1x}+\pi_h p_{h0}^{-1}s_{h0x}.
$$
Therefore, $\htaustr(\betastr)$ can be estimated by $\hat{\tau}_{\textnormal{str}}^{\textnormal{plg}}=\htaustr-(\hat{\beta}^{\textnormal{plg}}_{\textnormal{str}})^\top \tauxstr$. 
The plug-in principle is also used to estimate the normal component's variance in the asymptotic distribution of $\htaustr$ under stratified rerandomization \citep{wang2021rerandomization}. This is equal to the variance of the optimal linearly adjusted estimator $\htaustr(\betastr)$. We follow their procedure to derive a conservative variance estimator of $\hat{\tau}_{\textnormal{str}}^{\textnormal{plg}}$,
$$
\hat{V}^{\textnormal{plg}}_{\textnormal{str}} = \nstr^{-1}(\hat{V}_{\textnormal{str},\tau\tau}-\hat{V}_{\textnormal{str},\tau x}\vstrxx^{-1}\hat{V}_{\textnormal{str},x\tau}),\quad \hat{V}_{\textnormal{str},\tau\tau} = \sumh \pih\{\shyt\pht^{-1}+\shyc\phc^{-1}\}.
$$

\begin{figure}[ht]
  \includegraphics[width = \textwidth]{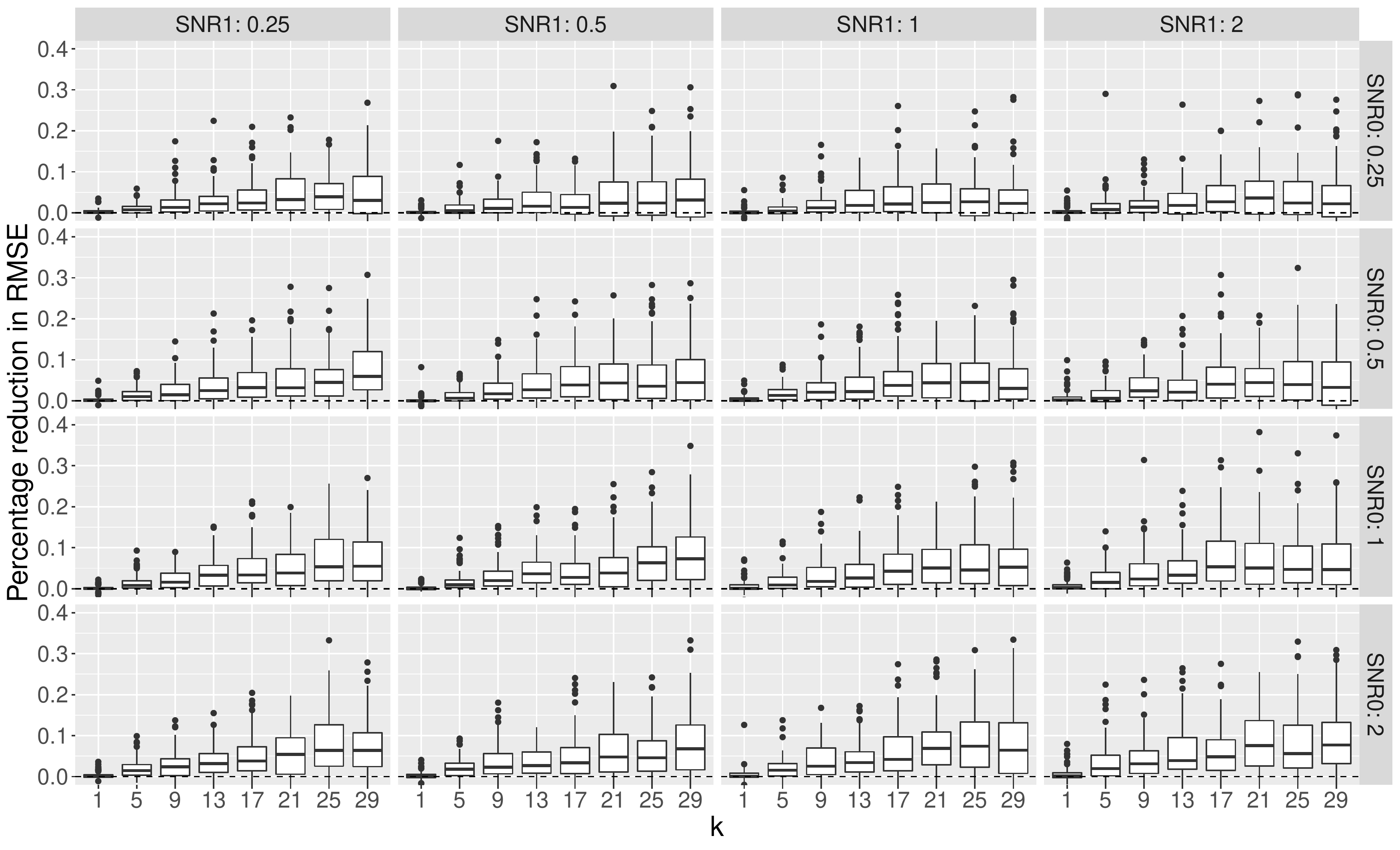}
  \caption{Percentage reduction in RMSE of $\taustrtyr$ versus $\hat{\tau}_{\textnormal{str}}^{\textnormal{plg}}$ in stratified randomized experiments with many small strata.}
  \label{figure:stra-prrmse}
\end{figure}

\begin{figure}[ht]
  \includegraphics[width = \textwidth]{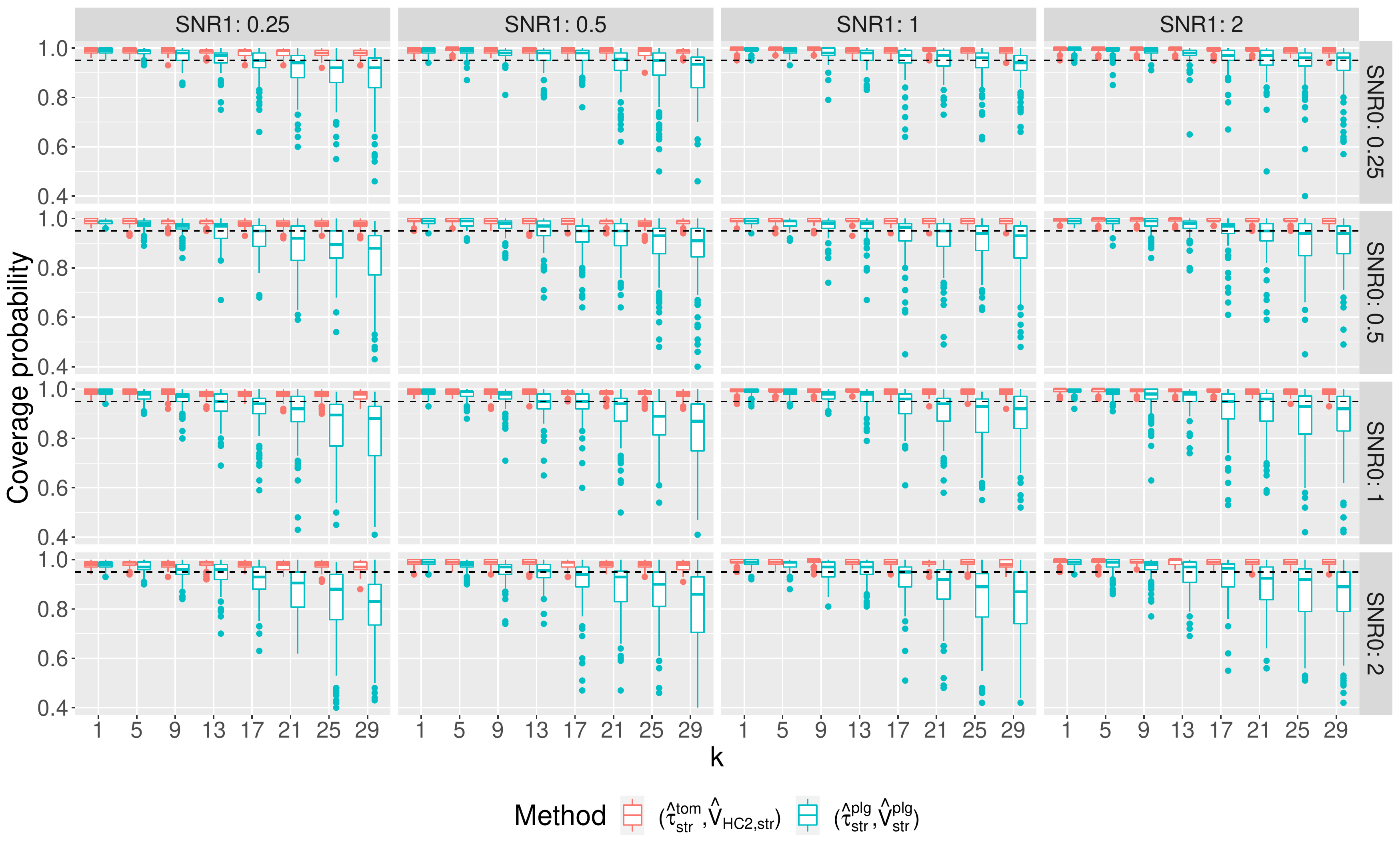}
  \caption{Coverage probabilities in stratified randomized experiments with many small strata.}
  \label{figure:stra-coverage}
\end{figure}

Figure~\ref{figure:stra-prrmse} depicts the percentage reduction in RMSE of $\taustrtyr$ versus $\hat{\tau}_{\textnormal{str}}^{\textnormal{plg}}$.
Figure~\ref{figure:stra-coverage} presents the empirical coverage probabilities of $95\%$ confidence intervals constructed by  $(\taustrtyr, \hat{V}_{\textnormal{HC}2,\textnormal{str}})$ 
and $(\hat{\tau}_{\textnormal{str}}^{\textnormal{plg}},\hat{V}^{\textnormal{plg}}_{\textnormal{str}})$. Both figures are results of many small strata scenario. The results of other scenarios are similar so we degrade them to the Supplementary Material.
It can be observed that, $\taustrtyr$ dominates the plug-in estimator under all scenarios, especially when the dimension of covariates grows. The increasing outliers in the boxplot as $\textnormal{SNR}z$ and dimension of covariates grow
imply that $\taustrtyr$ is more robust than the plug-in estimator under these scenarios. Moreover, Figure~\ref{figure:stra-coverage} shows that the plug-in variance estimator tends to underestimate
the true sampling variance and produce confidence intervals with coverage probabilities lower than the nominal level when the dimension of covariates is large. Therefore, we recommend $(\taustrtyr, \hat{V}_{\textnormal{HC}2,\textnormal{str}})$ for stratified randomized experiments.

\subsection{Completely randomized survey experiments}
We set the population size $N=10000$ and sampling fraction $f=0.01$ to generate data using the same model as \eqref{model:crt}.
Let $v_i=(x_{i1},x_{i2})$ be the covariates available at the sampling stage.
We use $(x_i,(Z_i-p_0)v_i)$ in ToM regression adjustment, with $k+2$ dimensions. 
We set $n=Nf$ for the sampling stage and $p=0.3$ for the treatment assignment stage. We simulate the
completely randomized survey experiments $1000$ times to compute the empirical RMSE of point estimators and empirical 
coverage probabilities of $95\%$ confidence intervals.
We vary the $\textnormal{SNR}z$ and $k$ in each scenario.  Table~\ref{tab:params} presents the values of these factors
 considered in the simulation. Each scenario
is repeated under $100$ different random seeds.

  Let $s^2_{x(z)}$ and $s^2_{v(z)}$ be the sample covariances of covariates under treatment arm $z$. Let 
  $s^2_z$ be the sample variance of $Y_i(z)$ and $s_{vz}$, $s_{xz}$ $(z=0,1)$ be the sample covariances between covariates and outcomes. \cite{yang2021rejective} used the plug-in principle to derive linearly adjusted point and variance estimators. The point estimator is derived by replacing the optimal projection coefficients $(\betacrs,\gammacrs)$ in the optimal linearly adjusted estimator
 with their consistent estimators 
$(\hbetacrs^{\textnormal{plg}},\hgammacrs^{\textnormal{plg}})$, where
\begin{align*}
  \hbetacrs^{\textnormal{plg}} = p_0\{\sxt\}^{-1}\sxyt + p_1\{\sxc\}^{-1}\sxyc ,\quad \hgammacrs^{\textnormal{plg}} = \{\svt\}^{-1}\svyt -\{\svc\}^{-1}\svyc.
\end{align*}
The variance estimator $\hat{V}^{\textnormal{plg}}_{\textnormal{crs}}$ is derived using 
the estimated adjusted potential outcomes $ Y_i(z;\hbetacrs^{\textnormal{plg}},\hgammacrs^{\textnormal{plg}})$ to replace $Y_i(z)$ in $n^{-1}(\syt p_1^{-1}+\syc p_0^{-1})$.
Both $\hat{V}_{\textnormal{HC}0,\textnormal{crs}}$ and Yang et al.'s variance estimator tend to underestimate the true sampling variance for large $k$ in finite samples. 
To remedy this issue, we use the $\textnormal{HC}_3$ type estimator $\hat{V}_{\textnormal{HC}3,\textnormal{crs}}$ suggested by \cite{lei2021regression}.

\begin{figure}[ht]
  \includegraphics[width = \textwidth]{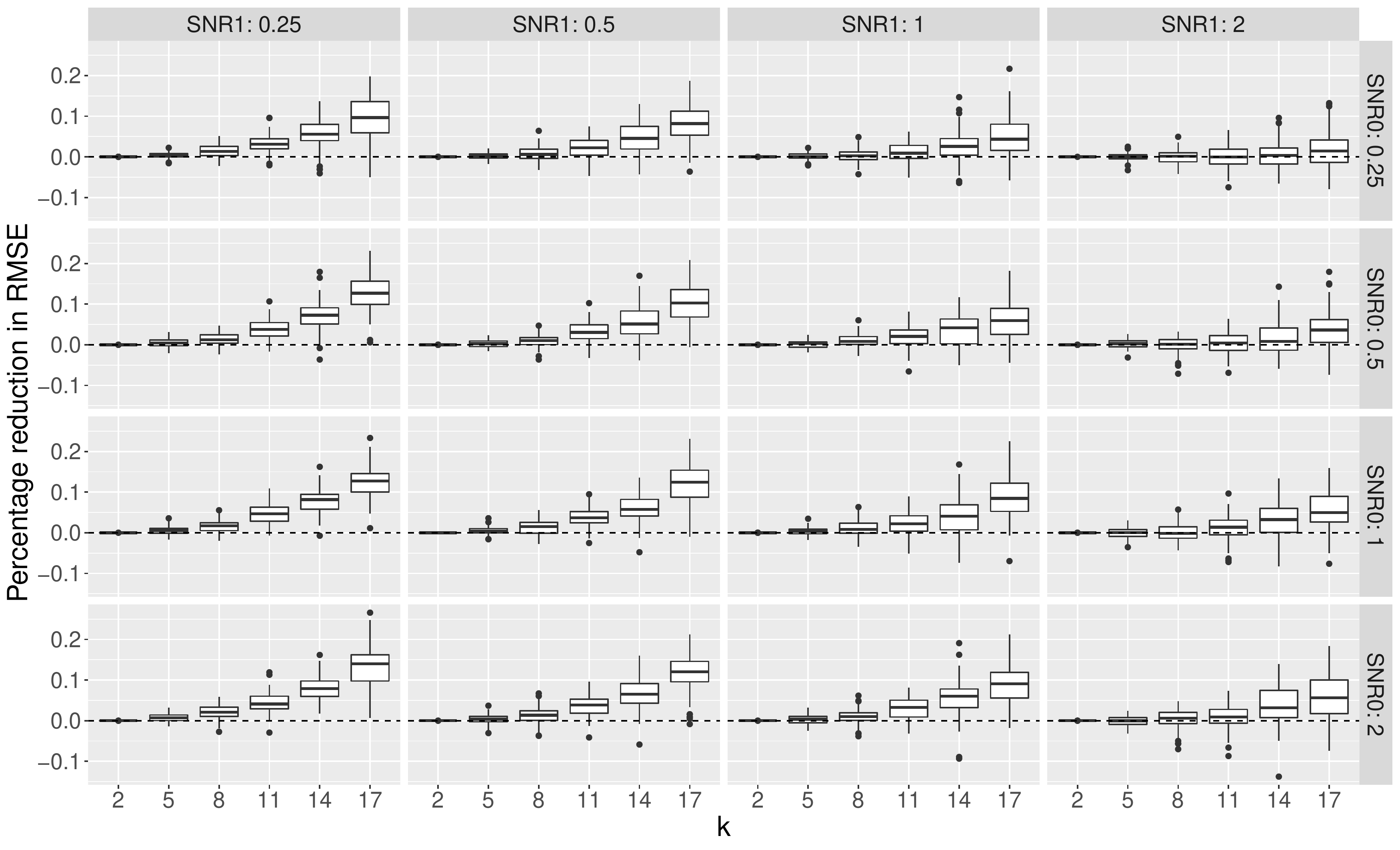}
  \caption{Percentage reduction in RMSE of $\taucrstyr$ versus $\hat{\tau}_{\textnormal{crs}}^{\textnormal{plg}}$ for completely randomized survey experiments.}
  \label{figure:crs-prrmse}
\end{figure}

\begin{figure}[ht]
  \includegraphics[width = \textwidth]{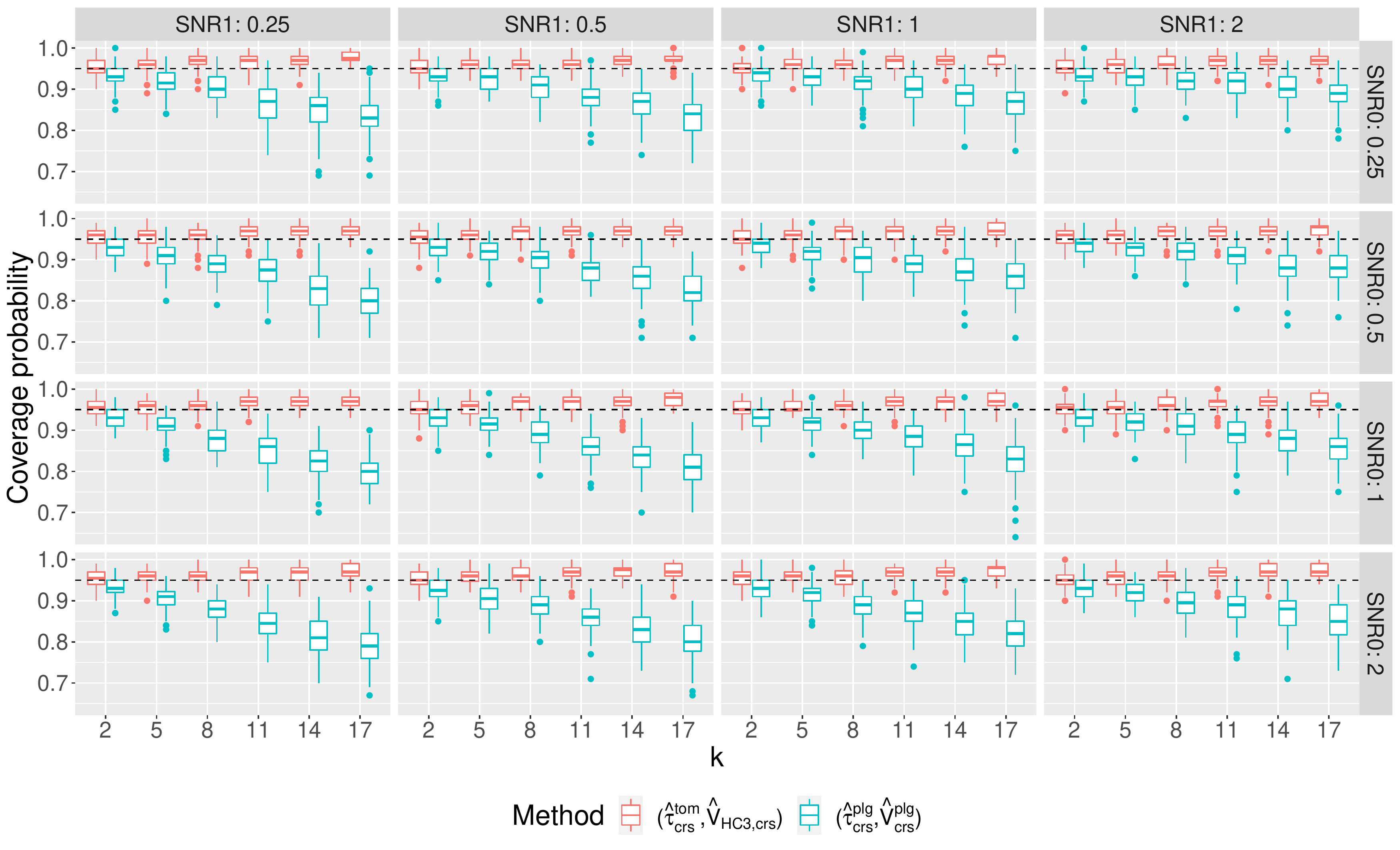}
  \caption{Coverage probabilities in completely randomized survey experiments.}
  \label{figure:crs-coverage}
\end{figure}

Figure~\ref{figure:crs-prrmse} depicts the percentage reduction in RMSE of $\taucrstyr$ versus $\hat{\tau}_{\textnormal{crs}}^{\textnormal{plg}}$. Similar to the completely randomized experiments, $\taucrstyr$ outperforms  $\hat{\tau}_{\textnormal{crs}}^{\textnormal{plg}}$ when the dimension of covariates grows. The trend 
becomes more evident when the majority group has a larger SNR and the minority group has a smaller SNR. Figure~\ref{figure:crs-coverage} depicts the coverage probabilities of $95\%$ confidence intervals constructed by
$(\taucrstyr,\hat{V}_{\textnormal{HC}3,\textnormal{crs}})$ and $(\hat{\tau}_{\textnormal{crs}}^{\textnormal{plg}},\hat{V}^{\textnormal{plg}}_{\textnormal{crs}})$. 
It can be observed that the combination of $(\taucrstyr,\hat{V}_{\textnormal{HC}3,\textnormal{crs}})$ maintains an average of $95\%$ coverage probabilities, while the combination of $(\hat{\tau}_{\textnormal{crs}}^{\textnormal{plg}},\hat{V}^{\textnormal{plg}}_{\textnormal{crs}})$
tends to have low coverage probabilities for large $k$ and performs  worse when the majority group has a larger SNR and the minority group has a smaller SNR. Therefore, we recommend $(\taucrstyr,\hat{V}_{\textnormal{HC}3,\textnormal{crs}})$ for analyzing completely randomized survey experiments. 

\section{Applications}

\subsection{The ``opportunity knocks" experiment}

The ``opportunity knocks" (OK) experiment \citep{angrist2014opportunity} was a stratified randomized experiment launched to evaluate the effect of financial incentive on college
students' academic performance. The experiment included first- and second-year students who applied for financial aid at a large Canadian commuter university.
Based on sex and discretized high school grades, the students were grouped into $8$ strata with strata sizes ranging from $46$ to $95$. In each stratum, approximately $25$ students received the treatment. Therefore, the $\pht$'s varied across strata. The grade point average (GPA) at the end of the fall semester was the outcome of interest. We consider $6$ covariates in ToM regression adjustment: high school grade, previous year GPA, age, whether the student's mother tongue is English,
whether the student lives at home, and whether the student has high concern about the funds.

Table~\ref{tab:unbalance-between-treatment-and-control-str} presents $\tauxstr$, the adjusted coefficient $\hbetastr$, and their hadamard product. We can see that $\taustrtyr$ adjusts $\htaustr$ because the treatment group's previous year GPA is lower on average, and more students live at home and have high concerns about the funds.

\begin{table}
  \caption{\label{tab:unbalance-between-treatment-and-control-str}$\tauxstr$, adjusted coefficient, and their hadamard product}
  \resizebox*{1\columnwidth}{!}{
    \begin{tabular}{@{}ccccccc@{}}
      \toprule
       & High school  & Previous year  & Age & Whether the student's mother  & Whether the student  & Whether the student has  \\ 
       &       grade  &     GPA        &     &          tongue is English    &   lives at home &    high concern about the funds  \\\midrule
       $\tauxstr$ &  0.003        &  -0.010        &  0.028  &  0.000   &     0.011    & 0.023\\ 
       $\hbetastr$&  0.186        &   7.543        & -0.089  & -0.201  &    -2.447         &  -1.914  \\
      $\tauxstr\circ\hbetastr$ &  0.001        &  -0.075        & -0.002  &  -0.000   &     -0.027    & -0.044\\ 
      \bottomrule
      \end{tabular}
  }
  \end{table}

Figure~\ref{fig:est-and-ci} depicts the average treatment effect estimators, standard errors, and $95\%$ confidence intervals. Both ToM regression-adjusted and unadjusted estimators show that the average treatment effect is insignificant. That is, we do not have sufficient evidence to support the following: financial incentive affects students' academic performance. However, it is interesting to see that ToM regression adjustment provides a larger average treatment effect estimator and decreases the estimated standard error by $22.7\%$.


\subsection{Social Trust in Polarized Times}

We re-analyze the experimental dataset from \cite{lee2022social} to evaluate the impact of 
perceived polarization on social trust levels. In this experiment, 1006 Americans over 18 years old were recruited from an online survey panel. We treat the experimental units as a simple random sample from the target population, that is, the entire American population
over 18 years old. The experimental units are randomly assigned to read one of the three news articles designed to either promote perceived polarization (more-polarization), 
reduce perceived polarization (less-polarization), or serve as a control article. We evaluate the treatment effects of more-polarization and less-polarization versus the control. The outcome is an index ranging from 0 to 1, with higher values indicating higher generalized social trust. The following types of covariates are used: 
\begin{itemize}
  \item $x_i$: whether the individual is white and non-Hispanic (race1), whether the individual is black or African American (race2), whether the individual is Hispanic (race3), whether the individual is female (sex), education type (education), household income type (income), marital status (marital), whether the individual does not go to college (nocollege), and age.
  \item $v_i$: race1, race2, race3, age, and sex. We obtain $\bv$ of the target population from the website of United States Census Bureau.
\end{itemize}

First, we add the main effect of $x_i$ and $v_i$, quadratic terms of the continuous covariates of $x_i$, and two-way interactions of $x_i$ in the full regression model, which produced a design matrix with $50$ columns. Then we use forward-backward stepwise regression to obtain a reduced model with $4$ and $9$ covariates entering ToM regression adjustment for the treatment effects of more-polarization and less-polarization versus the control, 
respectively. For both regression adjustments, none of the $v_i$ enters the model.

\begin{table}
  \caption{\label{tab:unbalance-crs-less-vs-control}$\tauxcrs$, adjusted coefficient, and their hadamard product (less-polarization vs control)}
    \centering
  \resizebox*{1\columnwidth}{!}{
    \begin{tabular}{cccccc}
      \toprule
       & age  &  age:education& age:marital & age:race3 & income:race1 \\
      \midrule
      $\tauxcrs$    & $-0.196$  & $-1.047$   & $-0.206$  &  $0.088$  & $0.018$ \\
      $\hbetacrs$       & $0.163$  & $-0.011$ & $-0.021$    & $-0.073$ & $ 0.021$ \\
      $\tauxcrs\circ\hbetacrs$ &$-0.032$&$0.012$&$0.004$&$-0.006$&$0.000$\\
      \bottomrule
      \toprule
      & income:race3  & sex:race2 & education:marital&nocollege:race1&   \\
      \midrule
      $\tauxcrs$    &$0.057$    & $0.024$  & $0.027$   & $0.004$& \\
      $\hbetacrs$   &$0.035$& $-0.178$  & $0.016$ & $-0.145$ &   \\
      $\tauxcrs\circ\hbetacrs$ &$0.002$&$-0.004$&$0.000$&$-0.001$&\\
      \bottomrule  
    \end{tabular}
  }
\end{table}

\begin{table}
  \caption{$\tauxcrs$, adjusted coefficient, and their hadamard product (more-polarization vs control)\label{tab:unbalance-crs-more-vs-control}}
\centering
  \resizebox*{0.9\columnwidth}{!}{
    \begin{tabular}{cccccc}
      \toprule
       & age:education  &  age:nocollege & income:race1 & education:nocollege &\\
      \midrule
      $\tauxcrs$   & $0.075$  & $0.009$   & $-0.047$  &  $0.041$ & \\
      $\hbetacrs$      & $0.007$  & $0.028$ & $0.018$    & $-0.038$ & \\
      $\tauxcrs\circ\hbetacrs$ &$ 0.001$&$0.000$&$-0.001$&$-0.002$ &\\
      \bottomrule
    \end{tabular}
  }
\end{table}


\begin{figure}[ht]
  \includegraphics[width = \textwidth]{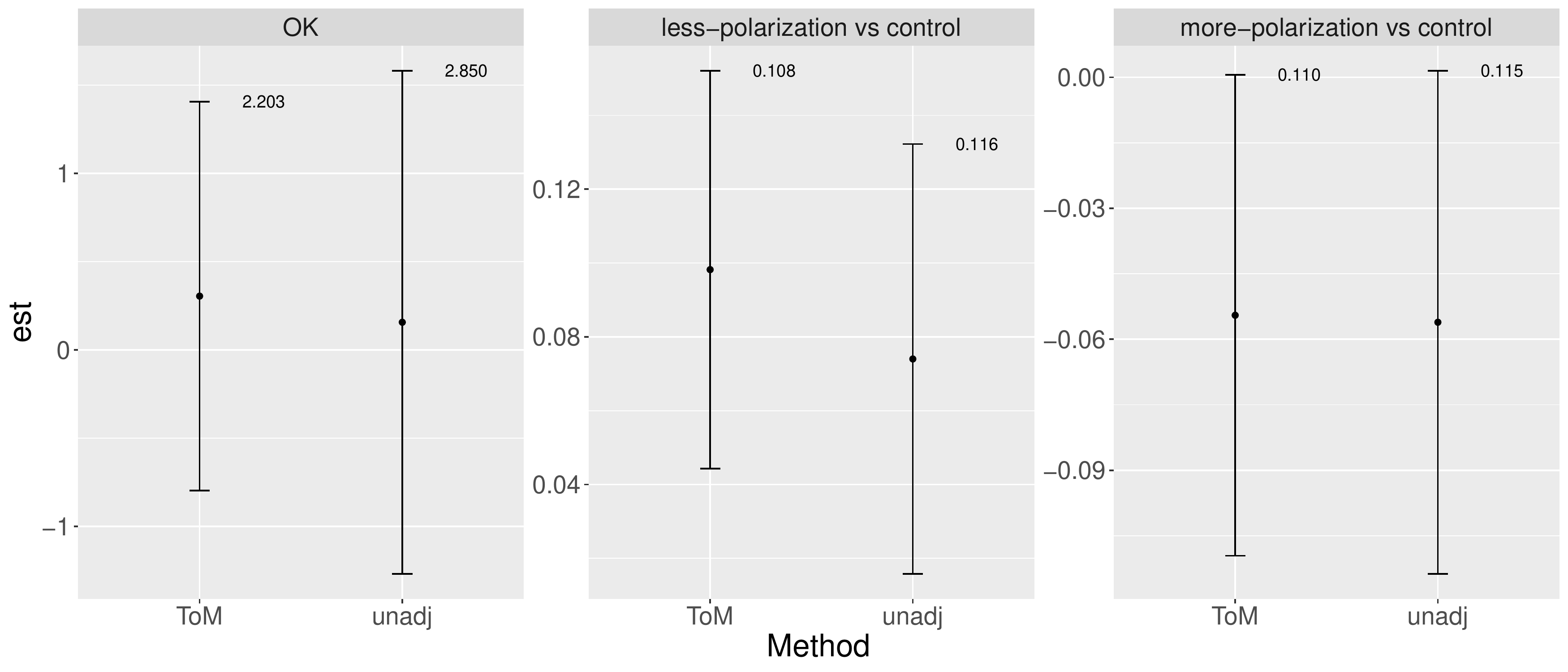}
    \caption{Point  estimators and 95\% confidence intervals of ToM regression-adjusted and unadjusted methods for two real datasets. The numbers on top-right of the confidence intervals are the corresponding confidence interval lengths.}
  \label{fig:est-and-ci}
\end{figure}

For the treatment effect of less-polarization versus control, Table~\ref{tab:unbalance-crs-less-vs-control} and Figure~\ref{fig:est-and-ci} show that ToM regression adjusts upwards $\htaucrs$ mainly because the treatment group is $0.2$ years younger than the control group. Both ToM regression-adjusted and unadjusted estimators indicate that the average treatment effect is significant, that is, less-polarization articles significantly affect people's social trust. In contrast, the treatment effect of more-polarization versus control is insignificant as presented by Figure~\ref{fig:est-and-ci}. ToM regression slightly adjusts $\htaucrs$; see Table~\ref{tab:unbalance-crs-more-vs-control}. Compared to the unadjusted estimator, ToM regression adjustment decreases the estimated standard error by $7.4\%$ and $4.6\%$, respectively, for the less-polarization versus control and more-polarization versus control.


\section{Extension to rerandomization}

Regression adjustment is used at the analysis stage to adjust for covariate imbalance. Rerandomization is an alternative approach  achieving covariate balance in the design stage \citep[see, e.g.,][]{morgan2012rerandomization,Morgan2015,Li9157,2020Rerandomization,Li2020factorial,wang2021rerandomization,zdrep,Lu2022}.  Recent work by \cite{2020Rerandomization}, \cite{wang2021rerandomization}, and \cite{zdrep} showd that the combination of rerandomization and Lin's with-interaction regression adjustment can further improve the efficiency if the analysis stage utilizes more covariate information than the design stage. The same conclusion holds true for the combination of rerandomization and ToM regression adjustment.

\liu{In randomized experiments, it is common that the covariates avaliable at the design stage are a subset or linear combinations of the covariates available at the analysis stage. In this case, the asymptotic normality and optimality of the ToM regression-adjusted estimator and the asymptotic properties of the heteroscedasticity-robust variance estimators still hold if} \liu{(1) rerandomization \citep{morgan2012rerandomization} is used in completely randomized experiments, or} \liu{(2) stratified rerandomization \citep{wang2021rerandomization} is used in stratified randomized experiments, or} \liu{(3) rejective sampling and reradnomization \citep{yang2021rejective} are used in completely randomized survey experiments, or} \liu{(4) rerandomization based on cluster-level covariates \citep{Lu2022} is used in cluster randomized experiments.}

\section{Discussion}

We re-examine ToM regression adjustment and justify its robustness compared to the with-interaction regression adjustment from three perspectives: first, ToM regression adjustment produces less extreme calibrated-weights; second, ToM regression adjustment produces smaller leverage scores; third, when the dimension of covariates is large or there is an imbalance in information between treatment and control groups, ToM regression adjustment produces estimator with smaller mean squared errors and better coverage probabilities. 
We proved the applicability of ToM regression adjustment to stratified randomized experiments, completely randomized survey experiments and cluster randomized experiments. Under each design, we showed that the ToM regression-adjusted average treatment effect estimator is asymptotically normal and optimal in the class of linearly adjusted estimators. We also studied the asymptotic properties of several heteroscedasticity-robust variance estimators derived from the ToM regression adjustment and found that some of these variance estimators may be anti-conservative. Our results
are design-based and allow model misspecification. Lastly, the inferential procedure can be easily implemented by standard statistical software packages.

The asymptotic theory may not be applicable when the number of experimental units is small. In such cases, we suggest using Fisher-randomization tests with studentized test statistics obtained from ToM regression adjustment \citep{zhao2021covariate}. The Fisher-randomization tests yield finite-sample exact $p$-values under the sharp null hypothesis and are asymptotically valid under the weak null hypothesis, with the average treatment effect as zero.

Our asymptotic analysis assumes that the number of covariates is fixed. However, in many randomized experiments, such as A/B tests, the number of covariates can be very large, even larger than the sample size \citep{bloniarz2016lasso,lei2021regression}. ToM regression adjustment can be easily extended to high-dimensional settings by adding an appropriate penalty on the adjusted coefficient. It would be interesting to study the design-based properties of this extension.

Finally, our theory focuses on experimental designs with binary treatment and perfect compliance. In practice, researchers may be interested in the effects of multiple-valued treatments in the presence of noncompliance. It is interesting to extend the applicability of ToM regression adjustment to analyze randomized experiments with multiple-valued treatments \citep{fisher1935design,liu-yang-ren-factorial,ye2022inference} and/or noncompliance \citep{imbens1994identification,angrist1995two,angrist1996identification,ding2017principal}.

\section*{Acknowledgement} 
This research is supported by the National Natural Science Foundation of China (12071242) and the Guo Qiang Institute of Tsinghua University.


\bibliographystyle{apalike}
\bibliography{causal}

\newpage

\appendix

\begin{center}
    \textbf{\centering \Large Supplementary Material}
\end{center}

Section~\ref{sec:A} provides parallel results for cluster randomized experiments.

Section~\ref{sec:B} provides additional simulation results.

Section~\ref{sec:C} provides formulas of the heteroskedasticity-robust standard errors $\textnormal{HC}_j$ $(j=0,1,2,3)$.

Section~\ref{sec:D} provides proofs for the results under completely randomized experiments.

Section~\ref{sec:E} provides proofs for the results   under stratified randomized experiments.

Section~\ref{sec:F} provides proofs for the results   under completely randomized survey experiments.

\section{ToM regression adjustment in cluster randomized experiments}
\label{sec:A}

Cluster randomized experiments randomly assign the treatment at the cluster level with units in the same cluster receiving the same treatment status \citep{hayes2017cluster}. Cluster randomized experiments have been widely used in empirical research when individual-level treatment assignment is infeasible or inconvenient. 



Consider $\ncl$ units nested in $m$ clusters of sizes $n_i$ ($i = 1,\ldots,m$, $\sumicl n_i = \ncl$). By design, $m_1$ clusters are randomly assigned to the treatment group and $m_0=m-m_1$ clusters are assigned to the control group. Let $Z_i$ be the treatment assignment indicator 
for cluster $i$. With a slight abuse of notation, let $p_z=m_z/m$. We use $ij$ to index unit $j$ in cluster $i$ ($i=1,\ldots,m$, $j=1,\ldots,n_i$). Let $x_{ij}$ and $Y_{ij}(z)$ ($z=0,1$) be the covariates and
potential outcomes for units $ij$. Let  $c_i$ be the cluster-level covariates. The average treatment effect is 
\begin{align}
  \label{eq:ATE:cl-individual-version}
  \taucl = \ncl^{-1} \sumicl\sumjcl \left\{Y_{ij}(1)- Y_{ij}(0)\right\}. \nonumber
\end{align}
Let $\bar{n}=\ncl/m$ be the average cluster size. Let $\tyidz=\bar{n}^{-1}\sumjcl Y_{ij}(z)$ ($z=0,1$) be the potential outcome total of cluster $i$
scaled by $\bar{n}^{-1}$ and $\check{Y}_{i\cdot} = Z_i\check{Y}_{i\cdot}(1)+(1-Z_i)\check{Y}_{i\cdot}(0)$ be the observed scaled potential outcome total.
Then, the average treatment effect can be rewritten as 
\begin{align*}
  \taucl = m^{-1}\sumicl \left\{\tyidt -\tyidc\right\}.
\end{align*}
Similarly, we define scaled covariate total $\txid$. We can view cluster randomized experiments 
as complete randomized experiments on the cluster level with cluster-level data 
$\{(\check{Y}_{i\cdot},c_i,\txid,Z_i)\}_{i=1}^m$ \citep{li2017general,middleton2015unbiased}. \cite{su2021modelassisted}
showed that regression adjustment using scaled covariate total together with cluster size $n_i$  leads to larger 
variance reduction compared with individual-level regression adjustment. Given assumption similar to Assumption~\ref{a:crt} on 
$\{(\tyidt,\tyidc,c_i,\txid,n_i)\}_{i=1}^m$, we have results in parallel with those in Section~\ref{sec:crt} in the main text.

Let $(\taucltyr,\vhwclj)$ be the estimated coefficient and heteroscedasticity-robust variance estimator of $Z_i$ in the following weighted regression: 
\begin{align*}
  \check{Y}_{i\cdot} \stackrel{w_i}{\sim} 1+ Z_i + c_i+\txid + n_i,
\end{align*}
where $w_i = Z_i/p_1^2+(1-Z_i)/p_0^2$. Corollary~\ref{cor::1} below is a direct result of Proposition~\ref{prop:CRT-consistency} and Theorem~\ref{prop:valid-CRT-confidence-interval}.

\begin{corollary}\label{cor::1}
Under Assumption~\ref{a:crt} with $n=m$, $Y_i(z) = \tyidz$ $(z=0,1)$, $x_i = (c_i,\txid,n_i)$, (i)
$\taucltyr$ is consistent for $\taucl$, asymptotically normal, and optimal in the class of linearly adjusted estimators, (ii) the probability limit of $m \vhwclj$ ($j=0,1,2,3$) is larger than or equal to the true asymptotic variance of $\sqrt{m} \taucltyr$, and (iii) the Wald-type $1-\alpha$ confidence intervals 
      \[
    \left[\taucltyr+ \vhwclj^{1/2}q_{\alpha/2}, \  \taucltyr+\vhwclj^{1/2}q_{1-\alpha/2}\right], \quad j=0,1,2,3,
  \]
  have asymptotic coverage rates greater than or equal to $ 1-\alpha$.
\end{corollary}

\section{Additional simulation results}
\label{sec:B}

\begin{figure}[ht]
  \includegraphics[width = \textwidth]{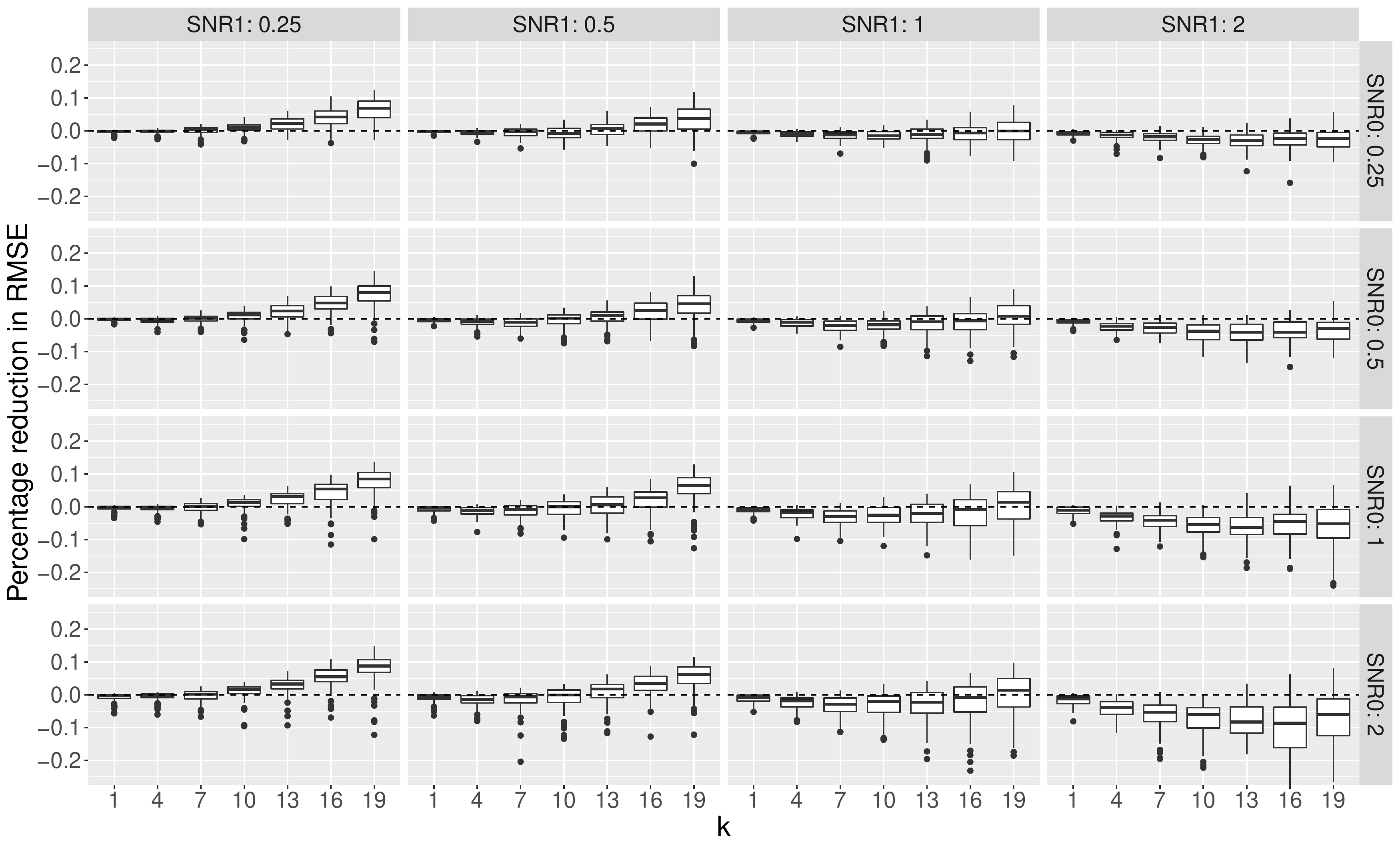}
  \caption{Percentage reduction in RMSE of $\tautyr$ versus $\taulin$ when $p_1=0.4$.}
  \label{figure:prmse0.4}
\end{figure}

\begin{figure}[ht]
  \includegraphics[width = \textwidth]{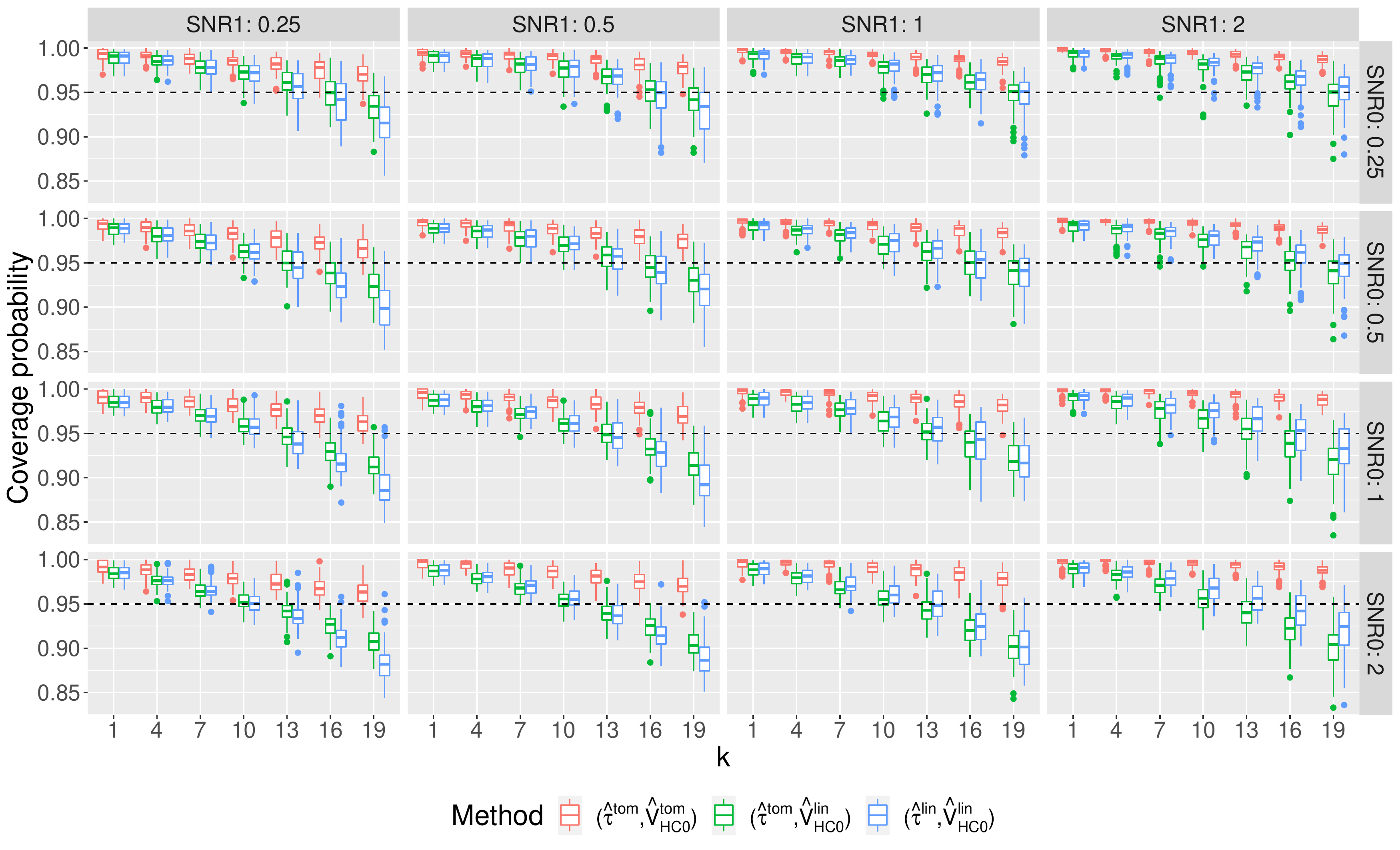}
  \caption{Coverage probabilities for $p_1=0.4$ in completely randomized experiments.}
  \label{figure:coverage0.4}
\end{figure}

\begin{figure}[ht]
  \includegraphics[width = \textwidth]{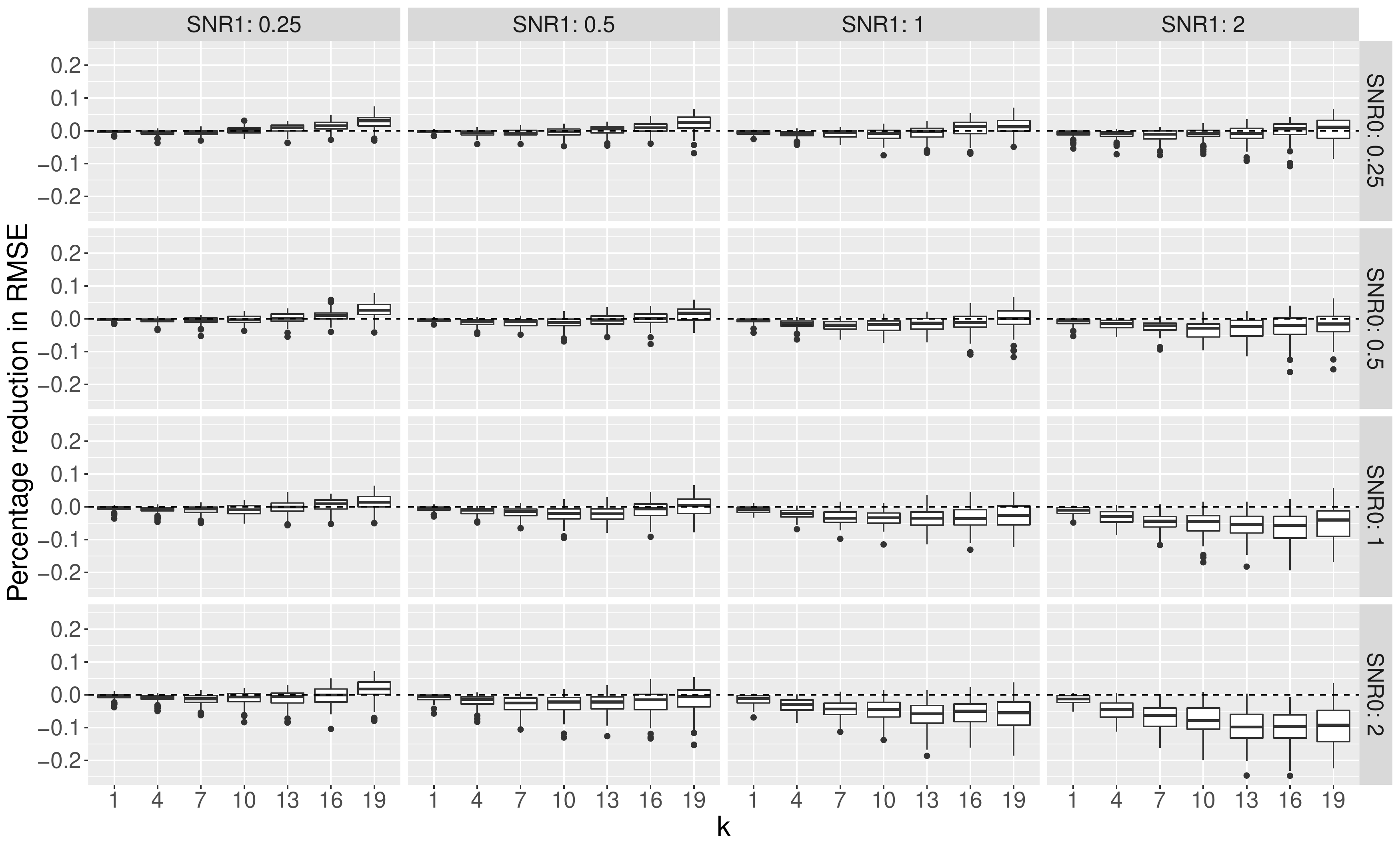}
  \caption{Percentage reduction in RMSE of $\tautyr$ versus $\taulin$ when $p_1=0.5$.}
  \label{figure:prmse0.5}
\end{figure}

\begin{figure}[ht]
  \includegraphics[width = \textwidth]{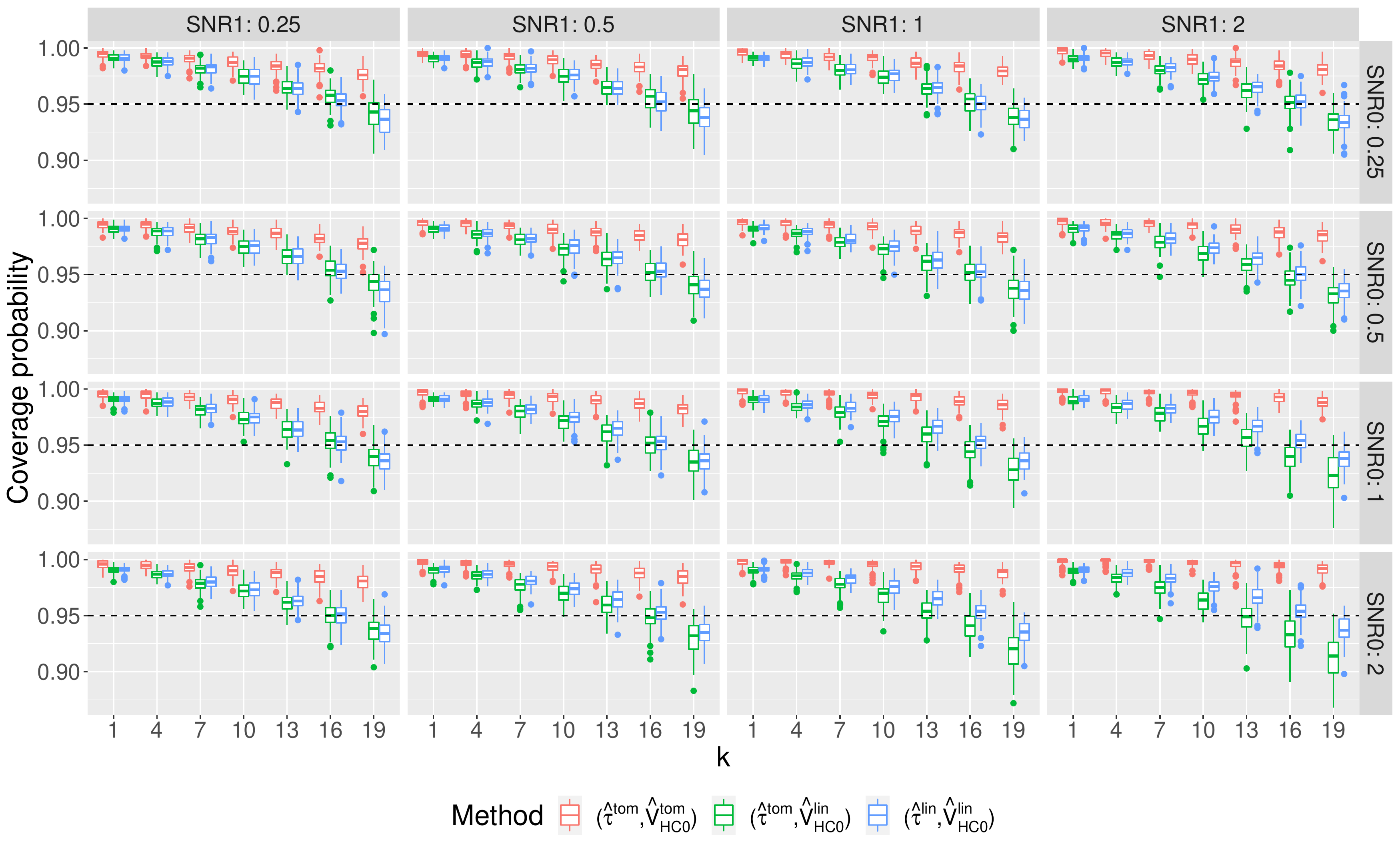}
  \caption{Coverage probabilities for $p_1=0.5$ in completely randomized experiments.}
  \label{figure:coverage0.5}
\end{figure}

\begin{figure}[ht]
  \includegraphics[width = \textwidth]{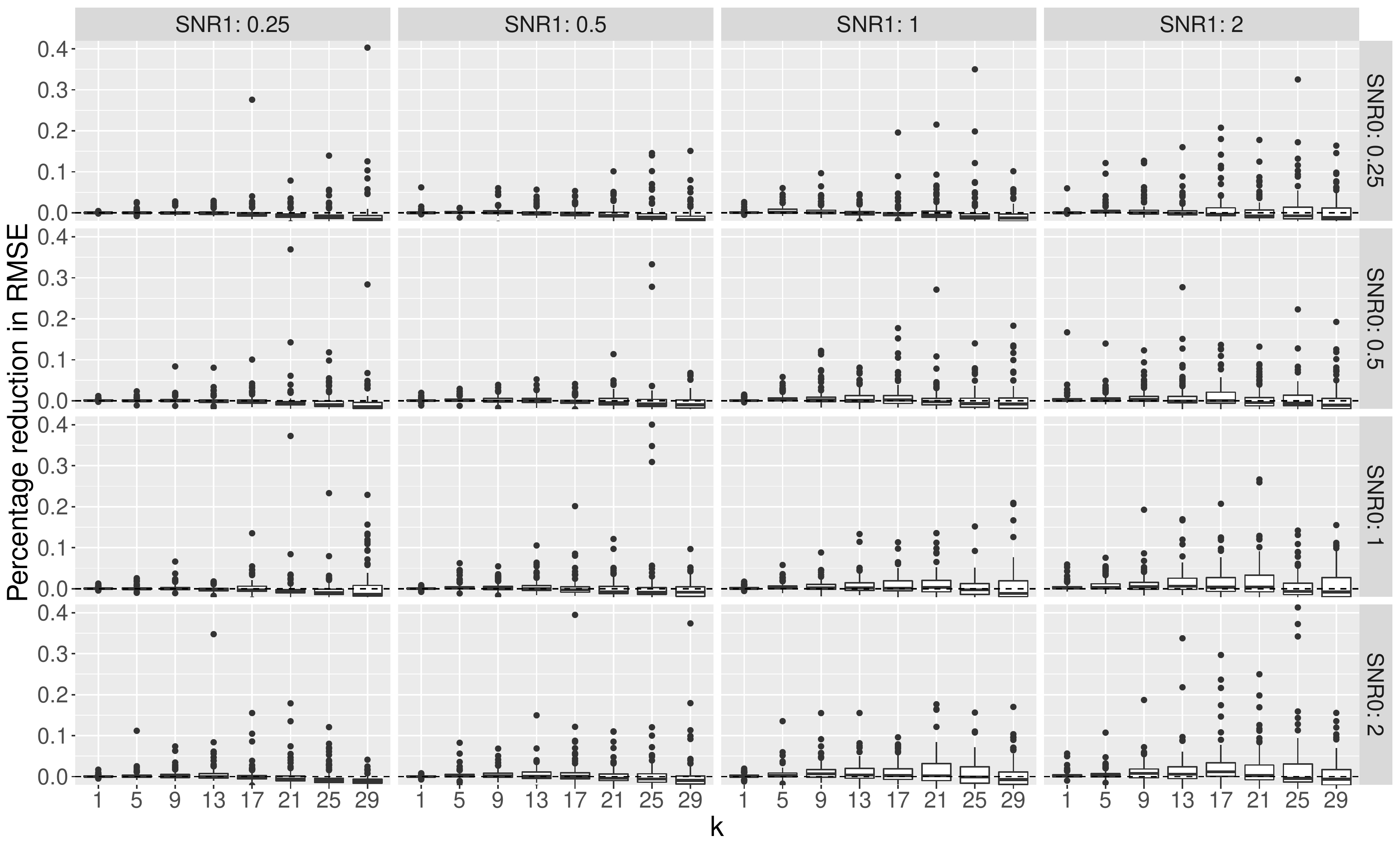}
  \caption{Percentage reduction in RMSE of $\taustrtyr$ versus $\hat{\tau}_{\textnormal{str}}^{\textnormal{plg}}$ in stratified randomized experiments when there are a few large strata.}
  \label{figure:stra-prrmse-FL}
\end{figure}

\begin{figure}[ht]
  \includegraphics[width = \textwidth]{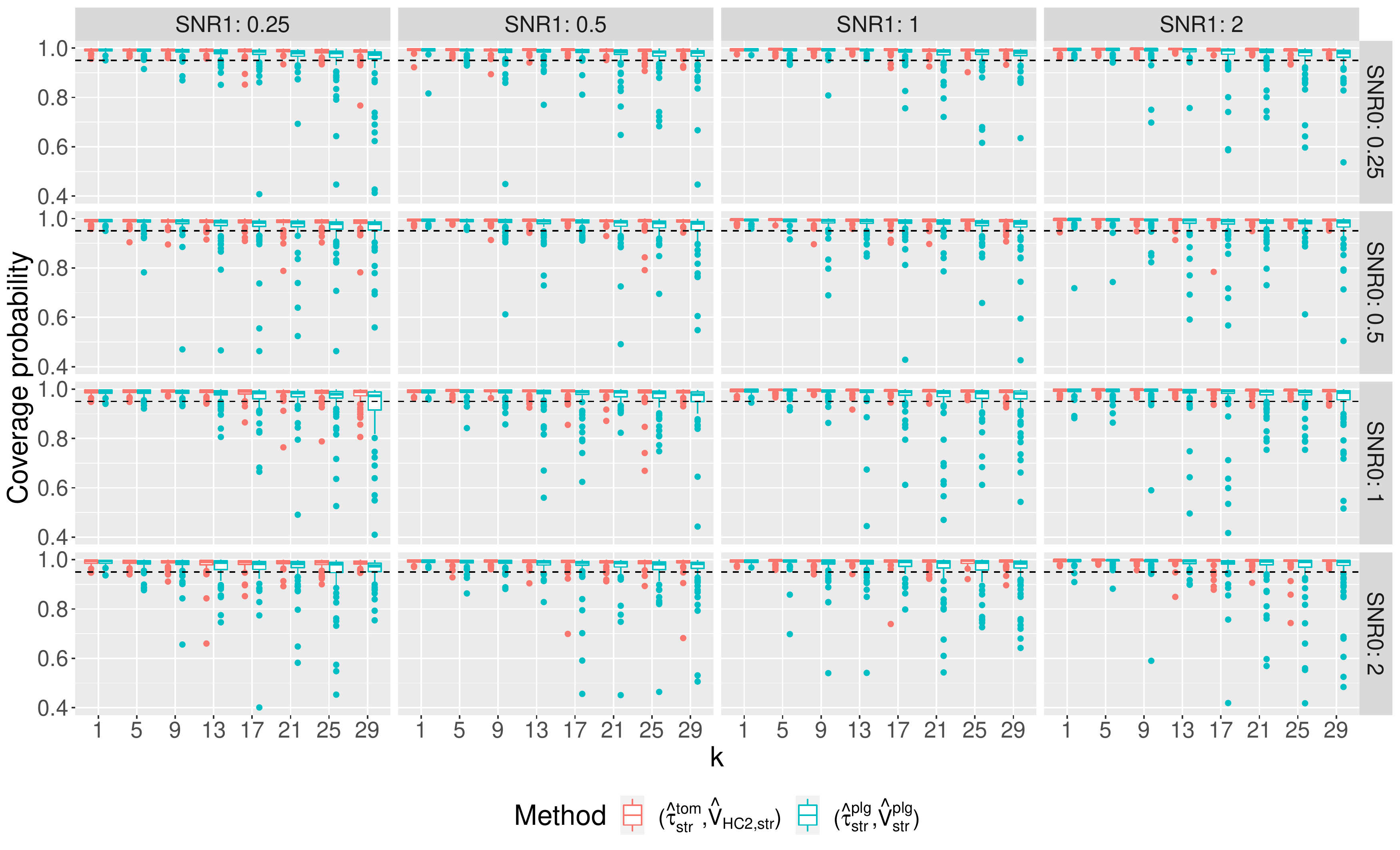}
  \caption{Coverage probabilities in stratified randomized experiments when there are a few large strata.}
  \label{figure:stra-coverage-FL}
\end{figure}

\begin{figure}[ht]
  \includegraphics[width = \textwidth]{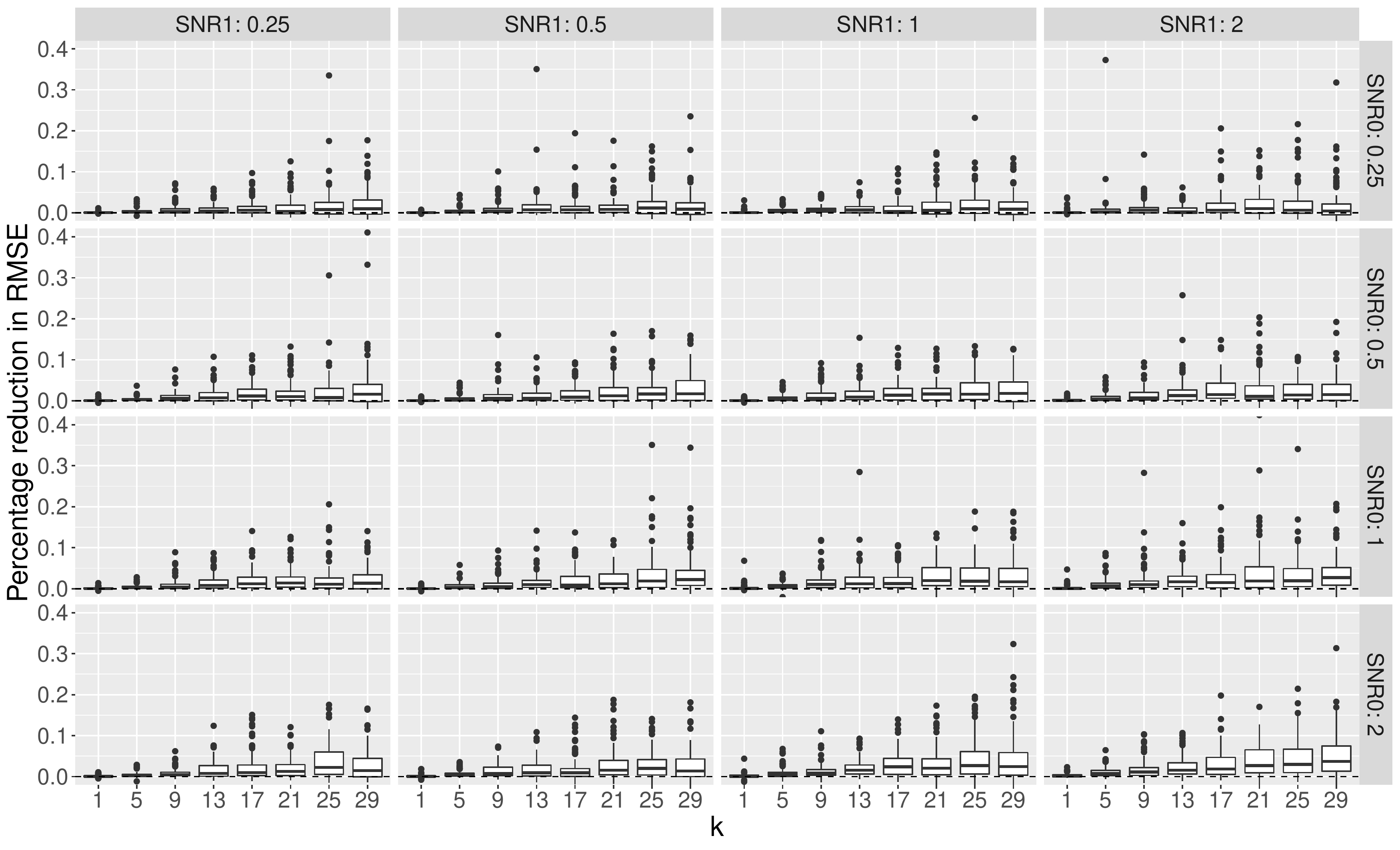}
  \caption{Percentage reduction in RMSE of $\taustrtyr$ versus $\hat{\tau}_{\textnormal{str}}^{\textnormal{plg}}$ in stratified randomized experiments when there are many small
  strata compounded with a few large strata.}
  \label{figure:stra-prrmse-MS+FL}
\end{figure}

\begin{figure}[ht]
  \includegraphics[width = \textwidth]{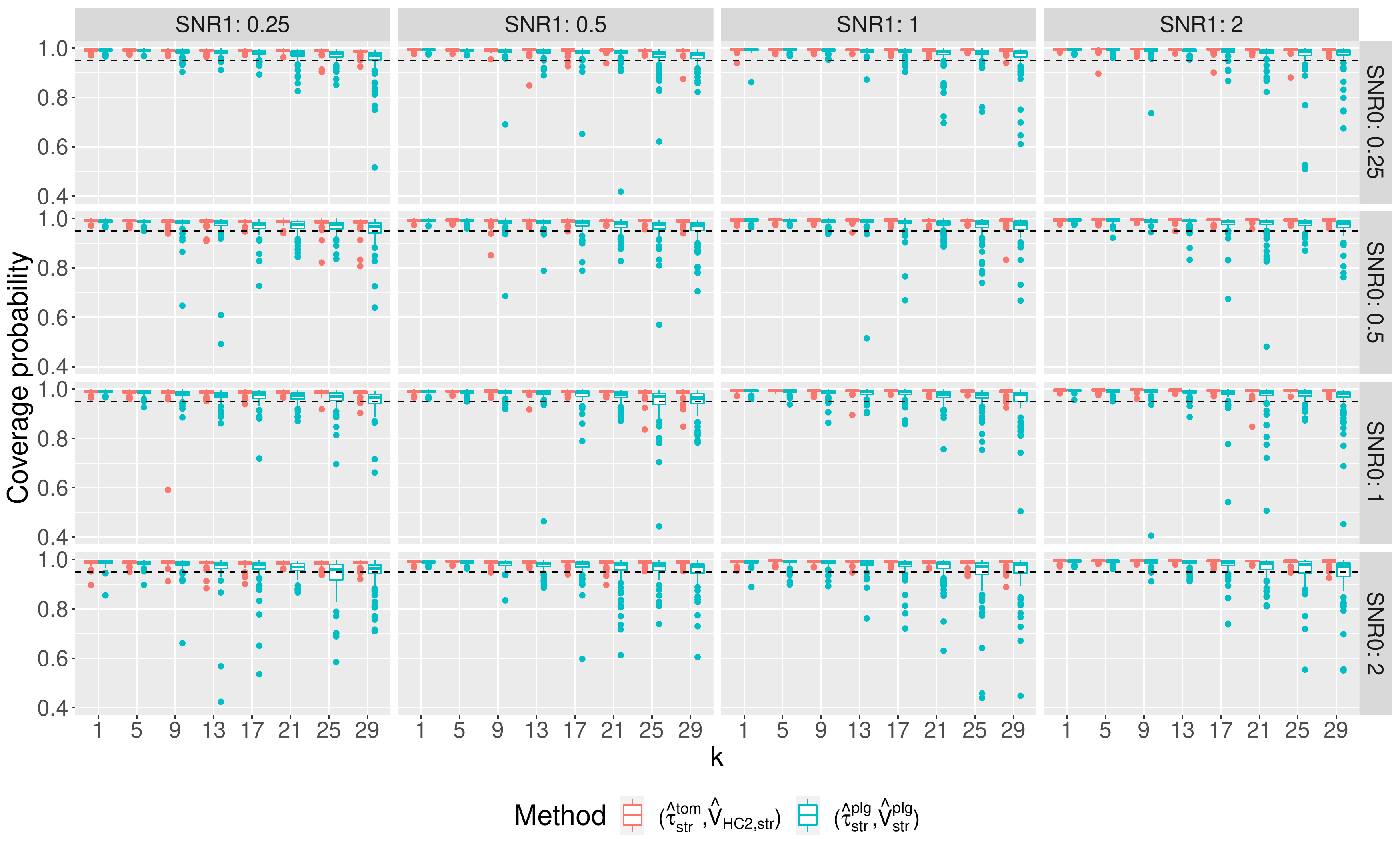}
  \caption{Coverage probabilities in stratified randomized experiments when there are many small
  strata compounded with a few large strata.}
  \label{figure:stra-coverage-MS+FL}
\end{figure}

Figures~\ref{figure:prmse0.4}--\ref{figure:coverage0.5} show the simulation results for completely randomized experiments when $p_1=0.4$ and $p_1=0.5$. Under these two more balanced scenarios, the advantages of $\tautyr$ over $\taulin$ are not as significant as that when $p=0.3$. In particular, when both SNR1 and SNR0 are large, $\tautyr$ performs worse than $\taulin$. This may be because the adjusted coefficients in both the treatment and control groups are  well estimated by $\taulin$. Therefore, for a nearly balanced design, we still recommend the use of $(\taulin,\hat{V}_{\textnormal{HC0}}^{\textnormal{lin}})$.

Figures~\ref{figure:stra-prrmse-FL}--\ref{figure:stra-coverage-MS+FL} show the simulation results for stratified randomized experiments with a few large strata and many small strata compounded with a few large strata. The conclusions are similar to those in the main text.

We also conduct simulation for cluster randomized experiments. The potential outcomes are generated by the following random effect model:
\begin{equation}
  \label{model:cluster-randomized-experiment}
  Y_{ij}(z) = f_{zi}(x_{ij})+e_{ij}(z),~ \text{with}~ f_{zi}(x_{ij})=\alpha_{zi} + x_{ij}^
\top\beta_{zi},\ z=0,1,\ i=1,\ldots,m,\ j=1,\ldots,n_i. \nonumber
\end{equation}
We set the number of clusters $m=50$. The cluster sizes $\{n_i\}_{i=1}^m$ are generated uniformly from the set
$\{n \in \mathbb{N} \mid 4 \leq n \leq 10\}$. The intercepts and slopes are generated by
 $\beta_{zi}=\beta_{z}+\zeta_{zi}$ and $\alpha_{zi} = \alpha_{z}+\eta_{zi}$, where $(\alpha_{z},\beta_{z})$ and $(\eta_{zi},\zeta_{zi})$  have i.i.d. entries generated from $t_3$ and standard normal distribution, respectively. The covariates $x_{ij}$'s
are realizations of independent random vectors of length $k$ from ${N}(0,\Sigma)$ with 
$\Sigma_{ij} = 0.6\delta_{ij}+0.4$, and $e_{ij}(z)$'s  are realizations of 
i.i.d. normal random variables with zero mean and variance fulfilling a given 
signal-to-noise ratio $\textnormal{SNR}z$, i.e., the ratio of the finite-population variance of 
$f_{zi}(x_{ij})$ to that of $e_{ij}(z)$. 

We set the proportion of clusters assigned to the treatment group $p_1=0.3$. After we have generated the data,
we use the scaled cluster totals in the analysis stage. We use $k+1$ covariates $(\tilde{x}_{i\cdot},n_i)$ in the regression adjustment as suggested by \cite{su2021modelassisted}. 
Again $1000$ cluster randomized experiments are simulated and empirical RMSE and coverage probabilities are computed.
 We consider scenarios with all parameter values presented in Table~\ref{tab:cluster-params}.
\begin{table}
  \caption{\label{tab:cluster-params}Parameters in simulation under cluster randomized experiments}
    \centering
  \begin{tabular}{cc}
    \toprule
    random seed      &  $1:100$ \\
    $k$        & $\left\{1,3,5,7,9\right\}$ \\
    $\textnormal{SNR}0$        &       $\left\{0.25,0.5,1,2\right\}$    \\
    $\textnormal{SNR}1$          &     $\left\{0.25,0.5,1,2\right\}$    \\
    \bottomrule
  \end{tabular}
\end{table}

\begin{figure}[ht]
  \includegraphics[width = \textwidth]{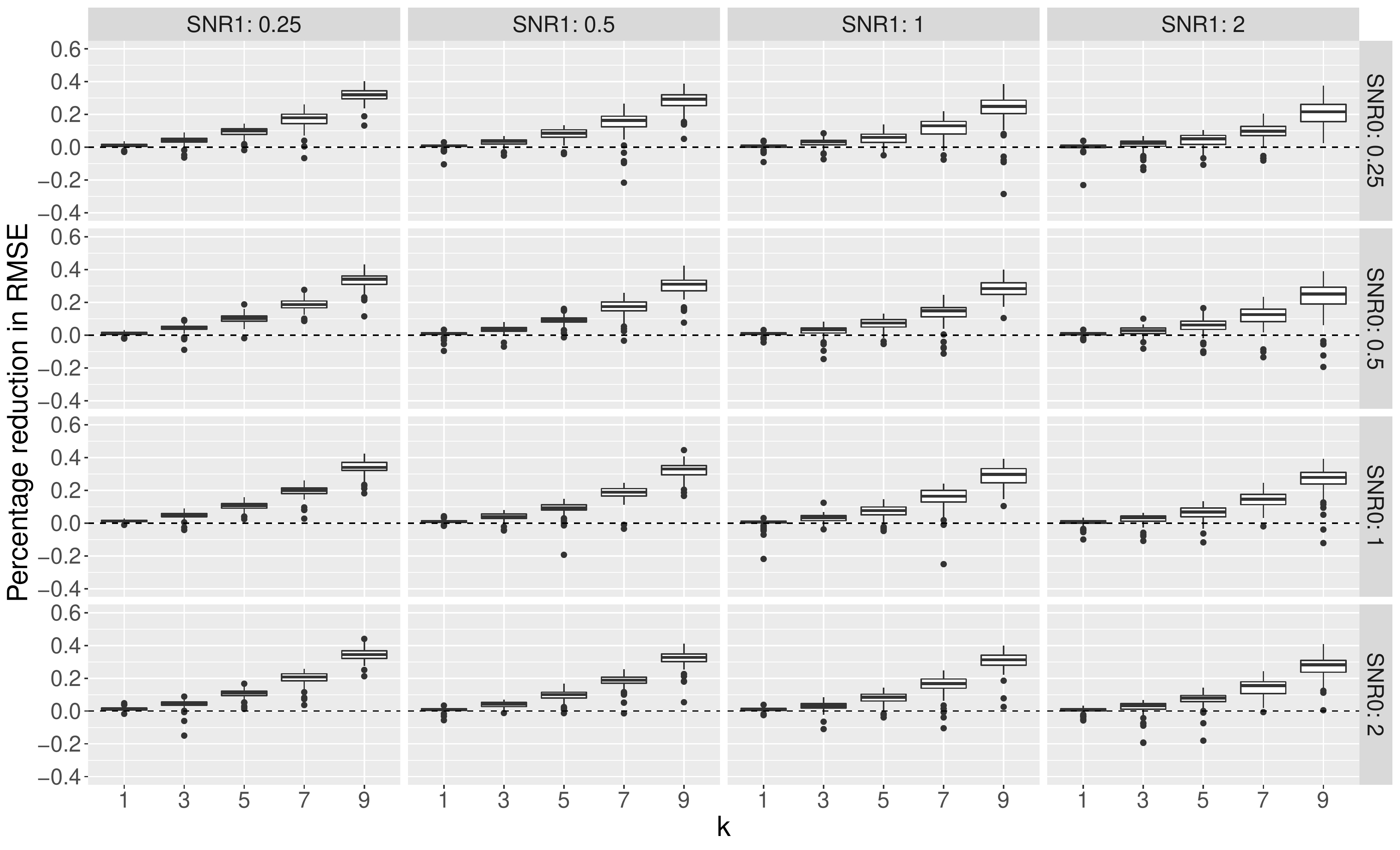}
  \caption{Percentage reduction in RMSE for cluster randomized experiments.}
  \label{figure:cluster-prrmse}
\end{figure}

\begin{figure}[ht]
  \includegraphics[width = \textwidth]{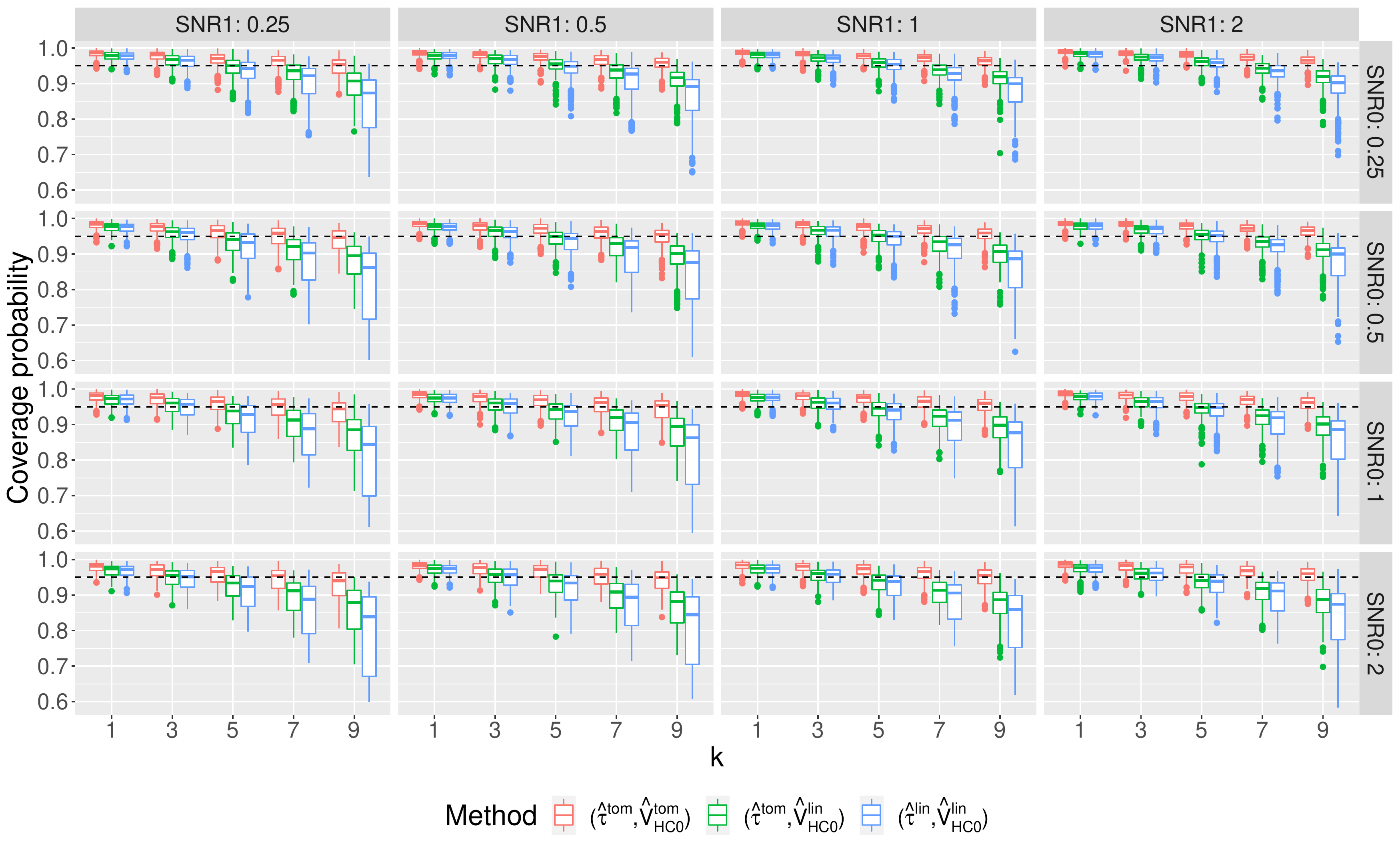}
  \caption{Coverage probabilities for cluster randomized experiments.}
  \label{figure:cluster-coverage}
\end{figure}

Figures~\ref{figure:cluster-prrmse} and \ref{figure:cluster-coverage} show the results. The conclusions are similar to those in completely randomized experiments. Despite a few outliers, the trend is more obvious when the data is generated with clustering feature.

\section{Heteroskedasticity-robust standard error and notation}
\label{sec:C}

Let $Y\in \mathbb{R}^{n}$ be the outcome vector, $X\in \mathbb{R}^{n\times k}$ be the covariate matrix and $W$ be a diagonal matrix. Consider a weighted regression with working model 
\begin{align*}
  Y=X\beta+e, \quad e \sim N(0, W^{-1}).
\end{align*}

The leverage score of the $i$th unit, denoted by $h_{i}$, is the $i$th diagonal entry  of the following matrix:
$$
X(X^\top WX)^{-1}X^\top W.
$$
Denote the estimated regression coefficient as $\hat{\beta}$, with
$$
\hat{\beta} = (X^\top WX)^{-1}X^\top WY.
$$
Let $\hei$ be the regression residual of unit $i$. Suppose that the target estimand is $d^\top \beta$, where $d$ is a known vector. Then the 
point estimator is $d^\top \hat{\beta}$ and the heteroskedasticity-robust variance estimator is 
$$
d^\top(X^\top W X)^{-1}X^\top W\Delta WX (X^\top W X)^{-1} d,
$$
where $\Delta$ is a diagonal matrix consisting of squared scaled residuals $\tei^2 = (\eta_i\hei)^2$, with $\eta_i$ varying for different estimating methods. In particular,
$\eta_i=1$ for $\textnormal{HC}_0$, $\eta_i=\{n/(n-k)\}^{-1/2}$ for $\textnormal{HC}_1$, $\eta_i=(1-h_i)^{-1/2}$ for $\textnormal{HC}_2$, and $\eta_i=(1-h_i)^{-1}$ for $\textnormal{HC}_3$.

We use lower case letter ``$s$" to denote sample variance and covariance. For example, $\sxyc$ is the sample covariance of $x_i$ and 
$Y_i(0)$, and $\syt$ is the sample variance of $Y_i(1)$. We use $``(z)"$ $(z=0,1)$ to denote sample mean, variance or covariance computed using samples from treatment arm $z$. 
For example, $\sxt$ is the sample covariance of $x_i$ in the treatment group and $\bhxt$ is the sample mean of $x_i$ in the treatment group. 
We use a hat to denote estimated quantity, such as $\bhxt$. 
Let $\xi_i$ denote a vector with $1$ at the $i$th dimension and $0$ at other dimensions. For square matrices $A$ and $B$, write $A>B$ if $A-B$ is positive definite and $A\geq B$ if $A-B$ is non-negative definite.
Let $[A]_{(i,j)}$ denote the $(i,j)$th element of matrix $A$. 
Let $\mathcal{S}_z=\{i:Z_i=z\}$ be the set of units under treatment arm $z$ (or $\mathcal{S}_z=\{hi:Z_{hi}=z\}$ for stratified experiment). 
Let $\|\cdot\|_{\textnormal{op}}$ and $\| \cdot \|_\infty$ denote the operator norm and infinity norm of a matrix, respectively.
For two random variables $U_1$ and $U_2$, write $U_1 \mathrel{\dot{\sim}} U_2$ if they have the same limiting distribution. Let $I_j$ be the identity matrix of dimension $j \times j$;
$1_j$ be the vector of all $1$'s of length $j$; and $0_j$ be the vector of all $0$'s of length $j$. We use $\max_{h,i}$ to denote $\max_{h=1}^H\max_{i=1}^{n_h}$ for short. We use $\max_{h,z}$ to denote $\max_{h=1}^H\max_{z\in\{1,0\}}$ for short.

\section{Proofs for the results under completely randomized experiments}
\label{sec:D}

\subsection{Preliminary results}

\begin{proposition}
  \label{prop:crt-beta-formula}
  $\tautyr = \htau-\hbetacr^\top\taux,$ where
\begin{align*}
  \hbetacr = 
    \{p_1^{-1}(1-n_1^{-1})\sxt+p_0^{-1}(1-n_0^{-1})\sxc\}^{-1}\{p_1^{-1}(1-n_1^{-1})\sxyt+p_0^{-1}(1-n_1^{-1})\sxyc\}.
\end{align*}
\end{proposition}
\begin{proof}
  Note that regression with weights $w_i$ is equivalent to OLS regression  with data multiplied by $w_i^{1/2}$.
   By Frisch--Waugh--Lovell (FWL) theorem \citep{ding2021frisch}, the estimated coefficient of $x_i$ in the weighted regression can be derived by 
   the OLS regression $w_i^{1/2}\tyi\mathrel{\sim}w_i^{1/2}\txi$, where
   $$\tyi=Y_i-Z_i\bhyt-(1-Z_i)\bhyc,\quad \txi = x_i-Z_i\bhxt-(1-Z_i)\bhxc.
   $$
Then 
$$
\hbetacr = \left(\sumicrt w_i \txi \txi^\top\right)^{-1}\left(\sumicrt w_i \txi\tyi\right).
$$
Simple algebra yeilds that 
\begin{align*}
  \sumicrt w_i \txi\txi^\top &= p_1^{-2}(n_1-1)\sxt+p_0^{-2}(n_0-1)\sxc,\\
  \sumicrt w_i \txi\tyi &= p_1^{-2}(n_1-1)\sxyt+p_0^{-2}(n_0-1)\sxyc.
\end{align*}
It follows that 
\begin{align*}
  \hbetacr = 
  \{p_1^{-1}(1-n_1^{-1})\sxt+p_0^{-1}(1-n_0^{-1})\sxc\}^{-1}\{p_1^{-1}(1-n_1^{-1})\sxyt+p_0^{-1}(1-n_0^{-1})\sxyc\}.
\end{align*}
By the property of OLS regression, $\tautyr$ is the estimated coefficient of $Z_i$ in the WLS regression of
$$
Y_i-x_i^\top\hbetacr\stackrel{w_i}{\sim} 1+Z_i.
$$
Therefore, $\tautyr = \htau-\hbetacr^\top\taux.$
\end{proof}

Lemma~\ref{lemma:var-cov-consistency} below is from \citet[][Lemma A16]{Li9157}.
\begin{lemma}
  \label{lemma:var-cov-consistency}
  Under Assumption~\ref{a:crt}, 
  \begin{align*}
    \syz-\Syz=\op(1),\quad \sxz-\Sx=\op(1),\quad \sxyz-\Sxyz = \op(1), \quad z=0,1.
  \end{align*} 
\end{lemma}
\begin{lemma}
  \label{lemma:crt-beta-consistency}
  Under Assumption~\ref{a:crt}, $\hbetacr - \betacr = \op(1).$
\end{lemma}
\begin{proof}
  By Proposition~\ref{prop:crt-beta-formula},
  \begin{align*}
    \hbetacr = 
    \{p_1^{-1}(1-n_1^{-1})\sxt+p_0^{-1}(1-n_0^{-1})\sxc\}^{-1}\{p_1^{-1}(1-n_1^{-1})\sxyt+p_0^{-1}(1-n_0^{-1})\sxyc\}.
  \end{align*}
  Under Assumption~\ref{a:crt} and by Lemma~\ref{lemma:var-cov-consistency}, we have
  \begin{align*}
    (p_1p_0)^{-1}\Sx-\{p_1^{-1}(1-n_1^{-1})\sxt+p_0^{-1}(1-n_0^{-1})\sxc\}=\op(1),\\
    p_1^{-1}\Sxyt+p_0^{-1}\Sxyc-\{p_1^{-1}(1-n_1^{-1})\sxyt+p_0^{-1}(1-n_0^{-1})\sxyc\} = \op(1).
  \end{align*}
  Therefore, 
  \begin{align*}
    \hbetacr - 
    \left\{(p_1p_0)^{-1}\Sx\right\} (p_1^{-1}\Sxyt+p_0^{-1}\Sxyc)=\op(1).
  \end{align*}
  By definition, the second term in the left-hand side of the above equation is equal to $\betacr$.
  Therefore, $\hbetacr-\betacr =\op(1)$.
\end{proof}


\begin{lemma}
  \label{lemma:crt-beta-sz}
  Under Assumption~\ref{a:crt},
  \begin{align*}
     \betacr = \argmin_{\beta}\left\{p_1^{-1}S^2_1(\beta)+
    p_0^{-1}S^2_0(\beta)\right\}.
  \end{align*}
\end{lemma}
\begin{proof}
  Note that 
  \begin{align*}
    p_1^{-1}S^2_1(\beta)+p_0^{-1}S^2_0(\beta) =& p_1^{-1}\left(\Syt-2\beta^\top \Sxyt + \beta^\top \Sx \beta\right) + p_0^{-1}\left(\Syc-2\beta^\top \Sxyc + \beta^\top \Sx \beta\right)\\
    =&\left(p_1^{-1}\Syt+p_0^{-1}\Syc\right) - 2\beta^\top \left(p_1^{-1}\Sxyt+p_0^{-1}\Sxyc\right)+(p_1p_0)^{-1}\beta^\top\Sx \beta \\
    =& \vtt + S_{\tau}^2 - 2\beta^\top\vxt + \beta^\top\vxx \beta,
  \end{align*}
where the last equality is due to the definition of $\vtt$, $\vxt$, and $\vxx$.

Taking derivative with respect to $\beta$, we have 
$$
\argmin_{\beta}\left\{p_1^{-1}S^2_1(\beta)+
    p_0^{-1}S^2_0(\beta)\right\} = \vxx^{-1}\vxt = \betacr.
$$
\end{proof} 

Let $\hei$ denote the residual of unit $i$ derived from the ToM regression adjustment. Let $s^2_{e(z)}$ denote the sample variance of the residuals corresponding to treatment arm $z$, i.e.,
\begin{align*}
  \set= (n_1-1)^{-1}\sumicrt Z_i\hei^2,\quad \sec= (n_0-1)^{-1} \sumicrt (1-Z_i)\hei^2.
\end{align*}
\begin{lemma}
  Under Assumption~\ref{a:crt},
  \label{lemma:crt-residual-variance-consistency}
  \begin{align*}
    \set- \Syt(\betacr) = \op(1),
    \quad \sec - \Syc(\betacr) = \op(1).
  \end{align*}
\end{lemma}
\begin{proof}
  Note that  $\hei = \tyi- \txi^\top\hbetacr$. 
  Therefore, 
  \begin{align*}
    (n_1-1)^{-1}\sumist \hei^2 &= \syt- 2\hbetacr^\top \sxyt + \hbetacr^\top \sxt \hbetacr
= \Syt - 2(\betacr)^\top \Sxyt + (\betacr)^\top \Sx \betacr + \op(1) \\
&= \Syt(\betacr)+\op(1).
  \end{align*}
  The second equality is obtained
  by Lemmas~\ref{lemma:var-cov-consistency} and \ref{lemma:crt-beta-consistency}.
  Similarly, we can prove the second half of Lemma~\ref{lemma:crt-residual-variance-consistency}.
\end{proof}

\begin{proposition}[\cite{li2017general}]
  \label{propA:crt-CLT}
  Under Assumption~\ref{a:crt},
  \begin{align*}
    n^{1/2}\begin{pmatrix}
      \htau - \tau \\
      \taux
    \end{pmatrix} \mathrel{\dot{\sim}} N\left(0,\begin{pmatrix}
      \vtt&\vtx\\
      \vxt & \vxx
      \end{pmatrix}\right).
  \end{align*}
\end{proposition}

\subsection{Proof of Proposition~\ref{prop:CRT-consistency}}
\begin{proof}
  Note that 
  \begin{align*}
    n^{1/2}(\tautyr - \tau) & = n^{1/2}\{\htau-\tau-(\betacr)^\top \taux\}+n^{1/2}(\betacr-\hbetacr)^\top\taux \\
    & =n^{1/2}\{\htau-\tau-(\betacr)^\top \taux\}+n^{1/2}\op(1) O_{\mathbb{P}}(n^{-1/2})\\
   & = n^{1/2}\{\htau-\tau-(\betacr)^\top \taux\}+ \op(1),
  \end{align*}
  where the first equality is due to Proposition~\ref{prop:crt-beta-formula} and the second equality is due to Lemma~\ref{lemma:crt-beta-consistency} and Proposition~\ref{propA:crt-CLT}.
  
  By Proposition~\ref{propA:crt-CLT} and the definition of $\betacr$, we have
  \begin{align*}
    n^{1/2}\{\htau-(\betacr)^\top \taux\} \mathrel{\dot{\sim}} N(0, \vtt-\vtx\vxx^{-1}\vxt).
  \end{align*}
  Compounded with Slusky's theorem, the conclusion follows.
\end{proof}

\subsection{Proof of Theorem~\ref{prop:valid-CRT-confidence-interval}}
  \begin{proof}
  Because completely randomized experiment is a special case of stratified randomized experiment with $H=1$. The conclusion for $j=2$ is a direct result of Theorem~\ref{thm:HC-str-limit}, so we omit its proof. The conclusions for $j=0,1,3$ can be proved with slight modifications of the proof of Theorem~\ref{thm:HC-str-limit}, so we omit them. 

\end{proof}

\subsection{Proof of Theorem~\ref{thm:total-distance-tyr-smaller-than-lin}}

\begin{proof}
By Proposition~\ref{prop:crt-beta-formula},
\begin{align*}
    \tautyr & = \htau-\taux^\top\hbetacr,\\
    \hbetacr &= \left\{\sumist p_1^{-2}\txi\txi^\top +\sumlsc p_0^{-2}\txl\txl^\top\right\}^{-1}\left\{\sumist p_1^{-2}\txi Y_i +\sumlsc p_0^{-2}\txl Y_i\right\}\\
    &= \left\{p_1^{-2}(n_1-1)\sxt + p_0^{-2}(n_0-1)\sxc\right\}^{-1}\left\{\sumist p_1^{-2}\txi Y_i +\sumlsc p_0^{-2}\txl Y_i\right\}.
\end{align*}
Rewritten $\tautyr$ as $
\tautyr = \sumist c^{\textnormal{tom}}_i Y_i-\sumlsc c^{\textnormal{tom}}_i Y_i $ , where
\begin{align*}
    c^{\textnormal{tom}}_i= 
      n_1^{-1} -\taux^\top\left\{p_1^{-2}(n_1-1)\sxt + p_0^{-2}(n_0-1)\sxc\right\}^{-1}p_1^{-2}\txi ,\quad i\in \mathcal{S}_1,\\
      c^{\textnormal{tom}}_i=n_0^{-1} +\taux^\top\left\{p_1^{-2}(n_1-1)\sxt + p_0^{-2}(n_0-1)\sxc\right\}^{-1}p_0^{-2} \txl , \quad i \in \mathcal{S}_0.
  \end{align*}
Note that
\begin{align*}
    \taulin & = \htau-\taux^\top(p_0\hbetat+p_1\hbetac),\\
    \hbetat & = \left\{\sumist\txi\txi^\top\right\}^{-1}\left\{\sumist \txi Y_i\right\}
     = \left\{(n_1-1) \sxt \right\}^{-1}\sumist \txi Y_i,
    \\
    \hbetac & = \left\{\sumlsc \txl\txl^\top\right\}^{-1}\left\{\sumlsc \txl Y_i\right\}
    = \left\{(n_0-1) \sxc\right\}^{-1} \sumlsc \txl Y_i.
\end{align*}
Rewritten $\taulin$ as $
\taulin = \sumist c^{\textnormal{lin}}_i Y_i-\sumlsc c^{\textnormal{lin}}_i Y_i $, where
  \begin{align*}
    c^{\textnormal{lin}}_i = 
      n_1^{-1} -p_0\taux^\top\left\{(n_1-1)\sxt\right\}^{-1}\txi,\quad i\in \mathcal{S}_1,\\
      c^{\textnormal{lin}}_i= n_0^{-1} +p_1\taux^\top\left\{(n_0-1)\sxc\right\}^{-1}\txl,\quad i\in \mathcal{S}_0.
  \end{align*}
  
  Next, we prove that $c^{\textnormal{tom}}$ minimizes the total distance 
  \begin{align*}
    F(c) = \sumist G(c_in_1) + \sumlsc G(c_in_0), \quad \text{where}\quad G(x) =  (x-1)^2/2,
  \end{align*}
  under the constraints \eqref{eq:constraint-total-weight} and \eqref{eq:constraint-exact-ATE-for-x} below.
  \begin{align}
  \label{eq:constraint-total-weight}
    \sumist c_i = 1, \quad \sumlsc c_i = 1,
  \end{align}
  \begin{align}
  \label{eq:constraint-exact-ATE-for-x}
      \sumist c_ix_i-\sumlsc c_ix_i=0.
  \end{align}
  In contrast, $c^{\textnormal{lin}}$ minimizes the total distance under the constraints \eqref{eq:constraint-total-weight} and \eqref{eq:constraint-exact-mean-for-x} below.
  \begin{align}
  \label{eq:constraint-exact-mean-for-x}
      \sumist c_ix_i=\bx,\quad \sumlsc c_ix_i=\bx,
  \end{align}
  Because \eqref{eq:constraint-exact-mean-for-x} implies \eqref{eq:constraint-exact-ATE-for-x}, $F(c^{\textnormal{tom}}) \leq F(c^{\textnormal{lin}})$.

  Denote $c$ the vector of $c_i$'s. Consider the following Lagrangian function:
  \begin{align*}
      \mathcal{L}^{\textnormal{tom}}(c,\lambda_1,\lambda_0,\lambda_x) &= \sumist 2^{-1}(c_in_1-1)^2+ \sumlsc 2^{-1}(c_in_0-1)^2-\\
      &\lambda_1\left(\sumist c_i-1\right)-\lambda_0\left(\sumlsc c_i-1\right)-\lambda_x^\top\left(\sumist c_ix_i-\sumlsc c_ix_i\right).
  \end{align*}
  Setting the gradient of $\mathcal{L}^{\textnormal{tom}}(c,\lambda_1,\lambda_0,\lambda_x)$ to $0$, we have
  \begin{align}
      n_1(c_in_1-1)-\lambda_1-\lambda_x^\top x_i=0, \quad i\in \mathcal{S}_1,\label{eq:gradient-c-i-tyr}\\
     n_0(c_in_0-1)-\lambda_0+\lambda_x^\top x_i=0, \quad i \in \mathcal{S}_0   \label{eq:gradient-c-l-tyr}.
  \end{align}
  Summarizing equation~\eqref{eq:gradient-c-i-tyr} for $i \in \mathcal{S}_1$ and by the constraint~\eqref{eq:constraint-total-weight}, we have
  \begin{align}
  \label{eq:lambda-t-lambdax}
     \lambda_1=-\lambda_x^\top \bhxt.
  \end{align}
  Summarizing equation~\eqref{eq:gradient-c-l-tyr} for $i \in \mathcal{S}_0$ and by the constraint~\eqref{eq:constraint-total-weight}, we have
    \begin{align}
    \label{eq:lambda-c-lambdax}
     \lambda_0=\lambda_x^\top \bhxc .
  \end{align}
  Plugging \eqref{eq:lambda-t-lambdax} into \eqref{eq:gradient-c-i-tyr} and \eqref{eq:lambda-c-lambdax} into \eqref{eq:gradient-c-l-tyr}, 
    \begin{align}
      n_1(c_in_1-1)-\lambda_x^\top \txi=0, \quad i\in \mathcal{S}_1, \label{eq:gradient-c-i-tyr-2}\\
      n_0(c_in_0-1)+\lambda_x^\top \txl=0, \quad i\in \mathcal{S}_0 \label{eq:gradient-c-l-tyr-2}.
  \end{align}
  Therefore, 
  \begin{align}
      c_i = n_1^{-1}+n_1^{-2}\lambda_x^\top \txi,\quad i\in \mathcal{S}_1,\label{eq:c-i-lambdax}\\
      c_i = n_0^{-1} -n_0^{-2}\lambda_x^\top \txl,\quad i\in \mathcal{S}_0\label{eq:c-l-lambdax}.
  \end{align}
  Plugging \eqref{eq:c-i-lambdax} and \eqref{eq:c-l-lambdax} into \eqref{eq:constraint-exact-ATE-for-x}, 
  \begin{align*}
      \taux+\left\{\sumist n_1^{-2} \txi\txi^\top+\sumlsc n_0^{-2} \txl\txl^\top\right\}\lambda_x=0.
  \end{align*}
  Therefore, 
  \begin{align}
  \label{eq:lambdax-formula}
    \lambda_x=  -\left\{\sumist n_1^{-2} \txi\txi^\top+\sumlsc n_0^{-2} \txl\txl^\top\right\}^{-1}\taux=-\left\{n_1^{-2}(n_1-1)\sxt + n_0^{-2}(n_0-1)\sxc\right\}^{-1}\taux.
  \end{align}
  Plugging \eqref{eq:lambdax-formula} into \eqref{eq:c-i-lambdax} and \eqref{eq:c-l-lambdax}, the minimizer of $F(c)$ under constraints \eqref{eq:constraint-total-weight} and \eqref{eq:constraint-exact-ATE-for-x} is
\begin{align*}
    c_i&= 
      n_1^{-1} -\taux^\top\left\{n_0^{-2}(n_0-1)\sxc+n_1^{-2}(n_1-1)\sxt\right\}^{-1}n_1^{-2}\txi \\
      &=n_1^{-1} -\taux^\top\left\{p_0^{-2}(n_0-1)\sxc+p_1^{-2}(n_1-1)\sxt\right\}^{-1}p_1^{-2}\txi \quad i\in \mathcal{S}_1,
      \\
      c_i&=n_0^{-1} +\taux^\top\left\{n_0^{-2}(n_0-1)\sxc+n_1^{-2}(n_1-1)\sxt\right\}^{-1}n_0^{-2} \txl\\
      &=n_0^{-1} +\taux^\top\left\{p_0^{-2}(n_0-1)\sxc+p_1^{-2}(n_1-1)\sxt\right\}^{-1}p_0^{-2} \txl, \quad i \in \mathcal{S}_0.
  \end{align*}

Similarly, consider the following lagrangian function:
\begin{align*}
    \mathcal{L}^{\textnormal{lin}}( & c,\lambda_1,\lambda_0,\lambda_{x1},\lambda_{x0}) = \sumist 2^{-1}(c_in_1-1)^2+ \sumlsc 2^{-1}(c_in_0-1)^2-\\
      &\lambda_1\left(\sumist c_i-1\right)-\lambda_0\left(\sumlsc c_i-1\right)-\lambda_{x1}^\top\left(\sumist c_ix_i-\bx\right)-\lambda_{x0}^\top\left(\sumlsc c_ix_i-\bx\right).
\end{align*}
Setting the gradient of $\mathcal{L}^{\textnormal{lin}}(c,\lambda_1,\lambda_0,\lambda_{x1},\lambda_{x0})$ to $0$, we have
  \begin{align}
      n_1(c_in_1-1)-\lambda_1-\lambda_{x1}^\top x_i=0, \quad i\in \mathcal{S}_1,\label{eq:gradient-c-i-lin}\\
     n_0(c_in_0-1)-\lambda_{0}-\lambda_{x0}^\top x_i=0, \quad i\in \mathcal{S}_0   \label{eq:gradient-c-l-lin}.
  \end{align}
  Summarizing equation~\eqref{eq:gradient-c-i-lin} for $i \in \mathcal{S}_1$ and by the constraint~\eqref{eq:constraint-total-weight}, we have
  \begin{align}
  \label{eq:lambda-t-lambdax-t}
     \lambda_{1}=-\lambda_{x1}^\top \bhxt.
  \end{align}
  Summarizing equation~\eqref{eq:gradient-c-l-lin} for $i \in \mathcal{S}_0$ and by the constraint~\eqref{eq:constraint-total-weight}, we have
    \begin{align}
    \label{eq:lambda-c-lambdax-c}
     \lambda_0=-\lambda_{x0}^\top \bhxc .
  \end{align}
  Plugging \eqref{eq:lambda-t-lambdax-t} into \eqref{eq:gradient-c-i-lin} and \eqref{eq:lambda-c-lambdax-c} into \eqref{eq:gradient-c-l-lin}, 
    \begin{align}
      n_1(c_in_1-1)-\lambda_{x1}^\top \txi=0, \quad i\in \mathcal{S}_1, \label{eq:gradient-c-i-lin-2}\\
      n_0(c_in_0-1)-\lambda_{x0}^\top \txl=0, \quad i\in \mathcal{S}_0 \label{eq:gradient-c-l-lin-2}.
  \end{align}
  Therefore, 
  \begin{align}
      c_i = n_1^{-1}+n_1^{-2}\lambda_{x1}^\top \txi,\quad i\in \mathcal{S}_1,\label{eq:c-i-lambdax-t}\\
      c_i = n_0^{-1} +n_0^{-2}\lambda_{x0}^\top \txl,\quad i\in \mathcal{S}_0\label{eq:c-l-lambdax-c}.
  \end{align}
  Plugging \eqref{eq:c-i-lambdax-t} and \eqref{eq:c-l-lambdax-c} into \eqref{eq:constraint-exact-mean-for-x}, 
  \begin{align*}
      \{ \bhxt-\bx \} +\left\{\sumist n_1^{-2} \txi\txi^\top\right\}\lambda_{x1}=0,\\
      \{ \bhxc-\bx \} +\left\{\sumlsc n_0^{-2} \txl\txl^\top\right\}\lambda_{x0}=0.
  \end{align*}
Therefore, 
  \begin{align}
    \lambda_{x1}=  -\left\{\sumist n_1^{-2} \txi\txi^\top\right\}^{-1}\{\bhxt-\bx\}=-\left\{n_1^{-2}(n_1-1)\sxt\right\}^{-1}p_0\taux,\label{eq:lambdax-t-formula}\\
     \lambda_{x0}=  -\left\{\sumlsc n_0^{-2} \txl\txl^\top\right\}^{-1}\{\bhxc-\bx\}=\left\{n_0^{-2}(n_0-1)\sxc\right\}^{-1}p_1\taux.\label{eq:lambdax-c-formula}
  \end{align}
  Plugging \eqref{eq:lambdax-t-formula} into \eqref{eq:c-i-lambdax-t} and \eqref{eq:lambdax-c-formula} into \eqref{eq:c-l-lambdax-c}, the minimizer of $F(c)$ under constraints \eqref{eq:constraint-total-weight} and \eqref{eq:constraint-exact-mean-for-x} is
  \begin{align*}
    c_i = 
      n_1^{-1} -p_0\taux^\top\left\{(n_1-1)\sxt\right\}^{-1}\txi,\quad i\in \mathcal{S}_1,\\
      c_i= n_0^{-1} +p_1\taux^\top\left\{(n_0-1)\sxc\right\}^{-1}\txl,\quad i\in \mathcal{S}_0.
  \end{align*}
 \end{proof}

 \subsection{Proof of Theorem~\ref{thm:leverage-score-tyr-smaller-than-lin}} 
 \begin{proof}
  By definition, the leverage score $h_{i}^{\textnormal{tom}}$ is the $i$th diagonal element of 
 $$
 X^{\textnormal{tom}}\left\{ (X^{\textnormal{tom}})^\top W X^{\textnormal{tom}}\right\}^{-1} (X^{\textnormal{tom}})^\top W,
 $$
 where $X^{\textnormal{tom}}$ is an ${n\times(2+k)}$ matrix with the $i$th row of $X^{\textnormal{tom}}$ being $(1,Z_i,x_i^\top)$.
 
 The leverage score $h_{i}^{\textnormal{lin}}$ is the $i$th diagonal element of 
 $$
 X^{\textnormal{lin}}\left\{ (X^{\textnormal{lin}})^\top  X^{\textnormal{lin}}\right\}^{-1} (X^{\textnormal{lin}})^\top,
 $$
 where $X^{\textnormal{lin}} \in \mathbb{R}^{n\times(2+2k)}$  with the $i$th row of $X^{\textnormal{lin}}$ being $(1,Z_i,(x_i-\bx)^\top,Z_i(x_i-\bx)^\top)$.
 
 Let $\breve{X}^{\textnormal{tom}} \in \mathbb{R}^{n\times(2+k)}$ with the $i$th row  being $(1-Z_i,Z_i,\txi^\top)$.
 Let $\breve{X}^{\textnormal{lin}}\in \mathbb{R}^{n\times(2+2k)}$ with the $i$th row  being $(1-Z_i,Z_i,(1-Z_i)\txi^\top,Z_i\txi^\top)$.
 Since
 \begin{align*}
  \breve{X}^{\textnormal{tom}} &= X^{\textnormal{tom}} \left(
    \begin{array}{ccc}
       1  & 0 & 0 \\ 
        -1 &  1 & 0\\ 
        0 & 0 & I_k
    \end{array}
    \right) \left(
      \begin{array}{ccc}
         1  & 0 & -\bhxc^\top \\ 
          0 &  1 & -\bhxt^\top\\  
          0 & 0 & I_k
      \end{array}
      \right),\\
  \breve{X}^{\textnormal{lin}} &= X^{\textnormal{lin}} \left(
    \begin{array}{cccc}
       1  & 0 & 0 &0\\ 
        -1 &  1 & 0 &0\\ 
        0 & 0 & I_k &0\\
        0 & 0 & -I_k &I_k\\  
    \end{array}
    \right) \left(
      \begin{array}{cccc}
        1  & 0 & \bx^\top-\bhxc^\top &0\\ 
         0 &  1 & 0 &\bx^\top-\bhxt^\top\\ 
         0 & 0 & I_k &0\\
         0 & 0 & 0 &I_k\\  
     \end{array}
      \right),
 \end{align*}
then
 \begin{align*}
      X^{\textnormal{lin}}\left\{ (X^{\textnormal{lin}})^\top  X^{\textnormal{lin}}\right\}^{-1} (X^{\textnormal{lin}})^\top&=      \breve{X}^{\textnormal{lin}}\left\{ (\breve{X}^{\textnormal{lin}})^\top  \breve{X}^{\textnormal{lin}}\right\}^{-1} (\breve{X}^{\textnormal{lin}})^\top,\\
 X^{\textnormal{tom}}\left\{ (X^{\textnormal{tom}})^\top W X^{\textnormal{tom}}\right\}^{-1} (X^{\textnormal{tom}})^\top W&=\breve{X}^{\textnormal{tom}}\left\{ (\breve{X}^{\textnormal{tom}})^\top W \breve{X}^{\textnormal{tom}}\right\}^{-1} (\breve{X}^{\textnormal{tom}})^\top W.
 \end{align*}
Note that 
 \begin{align*}
     (\breve{X}^{\textnormal{lin}})^\top  \breve{X}^{\textnormal{lin}} &= \left(
     \begin{array}{cccc}
        n_0  & 0 & 0 & 0\\ 
         0 &  n_1 & 0 & 0\\  
         0 & 0 & (n_1-1)\sxt & 0\\ 
         0 & 0 & 0 & (n_0-1)\sxc
     \end{array}\right),\\
     (\breve{X}^{\textnormal{tom}})^\top W \breve{X}^{\textnormal{tom}} &=  \left(
     \begin{array}{ccc}
        n_0p_0^{-2}  & 0 & 0\\ 
         0 &  n_1p_1^{-2} & 0 \\ 
         0 & 0 & p_1^{-2}(n_1-1)\sxt + p_0^{-2}(n_0-1)\sxc
     \end{array}\right).
 \end{align*}
 Therefore,
   \begin{align*}
    h_{i}^{\textnormal{lin}} = \begin{cases}
      n_1^{-1} + \txi^\top \left\{(n_1-1)\sxt\right\}^{-1}\txi, \quad \text{for}\quad i\in\mathcal{S}_1,\\
      n_0^{-1} + \txi^\top\left\{(n_0-1)\sxc\right\}^{-1}\txi, \quad \text{for}\quad i\in\mathcal{S}_0,
    \end{cases}
   \end{align*}
  \begin{align*}
  h_{i}^{\textnormal{tom}} = \begin{cases}
    n_1^{-1} + \txi^\top \left\{(n_1-1)\sxt+(p_1/p_0)^2(n_0-1)\sxc\right\}^{-1}\txi, \quad \text{for}\quad i\in\mathcal{S}_1,\\
    n_0^{-1} + \txi^\top\left\{(n_0-1)\sxc+(p_0/p_1)^2(n_1-1)\sxt\right\}^{-1}\txi, \quad \text{for}\quad i\in\mathcal{S}_0.
  \end{cases}
 \end{align*} 
 Since $\sxt \geq 0 $ and $\sxc \geq 0$, then
 \begin{align*}
  \left\{(n_1-1)\sxt\right\}^{-1} &\geq \left\{(n_1-1)\sxt+(p_1/p_0)^2(n_0-1)\sxc\right\}^{-1},\\
   \left\{(n_0-1)\sxc\right\}^{-1} & \geq \left\{(n_0-1)\sxc+(p_0/p_1)^2(n_1-1)\sxt\right\}^{-1}.
 \end{align*}
 Therefore, 
 $
   h_{i}^{\textnormal{tom}} \leq h_{i}^{\textnormal{lin}} .
$

\end{proof}

\section{Proofs for the results under stratified randomized experiments}
\label{sec:E}

\subsection{Preliminary results}

Let $\txhi=x_{hi}-Z_i\bhxht-(1-Z_i)\bhxhc$ and $\tyhi=Y_{hi}-Z_i\bhyht-(1-Z_i)\bhyhc$. Let $\mathcal{S}_{hz} = \{i=1,\ldots,n_h: Z_{hi}=z \}$ for $z=0,1$, $h=1,\ldots,H$. 
\begin{proposition}
  \label{prop:str-formula}
  $\taustrtyr = \htaustr-\hbetastr^\top \tauxstr$, where 
  \begin{eqnarray*}
    \label{eq:str-weight-2}
    \hbetastr= \left[ \sumh \big\{ \pih\pht^{-1}\shxt+\pih\phc^{-1}\shxc \big\} \right]^{-1}\left[\sumh \big\{ \pih\pht^{-1}\shxyt+\pih\phc^{-1}\shxyc \big\} \right]. 
  \end{eqnarray*}
\end{proposition}

\begin{proof}
  By FWL theorem, 
  the estimated coefficient of $x_{hi}$ in the weighted regression can be dervied as the OLS regression of 
  $w_{hi}^{1/2}\tyhi\mathrel{\sim} w_{hi}^{1/2}\txhi$.
  Therefore, 
  $$
  \hbetastr = \left(\sumh\sumistr \whi\txhi\txhi^\top\right)^{-1}\left(\sumh\sumistr \whi\txhi\tyhi\right).
  $$
  Simple algebra gives that 
  \begin{align*}
   \sumh\sumistr \whi\txhi\txhi^\top &= \sumh \big\{ \pht^{-2}\nht\shxt+\phc^{-2}\nhc\shxc \big\},\\
   \sumh\sumistr \whi\txhi\tyhi^\top &= \sumh \big\{ \pht^{-2}\nht\shxyt+\phc^{-2}\nhc\shxyc \big\}.
  \end{align*}
  Therefore,
  \begin{align*}
   \hbetastr= \left[ \sumh \big\{ \pih\pht^{-1}\shxt+\pih\phc^{-1}\shxc \big\} \right]^{-1}\left[\sumh \big\{ \pih\pht^{-1}\shxyt+\pih\phc^{-1}\shxyc \big\} \right].
  \end{align*}
  By the property of OLS, $\taustrtyr$ is the estimated coefficient of $Z_{hi}$ in the WLS regression: 
  $$
  Y_{hi}-x_{hi}^\top\hbetastr \stackrel{\whi}{\sim} 1 + Z_{hi}+\sum_{q=2}^{H} (\delta_{hq}-\pi_q)
+Z_{hi} \sum_{q=2}^{H} (\delta_{hq}-\pi_q).
  $$
  It follows that  $\taustrtyr = \htaustr-\hbetastr^\top \tauxstr$.
\end{proof}

\begin{lemma}
  \label{lemma:str-var-cov-consistency}
  Under Assumption~\ref{a:strata}, for $z=0,1$, we have
  \begin{align*}
    \sumh \pih\phz^{-1}\shxz - \sumh\pih\phz^{-1}\Shx = \op(1),\\
    \sumh \pih\phz^{-1}\shyz - \sumh\pih\phz^{-1}\Shyz=\op(1),\\
    \sumh \pih\phz^{-1}\shxyz - \sumh\pih\phz^{-1}\Shxyz=\op(1).
  \end{align*}
  \begin{proof}
    These are direct results of Lemma 7 in \cite{wang2021rerandomization}, although Assumption~\ref{a:strata} is slightly weaker than that used by \cite{wang2021rerandomization}. 
  \end{proof}
\end{lemma}

\begin{lemma}
  \label{lemma:str-beta-consistency}
  Under Assumption~\ref{a:strata}, 
  \begin{align*}
    \hbetastr - \betastr = \op(1).
  \end{align*}
\end{lemma}
\begin{proof}
  By Lemma~\ref{lemma:str-var-cov-consistency},
  \begin{align*}
    \sumh \pih\phz^{-1}\shxz - \sumh\pih\phz^{-1}\Shx = \op(1),\\
    \sumh \pih\phz^{-1}\shyz - \sumh\pih\phz^{-1}\Shyz=\op(1).\\
    \end{align*}
  Therefore, 
  \begin{align*}
    &\left\{\sumh \pih\pht^{-1}\shxt+\pih\phc^{-1}\shxc\right\}^{-1}\left(\sumh \pih\pht^{-1}\shxyt+\pih\phc^{-1}\shxyc\right)-\\ 
  &\left\{ \sumh\pih(\pht\phc)^{-1}\Shx\right\}^{-1}\left(\sumh\pih\pht^{-1}\Shxyt+\pih\phc^{-1}\Shxyc\right) = \op(1).
  \end{align*}
  The first term is $\hbetastr$ and the second term is $\betastr$. Therefore, the conclusion follows.
\end{proof}

Let $\hehi$ be the residuals from the weighted regression~\eqref{formula:stratified}. One of the variance estimator 
can be derived as 
\begin{align}
  \label{formula:variance-estimator-strata}
  \hvstr=\nstr^{-1}\sumh \left\{\pi_h p_{h1}^{-1}\shet +\pi_h p_{h0}^{-1}\shec\right\},
\end{align}
where
$$
\shet= (n_{h1}-1)^{-1} \sum_{i=1}^{n_h} Z_{hi} \hehi^2,\quad \shec= (n_{h0}-1)^{-1}  \sum_{i=1}^{n_h} (1 - Z_{hi})  \hehi^2.
$$
Lemma~\ref{lemma:strata-variance-estimator-consistent} below shows that \eqref{formula:variance-estimator-strata} is a
conservative estimator of the variance of $\taustrtyr$.
\begin{lemma}
  \label{lemma:strata-variance-estimator-consistent}
Under Assumption~\ref{a:strata},
\begin{equation}
  \label{eq:hvstr-limit}
  \nstr\hvstr =  \min_{\beta}\sum_{h=1}^H\left\{ 
\pi_h p_{h1}^{-1}S_{h1}^2(\beta)+\pih p^{-1}_{h0}S_{h0}^2(\beta)\right\} +o_{\mathbb{P}}(1), \nonumber
\end{equation}
where
$$
S^2_{hz}(\beta) = (n_h-1)^{-1}\sumistr \{Y_{hi}(z)-\byhz-(x_{hi}-\bxh)^\top \beta\}^2.
$$
\end{lemma}

\begin{proof}
  Note that 
  \begin{align*}
   & \nstr \hvstr\\
   = &\sumh \left\{\pi_h p_{h1}^{-1}\shet +\pi_h p_{h0}^{-1}\shec\right\} \\
    =&\sumh \pi_h p_{h1}^{-1}\left\{\shyt-2\hbetastr^\top \shxyt + \hbetastr^\top\shxt\hbetastr\right\} +\pi_h p_{h0}^{-1}\left\{\shyc-2\hbetastr^\top \shxyc + \hbetastr^\top\shxc\hbetastr\right\} \\
    =&\sumh \pi_h p_{h1}^{-1}\left\{\Shyt-2(\betastr)^\top \Shxyt + (\betastr)^\top\Shx\betastr\right\} \\
    & \quad +\pi_h p_{h0}^{-1}\left\{\Shyc-2(\betastr)^\top \Shxyc + (\betastr)^\top\Shx\betastr\right\}+\op(1)\\
    =& \sum_{h=1}^H \pi_h\left\{ 
       p_{h1}^{-1}S_{h1}^2(\betastr)+p^{-1}_{h0}S_{h0}^2(\betastr)\right\} + \op(1). \numberthis \label{eq:residual-consistency-1}
  \end{align*}
  The second equality is derived by Lemmas~\ref{lemma:str-var-cov-consistency} and \ref{lemma:str-beta-consistency}.
  By the optimality of $\betastr$,
  \begin{align}
    \label{eq:residual-consistency-2}
    \betastr = \argmin_{\beta}\sum_{h=1}^H
      \pih \left\{ \pht^{-1}S_{h1}^2(\beta)+\phc^{-1}S_{h0}^2(\beta)\right\}.
  \end{align}
  The conclusion follows from \eqref{eq:residual-consistency-1} and \eqref{eq:residual-consistency-2}.
\end{proof}
The following proposition is from \cite{wang2021rerandomization}. Assumption~\ref{a:strata} is slightly weaker than that used by \cite{wang2021rerandomization}, but it does not affect the conclusion.
\begin{proposition}[\cite{wang2021rerandomization}]
  \label{propA:str-CLT}
  Under Assumption~\ref{a:strata}, 
  \begin{align*}
    \nstr^{1/2}\begin{pmatrix}
      \htaustr-\taustr\\
      \tauxstr
    \end{pmatrix} \mathrel{\dot{\sim}} N\left(0,\begin{bmatrix}
      V_{\textnormal{str},\tau\tau}&V_{\textnormal{str},\tau x}\\
      V_{\textnormal{str},x\tau} & V_{\textnormal{str},xx}
      \end{bmatrix}\right).
  \end{align*}
\end{proposition}


\subsection{An equivalent form of regression formula}
\label{sec:equivalent-formula}
In this section, we prove that two regression formulas \eqref{formula:str-1} and \eqref{formula:str-2} below 
are equivalent in terms of point and variance estimators for the average treatment effect under stratified randomized experiments. It is useful for proving Theorem~\ref{thm:HC-str-limit}.

 Recall the regression formula we use in the main text
 \begin{align}
 \label{formula:str-1}
  Y_{hi} \stackrel{\whi}{\sim} 1 + Z_{hi}+\sum_{q=2}^{H} (\delta_{hq}-\pi_q)
+Z_{hi} \sum_{q=2}^{H} (\delta_{hq}-\pi_q)+
x_{hi},
\end{align}
where $\delta_{hq}=1$ if $q=h$ and $\delta_{hq} = 0$ otherwise, and 
$$\whi=\zhi\pht^{-2}\frac{\nht}{\nht-1}+(1-\zhi)\phc^{-2}\frac{\nhc}{\nhc-1}.$$
It is equivalent to the following weighed regression
\begin{align}
\label{formula:str-2}
  Y_{hi} \stackrel{\whi}{\sim}\sum_{q=1}^{H} Z_{hi}\delta_{hq}
+ \sum_{q=1}^{H} (1-Z_{hi})\delta_{hq}+(x_{hi}-\bxh).
\end{align}

Let $\Xstr = (X_{\textnormal{str},1}^\top,\ldots,X_{\textnormal{str},H}^\top)^\top \in \mathbb{R}^{\nstr\times (2H+k)}$ be the design matrix of regression \eqref{formula:str-1}
 with the $i$th row of $X_{\textnormal{str},h}$ being
\begin{align*}
    (1,Z_{hi},\delta_{h2}-\pi_2,\ldots,\delta_{hH}-\pi_H,Z_{hi}(\delta_{h2}-\pi_2),\ldots,Z_{hi}(\delta_{hH}-\pi_H),x_{hi}^\top).
\end{align*}
Let  $E = (E_1^\top,\ldots,E_H^\top)^\top \in \mathbb{R}^{\nstr\times (2H+k)}$ be the design matrix of regression \eqref{formula:str-2}
 with the $i$th row of $E_h$ being
\begin{align*}
    (Z_{hi}\delta_{h1},\ldots,Z_{hi}\delta_{hH},(1-Z_{hi})\delta_{h1},\ldots,(1-Z_{hi})\delta_{hH},(x_{hi}-\bxh)^\top).
\end{align*}
Let $W$ be the digonal matrix of $w_{hi}$'s and $Y$ be the vector of $Y_{hi}$'s $(h=1,\ldots,H,i=1,\ldots,n_h)$.
Let $\hat{\beta}_1$ and $\hat{\beta}_2$ be the estimated coefficients of regression~\eqref{formula:str-1} and \eqref{formula:str-2}, respectively. Then
\begin{align*}
    \hat{\beta}_1 = (\Xstr^\top W \Xstr)^{-1}\Xstr^\top W Y,\quad \hat{\beta}_2 = (E^\top W E)^{-1}E^\top W Y.
\end{align*}
Next, we prove some lemmas to build the equivalence between these two regressions.
Let $\ell = (0_{2H}^\top,1_{k}^\top)^\top$. Let $d=(\pi_1,,\ldots,\pi_H,-\pi_1,\ldots,-\pi_H,0_{k}^\top)^\top$ be a vector of length $2H+k$. Lemma~\ref{lemma:same-betax-coef-for-2-formula} below shows that they have the same estimated coefficient for the covariates.
\begin{lemma}
\label{lemma:same-betax-coef-for-2-formula}
\begin{align*}
    \ell^\top  \hat{\beta}_1 = \ell^\top\hat{\beta}_2 = \hbetastr.
\end{align*}
\end{lemma}
\begin{proof}
In the proof of Proposition~\ref{prop:str-formula}, we have shown that
   $$
   \ell^\top  \hat{\beta}_1  = \hbetastr.
   $$
Similar to the proof of Proposition \ref{prop:str-formula} with FWL theorem, we have
\begin{align*}
    \ell^\top\hat{\beta}_2 = \hbetastr.
\end{align*}
\end{proof}
Lemma~\ref{lemma:same-ATE-estimator-for-2-formula} below shows that we can derive the same average treatment effect estimator. Recall that $\xi_2 \in \mathbb R^{2H+k}$ is a vector with $1$ at the second dimension and $0$ at other dimensions.
\begin{lemma}
\label{lemma:same-ATE-estimator-for-2-formula}
\begin{align*}
    \xi_2^\top  \hat{\beta}_1 = d^\top\hat{\beta}_2 = \taustrtyr.
\end{align*}
\end{lemma}
\begin{proof}
In the proof of Proposition~\ref{prop:str-formula}, we have shown that
\begin{align*}
    \xi_2^\top  \hat{\beta}_1 = \taustrtyr.
\end{align*}
It suffices to show that
\begin{align*}
d^\top\hat{\beta}_2 = \taustrtyr.
\end{align*}
 By the property of OLS and Lemma~\ref{lemma:same-betax-coef-for-2-formula}, the estimated coefficient of $(1-Z_{hi})\delta_{hq}$ and $Z_{hi}\delta_{hq}$ $(q=1,\ldots,H)$ can be derived in the WLS regression of 
 
  $$
  Y_{hi}-(x_{hi}-\bxh)^\top\hbetastr \stackrel{\whi}{\sim} \sum_{q=1}^{H} Z_{hi}\delta_{hq}+\sum_{q=1}^{H} (1-Z_{hi})\delta_{hq}.
  $$
  It follows that  the estimated coefficients of $Z_{hi}\delta_{hq}$ and $(1-Z_{hi})\delta_{hq}$  are,  respectively, 
 \begin{align*}
   \hat{\bar{Y}}_{q}(1)- \{\hat{\bar{x}}_{q}(1)-\bar{x}_{q}\}^\top\hbetastr,  \quad \hat{\bar{Y}}_{q}(0)-\{\hat{\bar{x}}_{q}(0)-\bar{x}_{q}\}^\top\hbetastr,
 \end{align*}
for $q=1,\ldots,H$.
 Therefore, 
 \begin{align*}
    d^\top  \hat{\beta}_2 =& \sumh\pih \left[\bhyht-\{\bhxht-\bxh\}^\top\hbetastr-\bhyhc+\{\bhxhc-\bxh\}^\top\hbetastr\right]\\
    =&\sumh\pih \left[\bhyht-\bhyhc-\{\bhxht-\bhxhc\}^\top\hbetastr\right]=\taustrtyr.
\end{align*}
\end{proof}
\begin{lemma}
\label{lemma:same-residual-for-2-formula}
The residuals from regressions \eqref{formula:str-1} and \eqref{formula:str-2} are the same, which are equal to $\tyhi-\txhi^\top\hbetastr$ for unit $hi$.
\end{lemma}
\begin{proof}
By the property of OLS, the residuals of regression \eqref{formula:str-2}
are equal to those of the following regression:
  $$
  Y_{hi}-(x_{hi}-\bxh)^\top\hbetastr \stackrel{\whi}{\sim} \sum_{q=1}^{H} Z_{hi}\delta_{hq}+\sum_{q=1}^{H} (1-Z_{hi})\delta_{hq}.
  $$
  The residuals of regression \eqref{formula:str-1} are equal to those of the following regression:
  $$
  Y_{hi}-x_{hi}^\top\hbetastr \stackrel{\whi}{\sim} 1 + Z_{hi}+\sum_{q=2}^{H} (\delta_{hq}-\pi_q)
+Z_{hi} \sum_{q=2}^{H} (\delta_{hq}-\pi_q).
  $$
  Note that the fitted values of the above two regressions are the same for units in the same stratum under the same treatment arm. Therefore, the fitted value of unit $hi$ is the mean value over the units in the same stratum under the same treatment arm with $hi$. Thus, the residuals of unit $hi$ of regressions \eqref{formula:str-1} and \eqref{formula:str-2} are both equal to
$\tyhi-\txhi^\top\hbetastr$.
\end{proof}

The leverage scores of these two regressions are the diagonal elements of the following matrices
$$
 \Xstr\left(\Xstr^\top W \Xstr\right)^{-1} \Xstr^\top W, \quad E\left(E^\top W E\right)^{-1} E^\top W.
$$

As shown in the proof of Lemma~\ref{lemma:same-variance-estimator}, $E=\Xstr Q$ (The explicit formula of $Q$ can be found in the proof of Lemma~\ref{lemma:same-variance-estimator}). The fact that $Q$ is an invertible matrix indicates that 
\begin{align*}
  \Xstr\left(\Xstr^\top W \Xstr\right)^{-1} \Xstr^\top W= E\left(E^\top W E\right)^{-1} E^\top W.
\end{align*}
Therefore, the leverage scores of these two regression formulas are the same. We denote by $h_{hi,\textnormal{str}}$ the leverage score corresponding to unit $hi$. We will derive the formula of $h_{hi,\textnormal{str}}$
in Section~\ref{sec:leverage-stratified-experiment}.

Let
$\hehi = \tyhi-\txhi^\top\hbetastr$ be the regression residual of unit $hi$. Let $\tilde{e}_{hi}=\eta_{hi}\hehi$ be the scaled residual where $\eta_{hi}=1$  for $\textnormal{HC}_0$, $\eta_{hi} = \{\nstr/(\nstr-2H-k)\}^{1/2}$ for $\textnormal{HC}_1$, $\eta_{hi}  = (1-h_{hi,\textnormal{str}})^{-1/2}$ for $\textnormal{HC}_2$, $\eta_{hi}=(1-h_{hi,\textnormal{str}})^{-1}$ for $\textnormal{HC}_3$.
Let $\Delta$ be the diagonal matrix of $\tehi^2$ $(h=1,\ldots,H,i=1,\ldots,n_h)$. 

By Lemma~\ref{lemma:same-ATE-estimator-for-2-formula}, regressions \eqref{formula:str-1} and~\eqref{formula:str-2} lead to two variance estimators for $\taustrtyr$, which are derived as
\begin{align*}
&\xi_2^\top(\Xstr^\top W\Xstr)^{-1}\Xstr^\top W \Delta W\Xstr(\Xstr^\top W\Xstr)^{-1}\xi_2,\\
  & d^\top(E^\top WE)^{-1}E^\top W \Delta WE(E^\top WE)^{-1}d.
\end{align*}

Lemma~\ref{lemma:same-variance-estimator} below shows the equivalence of these two variance estimators.
\begin{lemma}
\label{lemma:same-variance-estimator}
\begin{align*}
&\xi_2^\top(\Xstr^\top W\Xstr)^{-1}\Xstr^\top W \Delta W\Xstr(\Xstr^\top W\Xstr)^{-1}\xi_2 \\
&=
   d^\top(E^\top WE)^{-1}E^\top W \Delta WE(E^\top WE)^{-1}d.
\end{align*}
\end{lemma}
\begin{proof}
First, we give the explicit formula of $Q$ subject to $E=\Xstr Q$. Let $P_{i,j} \in \mathbb R^{(2H+k) \times (2H+k)}$ denote the matrix with the $(i,j)$th element being $1$ and the other elements being $0$.
Let $I$ denote identify matrix of size $2H+k$. We can verify that 
\begin{align*}
    Q &= \prod_{q=2}^{H}(I+\pi_{q}P_{1,q+1})\prod_{q=2}^{H}(I+\pi_{q}P_{2,q+H})\prod_{q=2}^{H}(I-P_{H+q,q+1})\prod_{q=2}^{H}(I-P_{H+q,2})\prod_{t=2}^{2H}(I-P_{t,1}) Q_1 Q_2,
\end{align*}
where 
\begin{align*}
        Q_1 &= (\xi_2,\xi_{H+2},\ldots,\xi_{2H},\xi_{1},\xi_{3},\ldots,\xi_{H+1}),\quad 
    Q_2 = \left(\begin{array}{cc}
    I_{2H} & A \\
    0 & I_k
    \end{array}\right),
\end{align*}
$$
A = \left(\bar{x}_{1},\ldots,{\bar{x}}_{H},{\bar{x}}_{1},\ldots,{\bar{x}}_{H}\right)^\top.
$$
Here $(I+\pi_{q}P_{1,q+1})$ $(q=2,\ldots,H)$ corresponds to the operation of changing $\delta_{hq}-\pi_q$ to $\delta_{hq}$; $(I+\pi_{q}P_{2,q+H})$ $(q=2,\ldots,H)$ corresponds to the operation of changing $Z_{hi}(\delta_{hq}-\pi_q)$ to $Z_{hi}\delta_{hq}$; $(I-P_{H+q,q+1})$ $(q=2,\ldots,H)$ corresponds to the operation of changing $\delta_{hq}$ to $(1-Z_{hi})\delta_{hq}$; $\prod_{q=2}^{H}(I-P_{H+q,2})$ corresponds the operation of changing $Z_{hi}$ to  $Z_{hi}\delta_{h1}$; $\prod_{t=2}^{2H}(I-P_{t,1})$ corresponds to the operation of changing $1$ to $(1-Z_{hi})\delta_{h1}$; $Q_1$ corresponds to the operation of reordering the positions of the regressors; and $Q_2$ corresponds to the operation of centering $x_{hi}$ at $\bar{x}_{h}$.

After some calculation, we can verify that
\begin{align*}
   \xi_2^\top Q =  d^\top.
\end{align*}
Therefore,
\begin{align*}
    & \xi_2^\top(\Xstr^\top W\Xstr)^{-1}\Xstr^\top W \Delta W\Xstr(\Xstr^\top W\Xstr)^{-1}\xi_2 \\
    =& \xi_2^\top Q (E^\top WE)^{-1}E^\top W \Delta WE(E^\top WE)^{-1} Q^\top\xi_2\\
    =&  d^\top(E^\top WE)^{-1}E^\top W \Delta WE(E^\top WE)^{-1}d.
\end{align*}
\end{proof}

\subsection{Leverage scores of ToM regression in stratified randomized experiments}
\label{sec:leverage-stratified-experiment}
Define $\hvxxstr = \sumh \pih \left\{\pht^{-1}\shxt+\phc^{-1}\shxc\right\}.$ Define $w_h(z)$ the regression weights for units in stratum $h$ under treatment arm $z$ with $$\whi=\zhi\wht+(1-\zhi)\whc.$$ 

Proposition~\ref{prop:leverage-formula-str} below provides the formula of leverage scores of ToM regression in stratified randomized experiments.
 \begin{proposition}
 \label{prop:leverage-formula-str}
   \begin{align*}
    h_{hi,\textnormal{str}} = \begin{cases}
    \nht^{-1}+\txhi^\top\hvxxstr^{-1}\txhi w_h(1) \nstr^{-1}, \quad i \in \sth,\\
    \nhc^{-1}+\txhi^\top\hvxxstr^{-1}\txhi w_h(0)\nstr^{-1}, \quad i \in \sch.
    \end{cases}
\end{align*}
 \end{proposition}
 \begin{proof}
    Let  $\tXstr = (\breve{X}_{\textnormal{str},1}^\top,\ldots,\breve{X}_{\textnormal{str},H}^\top)^\top \in \mathbb{R}^{\nstr\times (2H+k)}$ 
 with the $i$th row of $\breve{X}_{\textnormal{str},h}$ being
\begin{align*}
    (Z_{hi}\delta_{h1},\ldots,Z_{hi}\delta_{hH},(1-Z_{hi})\delta_{h1},\ldots,(1-Z_{hi})\delta_{hH},(\txhi)^\top).
\end{align*}
There exists a squared and invertible matrix $Q$ such that $\tXstr= \Xstr Q$. Therefore,
 \begin{align*}
 \tXstr\left( \tXstr^\top W \tXstr\right)^{-1} \tXstr^\top W = \Xstr\left( \Xstr^\top W \Xstr\right)^{-1} \Xstr^\top W.
 \end{align*}
 Note that 
 \begin{align*}
     &\tXstr^\top W \tXstr = \\
     & \left(
     \begin{array}{ccccccc}
        n_{11} w_1(1)  &  &  &&&&\\ 
                &\ddots&&&&&\\
        &&n_{H1} w_H(1)& &&&\\
          & & &n_{10} w_1(0)&&&\\ 
        &&&&\ddots&&\\
          &&&& &n_{H0} w_H(0) &\\
         &&&&&& \sumh\big\{\sumist w_{h}(1)\txhi\txhi^\top+\\
         &&&&&&\sumlsc w_{h}(0)\txhl\txhl^\top\big\}
     \end{array}
     \right).
 \end{align*}
Moreover,
\begin{align*}
   \frac{1}{\nstr} \sumh\bigg\{\sumist w_{h}(1)\txhi\txhi^\top+
   \sumlsc w_{h}(0)\txhl\txhl^\top\bigg\} = \sumh \pih \left\{\pht^{-1}\shxt+\phc^{-1}\shxc\right\} = \hvxxstr.
\end{align*}
Therefore,
\begin{align*}
    h_{hi,\textnormal{str}} = \begin{cases}
    \nht^{-1}+\txhi^\top\hvxxstr^{-1}\txhi w_h(1) \nstr^{-1}, \quad i\in \sth,\\
    \nhc^{-1}+\txhi^\top\hvxxstr^{-1}\txhi w_h(0)\nstr^{-1}, \quad i\in \sch.
    \end{cases}
\end{align*}
 \end{proof}

\begin{lemma}
\label{lemma:hvxxstr-Op}
Under Assumption~\ref{a:strata}, 
\begin{align*}
    \|\hvxxstr^{-1}\|_{\textnormal{op}} = \Op(1),\quad \|\hvxxstr^{-1}\|_\infty = \Op(1).
\end{align*}
\end{lemma}
\begin{proof}
Let 
\begin{align*}
    \hat{V}_1 = \sumh\pih \pht^{-1}\shxt,\quad V_1 =\sumh \pih \pht^{-1}\Shx, \\
    \hat{V}_0 = \sumh\pih \phc^{-1}\shxc,\quad  V_0=\sumh \pih \phc^{-1}\Shx.
\end{align*}
By Lemma~\ref{lemma:str-var-cov-consistency}, for $j=1,\ldots,k,\ j^\prime = 1,\ldots,k$,
\begin{align*}
    |[\hat{V_1}-V_1]_{(j,j^\prime)}|=\op(1),\quad  |[\hat{V_0}-V_0]_{(j,j^\prime)}|=\op(1).
\end{align*}
Therefore, 
\begin{align*}
  \|\hat{V_1}-V_1\|_\infty  = \max_{j,j^\prime}|[\hat{V_1}-V_1]_{(j,j^\prime)}|\leq 
  \sum_{j,j^\prime}|[\hat{V_1}-V_1]_{(j,j^\prime)}| = \op(1).
\end{align*}
Thus,
\begin{align*}
   \|\hat{V_1}-V_1\|_{\textnormal{op}}\leq \left[\operatorname{tr}\left\{(\hat{V_1}-V_1)^2\right\}\right]^{1/2} \leq  k\|\hat{V_1}-V_1\|_\infty = \op(1).
\end{align*}
Similarly, 
\begin{align*}
    \|\hat{V_0}-V_0\|_{\textnormal{op}} = \op(1).
\end{align*}
Thus, 
\begin{align*}
  \|\hvxxstr-\vxxstr\|_{\textnormal{op}} \leq \|\hat{V_1}-V_1\|_{\textnormal{op}} + \|\hat{V_0}-V_0\|_{\textnormal{op}} = \op(1).
\end{align*}
By Assumption~\ref{a:strata}, the limit of $\vxxstr$ is an invertible matrix. Let $\lambda_{\min}(\vxxstr)>0$ be the smallest  eigenvalue of $\vxxstr$ and there exists a constant $c$ such that $\lambda_{\min}(\vxxstr)>c$ for sufficiently large $\nstr$. By Weyl's inequality, with probability tending to one,
\begin{align*}
  \|\hvxxstr-\vxxstr\|_{\textnormal{op}}  < c/2 &\Longrightarrow \lambda_{\min}(\vxxstr)-\lambda_{\min}(\hvxxstr) < c/2\\&\Longrightarrow \lambda_{\min}(\hvxxstr) >\lambda_{\min}(\vxxstr)-\frac{c}{2} > \frac{c}{2}.
\end{align*}
Therefore, with probability tending to one,
\begin{align*}
    \|\hvxxstr^{-1}\|_{\operatorname{op}} =\lambda_{\min}(\hvxxstr)^{-1} <\frac{2}{c}.
\end{align*}
Thus,
    $\|\hvxxstr^{-1}\|_{\operatorname{op}} = \Op(1).$
Since $ \|\hvxxstr^{-1}\|_\infty \leq  \|\hvxxstr^{-1}\|_{\operatorname{op}}$, then $\|\hvxxstr^{-1}\|_{\infty} = \Op(1).$
\end{proof}

Define \begin{align*}
    g_{hi} = \begin{cases}
    \nht^{-1}, \quad i\in \sth,\\
    \nhc^{-1}, \quad i\in \sch.
    \end{cases}
\end{align*}

\begin{lemma}
  \label{lemma:str-leverage-maximum}
Under Assumption~\ref{a:strata}, 
\begin{align*}
     \max_{h,i}|h_{hi,\textnormal{str}} - g_{hi} | = \op(1).
\end{align*}
\end{lemma}
\begin{proof}
By Lemma~\ref{lemma:hvxxstr-Op}, 
\begin{align*}
    \max_{h,i}|h_{hi,\textnormal{str}} - g_{hi} | &\leq \|\hvxxstr^{-1}\|_{\operatorname{op}} \max_{h,z}|w_h(z)| \max_{h,i} \txhi^\top\txhi \nstr^{-1}
    \\& \leq \|\hvxxstr^{-1}\|_{\operatorname{op}} \max_{h,z}|w_h(z)| \max_{h,i} k \|x_{hi}-\bxh\|^2_\infty \nstr^{-1}\\
    &=\Op(1)O(1)o(\nstr)\nstr^{-1} = \op(1).
\end{align*}
\end{proof}

\subsection{Proof of Theorem \ref{thm:tyr-strata-consistensy}}

\begin{proof}
  Note that 
  \begin{align*}
    \nstr^{1/2}(\taustrtyr-\taustr) &= \nstr^{1/2}\{\htaustr-\taustr-(\betastr)^\top \tauxstr\}+\nstr^{1/2}(\betastr-\hbetastr)^\top\tauxstr
    \\ &=\nstr^{1/2}\{\htaustr-\taustr-(\betastr)^\top \tauxstr\}+\nstr^{1/2}\op(1) O_{\mathbb{P}}(\nstr^{-1/2}) \\
    &= \nstr^{1/2}\{\htaustr-\taustr-(\betastr)^\top \tauxstr\}+\op(1)\\
    &= \nstr^{1/2}\{\htaustr(\betastr)-\taustr\}+\op(1),
  \end{align*}
where the first equality is due to Proposition~\ref{prop:str-formula} and the second equality is due to Lemma~\ref{lemma:str-beta-consistency} and Proposition~\ref{propA:str-CLT}.
  
  By Proposition~\ref{propA:str-CLT} and the definition of $\betastr$, we have
  \begin{align*}
    n^{1/2}(\htaustr-\taustr-(\betastr)^\top \tauxstr) \mathrel{\dot{\sim}} N(0, \vttstr-\vtxstr\vxxstr^{-1}\vxtstr).
  \end{align*}
  Compounded with Slusky's theorem, the conclusion follows. 
\end{proof}

\subsection{Proof of Theorem~\ref{thm:HC-str-limit}}

\begin{proof}
We use the following formula of the variance estimator
\begin{align*}
    d^\top(E^\top WE)^{-1}E^\top W \Delta WE(E^\top WE)^{-1}d.
\end{align*}
Let $u_{hi} = x_{hi}-\bxh$.
Define $H$ by
\begin{align*}
    H=\left(\begin{array}{cc}
        H_{11} & H_{12} \\ 
       H_{21}  & H_{22}
    \end{array}\right)=E^\top W\Delta W E/\nstr,
\end{align*}
where
\begin{align*}
    H_{11} &= \nstr^{-1}\textnormal{diag}\left( w^2_1(1) \sum_{i \in \mathcal{S}_{11}} \tilde{e}^2_{1i},\ldots,w^2_H (1) \sum_{i \in \mathcal{S}_{H1}} \tilde{e}^2_{Hi},w^2_1(0) \sum_{i \in \mathcal{S}_{10}} \tilde{e}^2_{1i},\ldots,w_H^2 (0) \sum_{i \in \mathcal{S}_{H0}} \tilde{e}^2_{Hi}\right),\\
    H_{21}&=H_{12}^\top = \\
    &  \nstr^{-1}\left(w^2_1(1) \sum_{i \in \mathcal{S}_{11}} \tilde{e}^2_{1i}u_{1i} ,\ldots,w^2_H (1) \sum_{i \in \mathcal{S}_{H1}} \tilde{e}^2_{Hi}u_{Hi},w^2_1 (0) \sum_{i \in \mathcal{S}_{10}} \tilde{e}^2_{1i}u_{1i},\ldots,w^2_H (0)\sum_{i \in \mathcal{S}_{H0}} \tilde{e}^2_{Hi}u_{Hi}\right),\\
    H_{22} &= \nstr^{-1}\sumh \left\{w^2_h(1)\sumisth u_{hi}u_{hi}^\top\tehi^2+w^2_h(0)\sumlsch u_{hi} u_{hi}^\top\tehl^2\right\}.
\end{align*}

Define $G$ by
\begin{align*}
    &G=\left(\begin{array}{cc}
        G_{11} & G_{12} \\ 
       G_{21}  & G_{22}
    \end{array}\right)=E^\top W E/\nstr,
\end{align*}
where 
\begin{align*}
    G_{11} &= \nstr^{-1}\operatorname{diag}\left( n_{11}w_1(1),\ldots,n_{H1} w_H (1),n_{10}w_1(0),\ldots,n_{H0}w_H (0)\right),\\
    G_{21}&=G_{12}^\top = \nstr^{-1}\left(w_1(1) \sum_{i \in \mathcal{S}_{11}} u_{1i},\ldots, w_H (1) \sum_{i \in \mathcal{S}_{H1}} u_{Hi}, w_1(0)\sum_{i \in \mathcal{S}_{10}} u_{1i} , \hdots,  w_H (0) \sum_{i \in \mathcal{S}_{H0}} u_{Hi}\right),\\
    G_{22} &= \nstr^{-1}\sumh \left\{w_h(1)\sumisth u_{hi} u_{hi}^\top+w_h(0)\sumlsch u_{hi} u_{hi}^\top\right\}.
\end{align*}
Define $\Lambda$ by
\begin{align*}
     \Lambda=G^{-1}=\left(\begin{array}{cc}
        \Lambda_{11} & \Lambda_{12} \\ 
       \Lambda_{21}  & \Lambda_{22}
    \end{array}\right).
\end{align*}
By the formula of inverse of $2\times 2$ block matrix, we have
\begin{align*}
    \Lambda_{11} &= G_{11}^{-1}+ G_{11}^{-1} G_{12} (G_{22}-G_{21}G_{11}^{-1}G_{12})^{-1}G_{21}G_{11}^{-1},\\
    \Lambda_{21}^\top &= \Lambda_{12} = -G_{11}^{-1}G_{12}(G_{22}-G_{21}G_{11}^{-1}G_{12})^{-1}.
\end{align*}
Let $d_1 =(\pi_1,\ldots,\pi_H,-\pi_1,\ldots,-\pi_H), $ it is easy to see that 
\begin{align}
  \label{eq:vhwstr-formula}
    \nstr\vhwstrj = d_1^\top\left(\begin{array}{cc}
      \Lambda_{11}   &  \Lambda_{12}
    \end{array}\right)H\left(\begin{array}{c}
      \Lambda_{11} \\ 
      \Lambda_{21}
    \end{array}\right)d_1.
    \end{align}
To derive the formula of $\vhwstrj$, we calculate the following two quantities:
\begin{align*}
  \textnormal{ (i)}~ G_{11}^{-1}G_{12},\quad \textnormal{(ii)}~ G_{22}-G_{21}G_{11}^{-1}G_{12}.
\end{align*}

For (i), we have
\begin{align*}
  G_{11}^{-1}G_{12} = \left(\hat{\bar{u}}_{1}(1),\ldots,\hat{\bar{u}}_{H}(1),\hat{\bar{u}}_{1}(0),\ldots, \hat{\bar{u}}_{H}(0)\right)^\top.
\end{align*}
Denote $G_{11}^{-1}G_{12}$ by $\hat{U}$.

For (ii), we have
\begin{align*}
    G_{22}-G_{21}G_{11}^{-1}G_{12} &=\frac{1}{\nstr} \sumh\left\{ w_h(1)\sumisth u_{hi} u_{hi}^\top+w_h(0)\sumlsch u_{hi} u_{hi}^\top\right\}-\hat{U}^\top G_{12}\\
    &=\frac{1}{\nstr} \sumh\left\{ w_h(1)\sumisth u_{hi} u_{hi}^\top+w_h(0)\sumlsch u_{hi} u_{hi}^\top\right\}-\\ &\quad \quad \quad\frac{1}{\nstr}\sumh\left\{w_h(1)n_{h1}\bhuht\bhuht^\top+w_h(0)n_{h0}\bhuhc\bhuht^\top\right\}\\
    &=\frac{1}{\nstr} \sumh\left\{ w_h(1)\sumisth \txhi \txhi^\top+w_h(0)\sumlsch \txhl \txhl^\top\right\}\\
    &= \sumh \pih \left\{\pht^{-1}\shxt+\phc^{-1}\shxc\right\} = \hvxxstr.
\end{align*}

Expanding \eqref{eq:vhwstr-formula}, we have
\begin{align*}
   & d_1^\top\left(\begin{array}{cc}
      \Lambda_{11}   &  \Lambda_{12}
    \end{array}\right)H\left(\begin{array}{c}
      \Lambda_{11} \\ 
      \Lambda_{21}
    \end{array}\right)d_1\\&= d_1^\top\left(\begin{array}{cc}
      G_{11}^{-1}+\hat{U}\hvxxstr^{-1}\hat{U}^\top  &  -\hat{U}\hvxxstr^{-1} 
    \end{array}\right)H\left(\begin{array}{c}
      G_{11}^{-1}+\hat{U}\hvxxstr^{-1}\hat{U}^\top \\ 
      -\hvxxstr^{-1} \hat{U}^\top
    \end{array}\right)d_1\\
    &=d_1^\top  \left(G_{11}^{-1}+\hat{U}\hvxxstr^{-1}\hat{U}^\top\right) H_{11}  \left(G_{11}^{-1}+\hat{U}\hvxxstr^{-1}\hat{U}^\top\right) d_1 + d_1^\top \hat{U}\hvxxstr^{-1}  H_{22}  \hvxxstr^{-1}\hat{U}^\top  d_1\\
    &\qquad-2d_1^\top  \left(G_{11}^{-1}+\hat{U}\hvxxstr^{-1}\hat{U}^\top\right) H_{12}  \hvxxstr^{-1}\hat{U}^\top d_1.
\end{align*}
Let 
\begin{align*}
    T_1 &= d_1^\top  G_{11}^{-1} H_{11}  G_{11}^{-1} d_1,\quad T_2 = d_1^\top  G_{11}^{-1} H_{11}  \hat{U}\hvxxstr^{-1}\hat{U}^\top d_1,\\
    T_3 &= d_1^\top  \hat{U}\hvxxstr^{-1}\hat{U}^\top H_{11}  \hat{U}\hvxxstr^{-1}\hat{U}^\top d_1,\quad T_4=d_1^\top  G_{11}^{-1} H_{12}  \hvxxstr^{-1}\hat{U}^\top d_1,\\
    T_5 &= d_1^\top  \hat{U}\hvxxstr^{-1}\hat{U}^\top H_{12}  \hvxxstr^{-1}\hat{U}^\top  d_1,\quad  T_6 = d_1^\top  \hat{U}\hvxxstr^{-1} H_{22}  \hvxxstr^{-1}\hat{U}^\top  d_1.
\end{align*}
Next, we derive the formula related to $T_i$ $(i=1,\ldots,6)$. 
\begin{align*}
    T_1 &= d_1^\top  G_{11}^{-1} H_{11}  G_{11}^{-1} d_1\\
    &=\sumh\pih^2\sumisth\frac{1}{\nstr} \tehi^2 w^2_h(1)\left\{\frac{\nht}{\nstr}w_h(1)\right\}^{-2}+\sumh\pih^2\sumlsch\frac{1}{\nstr} \tehl^2 w^2_h(0)\left\{\frac{\nhc}{\nstr}w_h(0)\right\}^{-2}\\
    &=\sumh\pih^2\sumisth\nstr \tehi^2 \nht^{-2}+\sumh\pih^2\sumlsch\nstr \tehl^2 \nhc^{-2}\\
    &=\sumh\pih\left\{\left(\sumisth\tehi^2/\nht\right)\pht^{-1}+\left(\sumlsch\tehl^2/\nhc\right)\phc^{-1}\right\},\\
    d_1^\top  G_{11}^{-1} H_{11}\hat{U} &=\sumh \pih\left\{\frac{n_{h1}}{\nstr}w_h(1)\right\}^{-1}\sumisth w^2_h(1)\tehi^2\frac{1}{\nstr}\bhuht^\top \\
    & \quad -\sumh \pih\left\{\frac{n_{h0}}{\nstr}w_h(0)\right\}^{-1}\sumlsch w^2_h(0)\tehl^2\frac{1}{\nstr}\bhuhc^\top\\
    &=\sumh \pih\nht^{-1}\sumisth w_h(1)\tehi^2\bhuht^\top-\sumh \pih\nhc^{-1}\sumlsch w_h(0)\tehl^2\bhuhc^\top,\\
    \hat{U}^\top H_{11}  \hat{U} &= \sumh\left\{\bhuht\bhuht^\top\right\}\left\{\sumisth \tehi^2 w^2_h(1)\frac{1}{\nstr}\right\}+ \\
    & \quad \sumh\left\{\bhuhc\bhuhc^\top\right\}\left\{\sumlsch \tehl^2 w^2_h(0)\frac{1}{\nstr}\right\}, 
\end{align*}
\begin{align*}
    d_1^\top  G_{11}^{-1} H_{12} &= \sumh \pih \left\{\frac{\nht}{\nstr}w_h(1)\right\}^{-1}w^2_h(1)\sumisth \tehi^2 u_{hi}^{\top}\frac{1}{\nstr}\\
    & \quad - \sumh \pih \left\{\frac{\nhc}{\nstr}w_h(0)\right\}^{-1}w^2_h(0)\sumlsch \tehl^2 u_{hi}^{\top}\frac{1}{\nstr}\\
    &=\sumh \pih \nht^{-1}w_{h}(1)\sumisth \tehi^2 u_{hi}^{\top}-
    \sumh \pih \nhc^{-1}w_{h}(0)\sumlsch \tehl^2 u_{hi}^{\top}, \\
    \hat{U}^\top H_{12} &= \sumh \bhuht w^2_h(1)\sumisth \tehi^2 u_{hi}^{\top}\frac{1}{\nstr} +\sumh \bhuhc w^2_h(0)\sumlsch \tehl^2 u_{hi}^{\top}\frac{1}{\nstr}.
\end{align*}
Next, we prove that $T_1=\nstr\hvstr\{1+\op(1)\}$ and $T_i = \op(1)$ $(i=2,\ldots,6)$. 
Note that for $\textnormal{HC}_2$, $\eta_{hi}^2 = (1- h_{hi,\textnormal{str}})^{-1}$ and
by Lemma~\ref{lemma:str-leverage-maximum}, 
\begin{align*}
    \max_{h,i}|\tehi^2/\{\hehi^2(1-g_{hi})^{-1}\} = \max_{h,i}|\eta_{hi}^2(1-g_{hi})| = 1+\op(1).
\end{align*}
Therefore,
\begin{align*}
    T_1 =& \sumh\pih\left\{\sumisth\hehi^2(1-g_{hi})^{-1}/\nht\pht^{-1}+\sumlsch\hehl^2(1-g_{hi})^{-1}/\nhc\phc^{-1}\right\}\{1+\op(1)\}\\
    =& \sumh\pih\left\{(\nht-1)^{-1}\sumisth\hehi^2\pht^{-1}+(\nhc-1)^{-1}\sumlsch\hehl^2\phc^{-1}\right\}\{1+\op(1)\}\\
    =&\nstr\hvstr\{1+\op(1)\}.
\end{align*}
Note that $(1-g_{hi})^{-1} \leq 2$, $\max_{h,i}\eta_{hi}^2 = \Op(1)$, $\nstr\hvstr=\Op(1)$, $\max_{h,z}|p_{hz}^{-1}| = O(1)$, $\max_{h,i}\|u_{hi}\|_{\infty}^2= o(\nstr)$ , $\max_{h,z}|w_{h}(z)|\leq 2 \max_{h,z}|p_{hz}^{-2}| = O(1)$. 
Therefore,
we derive the following stochastic order for terms related to $T_i \ (i=2,\ldots,6)$,
\begin{align*}
   \| d_1^\top  G_{11}^{-1} H_{11}\hat{U}\|_{\infty}  &\leq \max_{h,i}\|u_{hi}\|_{\infty}\sumh\pih\left\{(\nht-1)^{-1}\sumisth\tehi^2\pht^{-2}+(\nhc-1)^{-1}\sumlsch\tehl^2\phc^{-2}\right\} \\
   &\leq  \max_{h,i}\|u_{hi}\|_{\infty}  \left|\sumh \pih\pht^{-2}\shet+\sumh \pih\phc^{-2}\shec\right|\max_{h,i}\eta^2_{hi}\\
   &\leq \max_{h,i}\|u_{hi}\|_{\infty} \max_{h,z}|p_{hz}^{-1}| \nstr\hvstr \max_{h,i}\eta^2_{hi}
   \\ &= o(\nstr^{1/2})O(1)\Op(1)\Op(1) = \op(\nstr^{1/2}),\\
      \| \hat{U}^\top H_{11}  \hat{U}\|_{\infty} \leq &k \max_{h,i}\|u_{hi}\|^2_{\infty} \bigg[ \sumh\bigg\{\sumisth \tehi^2 w^2_h(1)\frac{1}{\nstr}\bigg\}+\sumh\bigg\{\sumlsch \tehl^2 w^2_h(0)\frac{1}{\nstr}\bigg\} \bigg] \\
   \leq & k\max_{h,i}\|u_{hi}\|^2_{\infty}\max_{h,z}|w_{h}(z)| \left|\sumh \pih\pht^{-1}\shet+\sumh \pih\phc^{-1}\shec\right|\max_{h,i}\eta^2_{hi}\\
   =& o(\nstr)O(1)\Op(1)\Op(1) = \op(\nstr),\\
    \|d_1^\top  G_{11}^{-1} H_{12} \|_{\infty}
    \leq & \max_{h,i}\|u_{hi}\|_{\infty}\max_{h,z}|\phz^{-1}|\left|\sumh \pih\pht^{-1}\shet+\sumh \pih\phc^{-1}\shec\right|\max_{h,i}\eta^2_{hi}\\
    = & o(\nstr^{1/2})O(1)\Op(1)\Op(1)=\op(\nstr^{1/2}),\\
    \|\hat{U}^\top H_{12}\|_\infty  \leq & \max_{h,i}\|u_{hi}\|^2_{\infty} \max_{h,z}|w_{h}(z)| \left|\sumh \pih\pht^{-1}\shet+\sumh \pih\phc^{-1}\shec\right|\max_{h,i}\eta^2_{hi}\\
    =&o(\nstr)O(1)\Op(1)\Op(1)=\op(\nstr),\\
           \end{align*}
\begin{align*}   
    \|H_{22}\|_{\infty} \leq& \max_{h,i}\|u_{hi}\|^2_{\infty}\max_{h,z}|w_h(z)| \nstr^{-1}\sumh \left\{w_h(1)\sumisth \tehi^2+w_h(0)\sumlsch\tehl^2\right\} \\
    \leq& \max_{h,i}\|u_{hi}\|^2_{\infty}\max_{h,z}|w_h(z)|\left|\sumh \pih\pht^{-1}\shet+\sumh \pih\phc^{-1}\shec\right|\max_{h,i}\eta^2_{hi}\\
    =&o(\nstr)O(1)\Op(1)\Op(1) = \op(\nstr).
\end{align*}
Note that $d_1^\top \hat{U} = \tauxstr^\top$. Therefore,
\begin{align*}
    |T_2| =& |d_1^\top  G_{11}^{-1} H_{11}  \hat{U}\hvxxstr^{-1}\hat{U}^\top d_1|
    =|d_1^\top  G_{11}^{-1} H_{11}  \hat{U}\hvxxstr^{-1}\tauxstr| \\
    \leq& k^2\|\tauxstr\|_\infty \|d_1^\top  G_{11}^{-1} H_{11}  \hat{U}\|_\infty\|\hvxxstr^{-1}\|_\infty
     = \Op(\nstr^{-1/2})\op(\nstr^{1/2})\Op(1) = \op(1),\\
  |T_3| =& | d_1^\top  \hat{U}\hvxxstr^{-1}\hat{U}^\top H_{11}  \hat{U}\hvxxstr^{-1}\hat{U}^\top d_1| =| \tauxstr^\top\hvxxstr^{-1}\hat{U}^\top H_{11}  \hat{U}\hvxxstr^{-1}\tauxstr|\\
  \leq& k^2 \| \tauxstr\|_\infty\|\hvxxstr^{-1}\hat{U}^\top H_{11}  \hat{U}\hvxxstr^{-1}\|_\infty\|\tauxstr\|_\infty = \Op(\nstr^{-1/2})\op(\nstr)\Op(\nstr^{-1/2})=\op(1),\\
|T_4|=&|d_1^\top  G_{11}^{-1} H_{12}  \hvxxstr^{-1}\hat{U}^\top d_1| = |d_1^\top  G_{11}^{-1} H_{12}  \hvxxstr^{-1}\tauxstr|
 \leq k^2\|d_1^\top  G_{11}^{-1} H_{12} \|_\infty \|\hvxxstr^{-1}\|_\infty\|\tauxstr\|_\infty\\
 = & \op(\nstr^{1/2})\Op(1)\Op(\nstr^{-1/2}) = \op(1),\\
|T_5| =& |d_1^\top  \hat{U}\hvxxstr^{-1}\hat{U}^\top H_{12}  \hvxxstr^{-1}\hat{U}^\top  d_1| = |\tauxstr^\top\hvxxstr^{-1}\hat{U}^\top H_{12}  \hvxxstr^{-1}\tauxstr|\\
\leq & k^4\|\tauxstr\|_\infty\|\hvxxstr^{-1}\|_\infty\|\hat{U}^\top H_{12}\|_\infty\|  \hvxxstr^{-1}\|_\infty\|\tauxstr\|_\infty
\\=&\Op(\nstr^{-1/2})\Op(1)\op(\nstr)\Op(1)\Op(\nstr^{-1/2})=\op(1),\\\
|T_6| =& |d_1^\top  \hat{U}\hvxxstr^{-1} H_{22}  \hvxxstr^{-1}\hat{U}^\top  d_1|
=|\tauxstr\hvxxstr^{-1} H_{22}  \hvxxstr^{-1}\tauxstr|\\
\leq& 
k^4\|\tauxstr\|_\infty\|\hvxxstr^{-1}\|_\infty\| H_{22}\|_\infty\|  \hvxxstr^{-1}\|_\infty\|\tauxstr\|_\infty\\ =& \Op(\nstr^{-1/2})\Op(1)\op(\nstr)\Op(1)\Op(\nstr^{-1/2})=\op(1).  
\end{align*}
Thus,
\begin{align*}
    \nstr\hat{V}_{\textnormal{HC}2,\textnormal{str}} = \nstr\hvstr\{1+\op(1)\}.
\end{align*}
Combining with Lemma~\ref{lemma:strata-variance-estimator-consistent}, we complete the proof.

\end{proof}

\subsection{Proof for Remark~\ref{remark:anti-conservative-stratified}}
We give an example to show that $\vhwstrj$ for $j=0,1$ are anti-conservative.
 Similar to the proof of Theorem~\ref{thm:HC-str-limit}, we have, for $j=0,1$,
\begin{align*}
  \hat{V}_{\textnormal{HC}j,\textnormal{str}} =& \sumh\pih\left\{\left(\sumisth\tehi^2/\nht\right)\pht^{-1}+\left(\sumlsch\tehl^2/\nhc\right)\phc^{-1}\right\} + \op(1).\\
\end{align*}
 Therefore
 \begin{align*}
  \hat{V}_{\textnormal{HC}0,\textnormal{str}} =& \sumh\pih\left\{\left(\sumisth\hehi^2/\nht\right)\pht^{-1}+\left(\sumlsch\hehl^2/\nhc\right)\phc^{-1}\right\} + \op(1),\\
  \hat{V}_{\textnormal{HC}1,\textnormal{str}} =& \frac{\nstr}{\nstr-2H-k}\sumh\pih\left\{\left(\sumisth\hehi^2/\nht\right)\pht^{-1}+\left(\sumlsch\hehl^2/\nhc\right)\phc^{-1}\right\} + \op(1). 
 \end{align*}
Let $n_{h1}=3$ and $n_{h0}=2$ for $h=1,\ldots,H$. By Lemmas~\ref{lemma:str-var-cov-consistency} and \ref{lemma:str-beta-consistency}, we have
\begin{align*}
  \hat{V}_{\textnormal{HC}0,\textnormal{str}} =& \sumh\pih\left\{\left(\sumisth\hehi^2/3\right)\pht^{-1}+\left(\sumlsch\hehl^2/2\right)\phc^{-1}\right\} + \op(1)\\
  =&  2/3\sumh\left\{ \pi_h p_{h1}^{-1}\shet\right\} + 1/2\sumh\left\{\pi_h p_{h0}^{-1}\shec \right\} + \op(1) \\
  =&2/3\sumh \left\{\pi_h p_{h1}^{-1}S_{h1}^2(\betastr)\right\}+1/2\sumh\left\{\pih p^{-1}_{h0}S_{h0}^2(\betastr)\right\} + \op(1), \\
  \hat{V}_{\textnormal{HC}1,\textnormal{str}} =& \frac{5H}{3H-k}\sumh\pih\left\{\left(\sumisth\hehi^2/3\right)\pht^{-1}+\left(\sumlsch\hehl^2/2\right)\phc^{-1}\right\} + \op(1)\\
  =&  10/9\sumh \left\{\pi_h p_{h1}^{-1}\shet\right\} + 5/6\sumh\left\{\pi_h p_{h0}^{-1}\shec \right\} + \op(1)\\
  =&10/9\sumh \left\{\pi_h p_{h1}^{-1}S_{h1}^2(\betastr)\right\}+5/6\sumh\left\{\pi_h p^{-1}_{h0}S_{h0}^2(\betastr)\right\} + \op(1)
\end{align*}

Therefore, $\hat{V}_{\textnormal{HC}0,\textnormal{str}}$ is anti-conservative when
\begin{align*}
  1/3\sumh \left\{\pi_h p_{h1}^{-1}S_{h1}^2(\betastr)\right\}+1/2\sumh\left\{\pih p^{-1}_{h0}S_{h0}^2(\betastr)\right\}-\sumh \pi_h S_{h
  \tau}^2>0;
\end{align*}
$\hat{V}_{\textnormal{HC}1,\textnormal{str}}$ is anti-conservative when
\begin{align*}
  -1/9\sumh \left\{\pi_h p_{h1}^{-1}S_{h1}^2(\betastr)\right\}+1/6\sumh\left\{\pih p^{-1}_{h0}S_{h0}^2(\betastr)\right\}-\sumh \pi_h S_{h
  \tau}^2>0.
\end{align*}
\section{Proofs for the results under completely randomized survey experiments}
\label{sec:F}

\subsection{Preliminary results}

\begin{proposition}
  \label{prop:crs-formula}
  $ \taucrstyr = \htaucrs-\tauxcrs^\top\hbetacrs-\deltav^\top\hgammacrs,$ where
  \begin{align*}
    \begin{pmatrix}
      \hbetacrs\\
      \hgammacrs
    \end{pmatrix}= &
    \begin{pmatrix}
       p_1^{-1}(1-n_1^{-1})\sxt+ p_0^{-1}(1-n_0^{-1})\sxc &(1-n_1^{-1})\sxvt-(1-n_0^{-1})\sxvc\\
      (1-n_1^{-1})\svxt-(1-n_0^{-1})\svxc & (p_1-n^{-1})\svt+ (p_0-n^{-1})\svc
    \end{pmatrix}^{-1}\\
   & \begin{pmatrix}
      p_1^{-1}(1-n_1^{-1})\sxyt+ p_0^{-1}(1-n_0^{-1})\sxyc\\
        (1-n^{-1}_1)\svyt-(1-n_0^{-1})\svyc
    \end{pmatrix}.
  \end{align*}
\end{proposition}
\begin{proof}
  Recall the regression 
  \begin{align*}
    Y_i\stackrel{w_i}{\sim} 1+Z_i + x_i + (Z_i-p_0)(v_i-\bv),\quad \text{where}~w_i = p_1^{-2}Z_i+p_0^{-2}(1-Z_i).
  \end{align*}
  Let $\tvi = v_i-Z_i\bhvt-(1-Z_i)\bhvc$. Recall that $\mathcal{S}$ is the set of sampled units. By FWL theorem, 
  \begin{align*}
    \begin{pmatrix}
      \hbetacrs\\
      \hgammacrs
    \end{pmatrix}=
    \begin{pmatrix}
      \sumis w_i\txi\txi^\top &\sumis w_i(Z_i-p_0)\txi\tvi^\top\\
      \sumis w_i(Z_i-p_0)\tvi\txi^\top & \sumis w_i(Z_i-p_0)^2\tvi\tvi^\top
    \end{pmatrix}^{-1}\begin{pmatrix}
      \sumis w_i\txi\tyi\\
      \sumis w_i(Z_i-p_0)\tvi\tyi
    \end{pmatrix}.
  \end{align*}
Simple algebra gives that 
\begin{gather*}
  \sumis w_i\txi\txi^\top =p_1^{-2}(n_1-1)\sxt+ p_0^{-2}(n_0-1)\sxc, \\
  \sumis w_i\txi\tyi = p_1^{-2}(n_1-1)\sxyt+ p_0^{-2}(n_0-1)\sxyc,\\
  \sumis w_i(Z_i-p_0)^2\tvi\tvi^\top = (n_1-1)\svt+ (n_0-1)\svc,\\
  \sumis w_i(Z_i-p_0)\tvi\tyi = p_1^{-1}(n_1-1)\svyt-p_0^{-1}(n_0-1)\svyc,\\
  \sumis w_i(Z_i-p_0)\txi\tvi^\top = p_1^{-1}(n_1-1)\sxvt-p_0^{-1}(n_0-1)\sxvc.
\end{gather*}
Therefore,
\begin{align*}
  \begin{pmatrix}
    \hbetacrs\\
    \hgammacrs
  \end{pmatrix}= &
  \begin{pmatrix}
     p_1^{-1}(1-n_1^{-1})\sxt+ p_0^{-1}(1-n_0^{-1})\sxc &(1-n_1^{-1})\sxvt-(1-n_0^{-1})\sxvc\\
    (1-n_1^{-1})\svxt-(1-n_0^{-1})\svxc & (p_1-n^{-1})\svt+ (p_0-n^{-1})\svc
  \end{pmatrix}^{-1}\\
  & \begin{pmatrix}
    p_1^{-1}(1-n_1^{-1})\sxyt+ p_0^{-1}(1-n_0^{-1})\sxyc\\
      (1-n^{-1}_1)\svyt-(1-n_0^{-1})\svyc
  \end{pmatrix}.
\end{align*}
\end{proof}

The following lemma is from Lemma B16 in \cite{yang2021rejective}.
\begin{lemma}
  \label{lemma:crs-var-cov-consistency}
  Under Assumption~\ref{a:crs}, for $z=0,1$,
  \begin{align*}
    \syz-\Syz=\op(1),\quad \sxz-\Sx=\op(1),\quad \sxyz-\Sxyz = \op(1),\\
    \svz-\Sv = \op(1),\quad \svxz-\Svx = \op(1), \quad \svyz-\Svyz = \op(1).
  \end{align*}
\end{lemma}
\begin{lemma}
  \label{lemma:crs-beta-gamma-consistency}
  Under Assumption~\ref{a:crs},
  \begin{align*}
    \hbetacrs = \betacrs+\op(1),\quad
    \hgammacrs = \gammacrs+\op(1).
  \end{align*}
  \begin{proof}
    By Lemma~\ref{lemma:crs-var-cov-consistency}, we have
    \begin{align*}
      (1-n_1^{-1})\svxt-(1-n_0^{-1})\svxc =\op(1),\\
       \left\{p_1^{-1}(1-n_1^{-1})\sxt+ p_0^{-1}(1-n_0^{-1})\sxc\right\}-(p_1p_0)^{-1}\Sx = \op(1) \\
      \left\{(p_1-n^{-1})\svt+ (p_0-n^{-1})\svc\right\} - \Sv = \op(1),\\
      \quad p_1^{-1}(1-n_1^{-1})\sxyt+ p_0^{-1}(1-n_0^{-1})\sxyc -\left(p_1^{-1}\Sxyt+ p_0^{-1}\Sxyc\right)=\op(1),\\
      (1-n^{-1}_1)\svyt-(1-n_0^{-1})\svyc -\left(\Svyt-\Svyc\right) = \op(1).
    \end{align*}
    By Proposition~\ref{prop:crs-formula}, 
    \begin{align*}
      \begin{pmatrix}
        \hbetacrs\\
        \hgammacrs
      \end{pmatrix}-
      \begin{pmatrix}
         (p_1p_0)^{-1}\Sx &0\\
        0 & \Sv
      \end{pmatrix}^{-1}
      \begin{pmatrix}
        p_1^{-1}\Sxyt+ p_0^{-1}\Sxyc\\
          \Svyt-\Svyc
      \end{pmatrix}=\op(1).
    \end{align*}
    Recall the definition of $\betacrs$ and $\gammacrs$ and $\Svyt-\Svyc = S_{v\tau}$. The conclusion follows.
  \end{proof}
\end{lemma}

Proposition~\ref{propA:crs-CLT} below is from \cite{yang2021rejective}.
\begin{proposition}
  \label{propA:crs-CLT}
  Under Assumption~\ref{a:crs}, 
  $\sqrt{n}\left(\htaucrs-\tau_{\textnormal{crs}}, \tauxcrs^\top, \deltav^\top \right)^{\top}$ is asymptotically normal 
  with zero mean and covariance
  $$
  \left(\begin{array}{ccc}
  \vcrstt & \vcrstx & \vcrstv \\
  \vcrsxt & \vcrsxx  & \vcrsxv\\
  \vcrsvt & \vcrsvx & \vcrsvv
  \end{array}\right)=\left(\begin{array}{ccc}
  p_{1}^{-1} \Syt+p_{0}^{-1} \Syc-f S_{\tau}^{2} & p_{1}^{-1} \Syxt+p_{0}^{-1} \Syxc & (1-f) S_{v\tau} \\
  p_{1}^{-1} \Sxyt+p_{0}^{-1} \Sxyc & \left(p_{1} p_{0}\right)^{-1} \Sx & 0 \\
  (1-f) S_{v\tau} & 0& (1-f) \Sv
  \end{array}\right).
  $$
\end{proposition}

\subsection{Proof of Theorem~\ref{thm:tyr-crs-consistensy}}

\begin{proof}
  Note that 
  \begin{align*}
    & n^{1/2}(\taucrstyr-\taucrs ) \\
    &= n^{1/2}\{\htaucrs-\taucrs-(\betacrs)^\top \tauxcrs-(\gammacrs)^\top \deltav\}+n^{1/2}(\betacrs-\hbetacrs)^\top\tauxcrs + n^{1/2}(\gammacrs-\hgammacrs)^\top\deltav
    \\&=n^{1/2}\{\htaucrs-\taucrs-(\betacrs)^\top \tauxcrs-(\gammacrs)^\top \deltav\}+n^{1/2}\op(1) \Op(n^{-1/2})+n^{1/2}\op(1) \Op(n^{-1/2})
    \\&= n^{1/2}\{\htaucrs-\taucrs-(\betacrs)^\top \tauxcrs-(\gammacrs)^\top \deltav\}+\op(1),
  \end{align*}
where the first equality is due to Proposition~\ref{prop:crs-formula} and the second equality is due to Propositions~\ref{propA:crs-CLT} and \ref{lemma:crs-beta-gamma-consistency}.
By Proposition~\ref{propA:crs-CLT} and the definition of $\betacrs$ and $\gammacrs$, we have
  \begin{align*}
    n^{1/2}\{\htaucrs-(\betacrs)^\top \tauxcrs-(\gammacrs)^\top \deltav\} \mathrel{\dot{\sim}} N(0, \vcrstt-\vcrstx\vcrsxx^{-1}\vcrsxt-\vcrstv\vcrsvv^{-1}\vcrsvt).
  \end{align*}
  Compounded with Slusky's theorem, the conclusion follows.
\end{proof}

\subsection{A plug-in variance estimator}
With a slight abuse of notation, let $\hei$ be the residual of unit $i$ from the WLS regression~\eqref{formula:crs}. One of the variance estimators of $\taucrstyr$ can be derived by
\begin{equation}
  \label{eq:plug-in-variance-estimator-two-stage}
  \hvcrs = n^{-1}\left\{p_{1}^{-1}\set + p_{0}^{-1}\sec\right\},
\end{equation}
where
$$
\set=(n_{1}-1)^{-1} \sumist \hei^2,\quad \sec= (n_{0}-1)^{-1} \sumlsc \hel^2.
$$
Proposition~\ref{prop:hvcrs-consistent} below demonstrates the asymptotic conservativeness of $\hvcrs$.  
\begin{proposition}
  \label{prop:hvcrs-consistent}
Under Assumption~\ref{a:crs},
\begin{equation}
  \label{eq:hvcrs-limit}
  \hvcrs  =  n^{-1} \min_{\beta,\gamma} \left\{ p_{1}^{-1}S_{1}^2(\beta,\gamma)+p^{-1}_{0}S_{0}^2(\beta,\gamma)\right\} +o_{\mathbb{P}}(n^{-1}),
\end{equation}
where
$$
S^2_{z}(\beta,\gamma) = (N-1)^{-1}\sum_{i=1}^{N} \{Y_{i}(z)-\byz-(x_{i}-\bx)^\top \beta-(z-p_0) (v_i-\bv)^\top \gamma\}^2, \quad z=0,1.
$$
\end{proposition}

\begin{proof}
   By Lemmas~\ref{lemma:crs-var-cov-consistency} and \ref{lemma:crs-beta-gamma-consistency}, and similar to the proof of 
   Proposition~\ref{lemma:strata-variance-estimator-consistent}, we have 
   \begin{align}
    \label{eq:proof-crs-variance-consistency-1}
    n\hvcrs =  p_{1}^{-1}S_{1}^2(\betacrs,\gammacrs)+p^{-1}_{0}S_{0}^2(\betacrs,\gammacrs)+\op(1).
   \end{align}
   Next, we show that 
   \begin{align}
    \label{eq:proof-crs-variance-consistency-2}
    (\betacrs,\gammacrs)= \argmin_{(\beta,\gamma)} \big\{ p_{1}^{-1}S_{1}^2(\beta,\gamma)+p^{-1}_{0}S_{0}^2(\beta,\gamma) \big\}.
   \end{align}
   Note that $\var\{\htaucrs(\betacrs,\gammacrs)\}$ can be derived by replacing $Y_{i}(z)$ by the adjusted potential outcome 
  $Y_i(z;\betacrs,\gammacrs)$
  in the formula of $\vttstr$.
   The optimality of $(\betacrs,\gammacrs)$ implies that
   \begin{equation}
     \label{eq:beta-gamma-argmin}
     (\betacrs,\gammacrs)= \argmin_{\beta,\gamma} \big\{ p_{1}^{-1}S_{1}^2(\beta,\gamma)+p^{-1}_{0}S_{0}^2(\beta,\gamma)-fS^2_{\tau}(\gamma) \big\},
   \end{equation} 
  Since $(\betacrs,\gammacrs)$ does not depend on $f$, then \eqref{eq:beta-gamma-argmin} holds for any $f$. Let $f=0$, we have
  \begin{align*}
      (\betacrs,\gammacrs)= \argmin_{\beta,\gamma} \big\{ p_{1}^{-1}S_{1}^2(\beta,\gamma)+p^{-1}_{0}S_{0}^2(\beta,\gamma) \big\}.
  \end{align*}
  The conclusion follows from~\eqref{eq:proof-crs-variance-consistency-1} and \eqref{eq:proof-crs-variance-consistency-2}.
\end{proof}
\subsection{Leverage scores of ToM regression in completely randomized survey experiments}
Recall that $\txi = x_i-Z_i\bhxt-(1-Z_i)\bhxc$ and we similarly define $\tvi = v_i-Z_i\bhvt-(1-Z_i)\bhvc$. Define $\hvxvxv$ by
\begin{align*}
  \hvxvxv =& \begin{pmatrix}
    p_1^{-1}(1-n_1^{-1})\sxt+ p_0^{-1}(1-n_0^{-1})\sxc &(1-n_1^{-1})\sxvt-(1-n_0^{-1})\sxvc\\
   (1-n_1^{-1})\svxt-(1-n_0^{-1})\svxc & (p_1-n^{-1})\svt+ (p_0-n^{-1})\svc
 \end{pmatrix}\\ =& n^{-1}\begin{pmatrix}
  \sumis w_i\txi\txi^\top &\sumis w_i(Z_i-p_0)\txi\tvi^\top\\
  \sumis w_i(Z_i-p_0)\tvi\txi^\top & \sumis w_i(Z_i-p_0)^2\tvi\tvi^\top
\end{pmatrix}.
\end{align*}
We define the weights for treatment arm $z$ as $w(z) = p_z^{-2}$.
\begin{proposition}
  \label{prop:leverage-crs}
  The leverage score of ToM regression for unit $i$ under completely randomized survey experiments is 
  \begin{align*}
    h_{i,\textnormal{crs}} = \begin{cases}
    n_1^{-1}+(\txi^\top,p_1\tvi^\top)^\top\hvxvxv^{-1}(\txi^\top,p_1\tvi^\top) w(1) n^{-1}, \quad i\in \st,\\
    n_0^{-1}+(\txi^\top,-p_0\tvi^\top)^\top\hvxvxv^{-1}(\txi^\top,-p_0\tvi^\top) w(0)n^{-1}, \quad i\in \sc.
    \end{cases}.
  \end{align*}
\end{proposition}
\begin{proof}
  Let  $\tXcrs \in \mathbb{R}^{n\times (k_1+k_2+2)}$ 
with the $i$th row of $\tXcrs$ being
\begin{align*}
  (Z_i,1-Z_i,\txi^\top,(Z_i-p_0)\tvi^\top).
\end{align*}
There exists an invertible matrix $Q$ such that $\tXcrs = \Xcrs Q$. Therefore,
\begin{align*}
\Xcrs\left( \Xcrs^\top W \Xcrs\right)^{-1} \Xcrs^\top W = \tXcrs\left( \tXcrs^\top W \tXcrs\right)^{-1} \tXcrs^\top W.
\end{align*}
Note that 
\begin{align*}
   \tXcrs^\top W \tXcrs/n = \left(
   \begin{array}{ccc}
    p_1w(1)&&\\
    & p_0w(0)&\\
    && \hvxvxv
   \end{array}
   \right).
\end{align*}
Therefore,
\begin{align*}
  h_{i,\textnormal{crs}} = \begin{cases}
  n_1^{-1}+(\txi^\top,p_1\tvi^\top)^\top\hvxvxv^{-1}(\txi^\top,p_1\tvi^\top) w(1) n^{-1}, \quad i\in \st,\\
  n_0^{-1}+(\txi^\top,-p_0\tvi^\top)^\top\hvxvxv^{-1}(\txi^\top,-p_0\tvi^\top) w(0)n^{-1}, \quad i\in \sc.
  \end{cases}
\end{align*}
\end{proof}

\begin{lemma}
  \label{lemma:hvxvxv-op}
  Under Assumption~\ref{a:crs}, 
  \begin{align*}
    \|\hvxvxv^{-1}\|_\infty = \Op(1), \quad \|\hvxvxv^{-1}\|_{\textnormal{op}} = \Op(1).
  \end{align*}
\end{lemma}
The proof of Lemma~\ref{lemma:hvxvxv-op} is similar to that of Lemma~\ref{lemma:hvxxstr-Op}, so we omit it.
\begin{lemma}
  \label{lemma:crs-leverage-maximum}
Under Assumption~\ref{a:crs}, 
\begin{align*}
     \max_{i}h_{i,\textnormal{crs}} = \op(1).
\end{align*}
\end{lemma}
The proof of Lemma~\ref{lemma:crs-leverage-maximum} is similar to that of~\ref{lemma:str-leverage-maximum}, so we omit it.

\subsection{Proof of Theorem~\ref{prop:vhwcrs-consistent}}

  
Let $\hei$ be the residual of unit $i$. Let $\tei$ be the scaled residual with $\tei = \eta_i\hei$, where $\eta_i = 1$ for $\textnormal{HC}_0$,
$\eta_i = \{n/(n-k_1-k_2-2)\}^{1/2}$ for $\textnormal{HC}_1$, $\eta_i = (1-h_{i,\textnormal{crs}})^{-1/2}$ for $\textnormal{HC}_2$, and $\eta_i = (1-h_{i,\textnormal{crs}})^{-1}$ for $\textnormal{HC}_3$. 
The variance estimator $\textnormal{HC}_j$ $(j=0,1,2,3)$ derives as
\begin{align*}
  \xi_2^\top(\Xcrs^\top W\Xcrs)^{-1}\Xcrs^\top W \Delta W\Xcrs(\Xcrs^\top W\Xcrs)^{-1}\xi_2,
\end{align*}
where $\Xcrs \in\mathbb{R}^{n\times(2+k_1+k_2)}$ with the $i$th row being $(1, Z_i, x_i^\top, (Z_i-p_0)(v_i-\bv)^\top)$, $W$ is the diagonal matrix of $w_i$, and
$\Delta$ is the diagonal matrix of scaled residual squares $\tei^2$.

Motivated by the following equivalent regression
\begin{align*}
  Y_i \stackrel{w_i}{\sim} Z_i+(1-Z_i) + (x_i-\bx) + (Z_i-p_0)(v_i-\bv)
\end{align*}
An equivalent variance estimator derives as 
\begin{align*}
  d^\top(E^\top WE)^{-1}E^\top W \Delta WE(E^\top WE)^{-1}d,
\end{align*}
where $d=(1,-1,\mathbf{0}^\top_{k_1+k_2})^\top$, $E \in\mathbb{R}^{n\times(2+k_1+k_2)}$ with the $i$th row being $(Z_i,1-Z_i, (x_i-\bx)^\top, (Z_i-p_0)(v_i-\bv)^\top)$.
Note that $\bx$ is unknown, and therefore the regression is infeasible, but it is useful for proving Theorem~\ref{prop:vhwcrs-consistent}.

The proof of equivalence is similar to that in Section~\ref{sec:equivalent-formula}, so we omit it.
We will base our proof of Theorem~\ref{prop:vhwcrs-consistent} on this equivalent variance estimator.

\begin{proof}
  Let $u_i = x_{i}-\bx$ and $r_i = v_i-\bv$.
Define $H$ by
\begin{align*}
    H=\left(\begin{array}{cc}
        H_{11} & H_{12} \\ 
       H_{21}  & H_{22}
    \end{array}\right)=E^\top W \Delta WE/n,
\end{align*}
where
\begin{align*}
    H_{11} &= n^{-1}\textnormal{diag}\left( w^2(1)\sumist \tei^2,w^2(0)\sumlsc \tel^2\right),\\
    H_{21}&=H_{12}^\top = n^{-1}\begin{pmatrix}
      w^2(1)\sumist \tei^2 u_i & w^2(0)\sumlsc \tel^2  u_i\\
      w^2(1) p_1\sumist \tei^2 r_i & -w^2(0) p_0\sumlsc \tel^2 r_i
    \end{pmatrix},\\
    H_{22} &= n^{-1}\left(\begin{array}{c;{2pt/2pt}c}
      w^2(1)\sumist \tei^2 u_i u_i^\top+ & w^2(1)p_1\sumist \tei^2 u_i r_i^\top- \\
      w^2(0)\sumlsc\tel^2  u_i u_i^\top& w^2(0)p_0\sumlsc \tel^2 u_i r_i^\top
      \\ \hdashline[2pt/2pt]
      w^2(1)p_1\sumist \tei^2  r_i u_i^\top-  & w^2(1)p_1^2\sumist \tei^2r_i r_i^\top +   \\
      w^2(0)p_0\sumlsc \tel^2 r_i u_i^\top & w^2(0)p_0^2\sumlsc\tel^2 r_i r_i^\top
    \end{array}\right).
\end{align*}
Define $G$ by
\begin{align*}
    &G=\left(\begin{array}{cc}
        G_{11} & G_{12} \\ 
       G_{21}  & G_{22}
    \end{array}\right)=E^\top W E/n,
\end{align*}
where 
\begin{align*}
  G_{11} &= n^{-1}\textnormal{diag}\left( w(1)n_1,w(0)n_0\right),\quad
  G_{21}=G_{12}^\top = n^{-1}\begin{pmatrix}
    w(1)\sumist u_i & w(0)\sumlsc u_i\\
    w(1) p_1\sumist r_i & -w(0) p_0\sumlsc r_i
  \end{pmatrix},\\
  G_{22} &= n^{-1}\begin{pmatrix}
    w(1)\sumist u_i u_i^\top+w(0)\sumlsc  u_i u_i^\top & w(1)p_1\sumist u_i r_i^\top- w(0)p_0\sumlsc u_i r_i^\top\\
    w(1)p_1\sumist  r_i u_i^\top- w(0)p_0\sumlsc r_i u_i^\top & w(1)p_1^2\sumist r_i r_i^\top + w(0)p_0^2\sumlsc r_i r_i^\top  
  \end{pmatrix}.
\end{align*}
Define $\Lambda$ by
\begin{align*}
     \Lambda=G^{-1}=\left(\begin{array}{cc}
        \Lambda_{11} & \Lambda_{12} \\ 
       \Lambda_{21}  & \Lambda_{22}
    \end{array}\right).
\end{align*}
By the formula of inverse of $2\times 2$ block matrix, we have
\begin{align*}
    \Lambda_{11} &= G_{11}^{-1}+ G_{11}^{-1} G_{12} (G_{22}-G_{21}G_{11}^{-1}G_{12})^{-1}G_{21}G_{11}^{-1},\\ \Lambda_{21}^\top &= \Lambda_{12} = -G_{11}^{-1}G_{12}(G_{22}-G_{21}G_{11}^{-1}G_{12})^{-1}.
\end{align*}
Let $d_1 =(1,-1)^\top, $ it is easy to see that 
\begin{align}
  \label{eq:vhwcrs-formula}
    n \vhwcrsj= d_1^\top\left(\begin{array}{cc}
      \Lambda_{11}   &  \Lambda_{12}
    \end{array}\right)H\left(\begin{array}{c}
      \Lambda_{11} \\ 
      \Lambda_{21}
    \end{array}\right)d_1.
    \end{align}

    Recall that
    \begin{align*}
      \hvxvxv = \begin{pmatrix}
        p_1^{-1}(1-n_1^{-1})\sxt+ p_0^{-1}(1-n_0^{-1})\sxc &(1-n_1^{-1})\sxvt-(1-n_0^{-1})\sxvc\\
       (1-n_1^{-1})\svxt-(1-n_0^{-1})\svxc & (p_1-n^{-1})\svt+ (p_0-n^{-1})\svc
     \end{pmatrix}.
    \end{align*}
    After some calculation, we have
    \begin{align*}
      \textnormal{ (i)}~ G_{11}^{-1}G_{12} = \hat{U},\quad \textnormal{(ii)}~ G_{22}-G_{21}G_{11}^{-1}G_{12} = \hvxvxv,
    \end{align*}
    where 
    \begin{gather*}
    \hat{U} = \begin{pmatrix}
      \bhut^\top& \bhrt^\top p_1\\
      \bhuc^\top& -\bhrc^\top p_0
    \end{pmatrix}.
    \end{gather*}

    We expand equation~\eqref{eq:vhwcrs-formula} as follows:
    \begin{align*}
       & d_1^\top\left(\begin{array}{cc}
          \Lambda_{11}   &  \Lambda_{12}
        \end{array}\right)H\left(\begin{array}{c}
          \Lambda_{11} \\ 
          \Lambda_{21}
        \end{array}\right)d_1\\&= d_1^\top\left(\begin{array}{cc}
          G_{11}^{-1}+\hat{U}\hvxvxv^{-1}\hat{U}^\top   & -\hat{U}\hvxvxv^{-1} 
        \end{array}\right)H\left(\begin{array}{c}
          G_{11}^{-1}+\hat{U}\hvxvxv^{-1}\hat{U}^\top \\ 
          -\hvxvxv^{-1} \hat{U}^\top
        \end{array}\right)d_1\\
        &=d_1^\top  \left(G_{11}^{-1}+\hat{U}\hvxvxv^{-1}\hat{U}^\top\right) H_{11}  \left(G_{11}^{-1}+\hat{U}\hvxvxv^{-1}\hat{U}^\top\right) d_1 + d_1^\top \hat{U}\hvxvxv^{-1}  H_{22}  \hvxvxv^{-1}\hat{U}^\top  d_1\\
        &\qquad-2d_1^\top  \left(G_{11}^{-1}+\hat{U}\hvxvxv^{-1}\hat{U}^\top\right) H_{12}  \hvxvxv^{-1}\hat{U}^\top d_1.
    \end{align*}
    Let 
    \begin{align*}
        T_1 &= d_1^\top  G_{11}^{-1} H_{11}  G_{11}^{-1} d_1,\quad T_2 = d_1^\top  G_{11}^{-1} H_{11}  \hat{U}\hvxvxv^{-1}\hat{U}^\top d_1,\\ 
        T_3 &= d_1^\top  \hat{U}\hvxvxv^{-1}\hat{U}^\top H_{11}  \hat{U}\hvxvxv^{-1}\hat{U}^\top d_1, \quad
    T_4=d_1^\top  G_{11}^{-1} H_{12}  \hvxvxv^{-1}\hat{U}^\top d_1,\\ 
    T_5 &= d_1^\top  \hat{U}\hvxvxv^{-1}\hat{U}^\top H_{12}  \hvxvxv^{-1}\hat{U}^\top  d_1,\quad  T_6 = d_1^\top  \hat{U}\hvxvxv^{-1} H_{22}  \hvxvxv^{-1}\hat{U}^\top  d_1.
    \end{align*}
    Then, 
    \begin{align*}
        T_1 &= d_1^\top  G_{11}^{-1} H_{11}  G_{11}^{-1} d_1\\
        &=\sumist n^{-1} \tei^2 w^2(1)\left\{ \frac{n_1}{n}w(1)\right\}^{-2}+\sumlsc n^{-1} \tel^2 w^2(0)\left\{\frac{n_0}{n}w(0)\right\}^{-2}\\
        &=\sumist n \tei^2 n_1^{-2}+\sumlsc n \tel^2 n_0^{-2}\\
        &=\left(\sumist\tei^2/n_1\right)p_1^{-1}+\left(\sumlsc\tel^2/n_0\right)p_0^{-1},\\
      d_1^\top  G_{11}^{-1} H_{11}\hat{U} &= \left\{\frac{n_{1}}{n}w(1)\right\}^{-1}n^{-1}\sumist w^2(1)\tei^2 \begin{pmatrix}\bhut\\ p_1\bhrt\end{pmatrix}^\top- \\
      & \quad \left\{\frac{n_{0}}{n}w(0)\right\}^{-1}n^{-1}\sumlsc w^2(0)\tel^2 \begin{pmatrix}\bhuc\\ -p_0\bhrc\end{pmatrix}^\top\\
      &=n_1^{-1}\sumist w(1)\tei^2\begin{pmatrix}\bhut \\ p_1\bhrt \end{pmatrix}^\top-n_0^{-1}\sumlsc w(0)\tel^2\begin{pmatrix}\bhuc\\ -p_0\bhrc\end{pmatrix}^\top\\
      &=n_1^{-1}\sumist p_1^{-2}\tei^2\begin{pmatrix}\bhut\\ p_1\bhrt\end{pmatrix}^\top-n_0^{-1}\sumlsc p_0^{-2}\tel^2\begin{pmatrix}\bhuc\\ -p_0\bhrc\end{pmatrix}^\top,\\
       \hat{U}^\top H_{11}  \hat{U} =& \begin{pmatrix}
        \bhut\\
        p_1\bhrt
      \end{pmatrix}\begin{pmatrix}
        \bhut\\
        p_1\bhrt
      \end{pmatrix}^\top\left\{n^{-1}\sumist \tei^2 w^2(1)\right\}+ \\
      & \quad \begin{pmatrix}
        \bhuc\\
        -p_0\bhrc
      \end{pmatrix}\begin{pmatrix}
        \bhuc\\
        -p_0\bhrc
      \end{pmatrix}^\top \left\{n^{-1}\sumlsc \tel^2 w^2(0)\right\},
      \end{align*}
      \begin{align*}
      d_1^\top  G_{11}^{-1} H_{12} =&  \left\{\frac{n_1}{n }w(1)\right\}^{-1}w^2(1)\sumist n^{-1}\tei^2 
      \begin{pmatrix}
         u_i\\
        p_1r_i
      \end{pmatrix}^{\top}- \left\{\frac{n_0}{n }w(0)\right\}^{-1}w^2(0)\sumlsc n^{-1}\tel^2 
      \begin{pmatrix}
         u_i\\
        -p_0r_i
      \end{pmatrix}^{\top}\\
      =& n_1^{-1}w(1)\sumist \tei^2 
      \begin{pmatrix}
         u_i\\
        p_1r_i
      \end{pmatrix}^{\top}-n_0^{-1}w(0)\sumlsc \tel^2       
      \begin{pmatrix}
         u_i\\
        -p_0r_i
      \end{pmatrix}^{\top}\\
      =& p_1^{-2}n_1^{-1}\sumist \tei^2       
      \begin{pmatrix}
         u_i\\
        p_1r_i
      \end{pmatrix}^{\top}- p_0^{-2}n_0^{-1}\sumlsc \tel^2       
      \begin{pmatrix}
         u_i\\
        -p_0r_i
      \end{pmatrix}^{\top},\\
      \hat{U}^\top H_{12} =& \begin{pmatrix}
        \bhut\\
        p_1\bhrt
      \end{pmatrix} w^2(1)\sumist \tei^2 
      \begin{pmatrix}
         u_i\\
        p_1r_i
      \end{pmatrix}^{\top}n^{-1} +\begin{pmatrix}
        \bhuc\\
        -p_0\bhrc
      \end{pmatrix} w^2(0)\sumlsc \tel^2 
      \begin{pmatrix}
         u_i\\
        -p_0r_i
      \end{pmatrix}^{\top}n^{-1}.
  \end{align*}    
  Note that $\{n/(n-2-k_1-k_2)\}^{1/2} = 1+\op(1)$, by Lemma~\ref{lemma:crs-leverage-maximum}, 
$$
\max_i (1-h_{i,\textnormal{crs}})^{-1/2} = 1+\op(1),\quad \max_i (1-h_{i,\textnormal{crs}})^{-1} = 1+\op(1).
$$
Therefore, $\max_i \eta_i^2 = 1+\op(1)$ for $\textnormal{HC}_j$ $(j=0,1,2,3)$.
Moreover, $\max_{z}n_z/(n_z-1) = 1+o(1)$ and by Proposition~\ref{prop:hvcrs-consistent}, $n\hvcrs = p_1^{-1}\set+p_0^{-1}\sec=\Op(1)$. Therefore, 
  \begin{align*}
      T_1 =& \left\{\sumist\hei^2/n_1 p_1^{-1}+\sumlsc\hel^2/n_0p_0^{-1}\right\} \{ 1+\op(1) \}\\
      =& \left\{(n_1-1)^{-1}\sumist\hei^2 p_1^{-1}+(n_0-1)^{-1}\sumlsc\hel^2 p_0^{-1}\right\}\{ 1+\op(1) \}\\
      =&p_1^{-1}\set +p_0^{-1}\sec +\op(1) = n \hvcrs+\op(1).
  \end{align*}

  Note that $\max_{z}|p_{z}^{-1}| = O(1)$, $\max_{i}\|u_{i}\|_{\infty}^2= o(n)$, $\max_{i}\|r_{i}\|_{\infty}^2= o(n)$, $\max_{z}|w(z)|\leq 2 \max_{z}|p_{z}^{-2}| = O(1)$. Therefore,

  \begin{align*}
     \| d_1^\top  G_{11}^{-1} H_{11}\hat{U}\|_{\infty}  &\leq  \max\left\{\left\|\bhut\right\|_{\infty},\left\|p_1\bhrt\right\|_{\infty}, \left\|\bhuc\right\|_{\infty}, \left\|p_0\bhrc\right\|_{\infty}\right\} \left(n_1^{-1}\sumist p_1^{-2}\tei^2+n_0^{-1}\sumlsc p_0^{-2}\tel^2\right)\\
     &\leq \max\left\{\max_{i}\|u_{i}\|_{\infty}, \max_{i}\|r_{i}\|_{\infty}\right\} \max_{z}p_{z}^{-1} n\hvcrs  \max_i \eta_i^2
     \\ &= o(n^{1/2})O(1)\Op(1)\Op(1) = \op( n^{1/2}),\\
     \| \hat{U}^\top H_{11}  \hat{U}\|_{\infty} \leq &\max\left\{\left\|\bhut\right\|_{\infty},\left\|p_1\bhrt\right\|_{\infty}, \left\|\bhuc\right\|_{\infty}, \left\|p_0\bhrc\right\|_{\infty}\right\}^2 n^{-1}\left(\sumist \tei^2 w^2(1)+\sumlsc \tel^2 w^2(0)\right)\\
     \leq&\max\left\{\max_{i}\|u_{i}\|_{\infty}^2, \max_{i}\|r_{i}\|_{\infty}^2\right\}\max_{z}|w(z)| n\hvcrs\max_i \eta_i^2\\
     =& o( n)O(1)\Op(1)\Op(1) = \op( n),\\
      \|d_1^\top  G_{11}^{-1} H_{12} \|_{\infty}
      \leq & \max\left\{\max_{i}\|u_{i}\|_{\infty}, \max_{i}\|r_{i}\|_{\infty}\right\}\max_{z}|p_z^{-1}|n\hvcrs\max_i \eta_i^2\\
      =&o( n^{1/2})O(1)\Op(1)\Op(1)=\op( n^{1/2}),\\
      \|\hat{U}^\top H_{12}\|_\infty \leq & \max\left\{\max_{i}\|u_{i}\|_{\infty}^2, \max_{i}\|r_{i}\|_{\infty}^2\right\}  \max_{z}|w(z)| n\hvcrs\max_i \eta_i^2\\
      =&o( n)O(1)\Op(1)\Op(1)=\op( n),\\
      \|H_{22}\|_{\infty} \leq& \max\left\{\max_{i}\|u_{i}\|^2_{\infty}, \max_{i}\|r_{i}\|^2_{\infty}\right\}   n\hvcrs\max_i \eta_i^2   \\
      =&o( n)\Op(1)\Op(1) = \op( n).
  \end{align*}
  
  \newpage
  Note that $d_1^\top \hat{U} = (\tauxcrs,\deltav)$. Therefore,
  \begin{align*}
      |T_2| =& |d_1^\top  G_{11}^{-1} H_{11}  \hat{U}\hvxvxv^{-1}\hat{U}^\top d_1| 
      \leq (k_1+k_2)^2\max\{\|\tauxcrs\|_\infty,\|\deltav\|_\infty\} \|d_1^\top  G_{11}^{-1} H_{11}  \hat{U}\|_\infty\|\hvxvxv^{-1}\|_\infty
      \\ =& \Op( n^{-1/2})\op( n^{1/2})\Op(1) = \op(1),\\
    |T_3| =& | d_1^\top  \hat{U}\hvxvxv^{-1}\hat{U}^\top H_{11}  \hat{U}\hvxvxv^{-1}\hat{U}^\top d_1| \\
    \leq & (k_1+k_2)^2 \max\{\|\tauxcrs\|_\infty,\|\deltav\|_\infty\}\|\hvxvxv^{-1}\hat{U}^\top H_{11}  \hat{U}\hvxvxv^{-1}\|_\infty\max\{\|\tauxcrs\|_\infty,\|\deltav\|_\infty\}\\ =& \Op( n^{-1/2})\op( n)\Op( n^{-1/2})=\op(1), \\
  |T_4|=&|d_1^\top  G_{11}^{-1} H_{12}  \hvxvxv^{-1}\hat{U}^\top d_1| 
   \leq (k_1+k_2)^2\|d_1^\top  G_{11}^{-1} H_{12} \|_\infty \|\hvxvxv^{-1}\|_\infty\max\{\|\tauxcrs\|_\infty,\|\deltav\|_\infty\}\\
   = & \op( n^{1/2})\Op(1)\Op( n^{-1/2}) = \op(1),\\
  |T_5| =& |d_1^\top  \hat{U}\hvxvxv^{-1}\hat{U}^\top H_{12}  \hvxvxv^{-1}\hat{U}^\top  d_1| \\
  \leq & (k_1+k_2)^4\max\{\|\tauxcrs\|_\infty,\|\deltav\|_\infty\}\|\hvxvxv^{-1}\|_\infty\|\hat{U}^\top H_{12}\|_\infty\|  \hvxvxv^{-1}\|_\infty\max\{\|\tauxcrs\|_\infty,\|\deltav\|_\infty\}
  \\=&\Op( n^{-1/2})\Op(1)\op( n)\Op(1)\Op( n^{-1/2})=\op(1),\\\
  |T_6| =& |d_1^\top  \hat{U}\hvxvxv^{-1} H_{22}  \hvxvxv^{-1}\hat{U}^\top  d_1|
  \\
  \leq& 
  (k_1+k_2)^4\max\{\|\tauxcrs\|_\infty,\|\deltav\|_\infty\}\|\hvxvxv^{-1}\|_\infty\| H_{22}\|_\infty\|  \hvxvxv^{-1}\|_\infty\max\{\|\tauxcrs\|_\infty,\|\deltav\|_\infty\}\\ =& \Op( n^{-1/2})\Op(1)\op( n)\Op(1)\Op( n^{-1/2})=\op(1).  
  \end{align*}
  Therefore,
  \begin{align*}
       n\hat{V}_{\textnormal{HC}j,\textnormal{crs}} =    n\hvcrs+\op(1), \quad j=0,1,2,3.
  \end{align*}
  Hence, combining with Lemma~\ref{prop:hvcrs-consistent}, we complete the proof. 

\end{proof}

\end{document}